\documentstyle[prb,aps,twocolumn]{revtex}

\begin{document}
\input{epsf.tex}
\draft
\wideabs{
\title{Thermodynamics of spin $\bbox{S = 1/2}$ antiferromagnetic uniform\protect\\
and alternating-exchange Heisenberg chains}
\author{D. C. Johnston}
\address{Ames Laboratory and Department of Physics and Astronomy, Iowa
State University, Ames, Iowa 50011}
\author{R. K. Kremer}
\address{Max-Planck-Institut f\"ur Festk\"orperforschung,
Heisenbergstrasse 1, Postfach 800665, D-70569 Stuttgart, Germany}
\author{M. Troyer}
\address{Institute for Solid State Physics, University of Tokyo, Roppongi
7-22-1, Tokyo 106, Japan\protect\\and Theoretische Physik,
Eidgen\"ossische Technische Hochschule-Z\"urich, CH-8093 Z\"urich,
Switzerland}
\author{X. Wang}
\address{Institut Romand de Recherche Num\'erique en Physique des
Materiaux, IN-Ecublens, CH-1015 Lausanne, Switzerland}
\author{A. Kl\"umper}
\address{Universit\"at zu K\"oln, Institut f\"ur Theoretische Physik,
Z\"ulpicher Strasse 77, D-50937, Germany}
\author{S. L. Bud'ko, A. F. Panchula,\protect\cite{Panchula} and
P. C. Canfield}
\address{Ames Laboratory and Department of Physics and Astronomy, Iowa
State University, Ames, Iowa 50011}
\date{To be published in Physical Review B}
\maketitle
\begin{abstract}\hglue0.15in The magnetic susceptibility $\chi^*(t)$ and
specific heat $C(t)$ versus temperature $t$ of the spin $S = 1/2$
antiferromagnetic (AF) alternating-exchange ($J_1$ and $J_2$) Heisenberg
chain are studied for the entire range $0 \leq \alpha \leq 1$ of the
alternation parameter $\alpha\equiv J_2/J_1$ ($J_1,\ J_2\geq 0,\ J_2 \leq J_1$,
$t=k_{\rm B}T/J_1$, $\chi^* = \chi J_1/Ng^2\mu_B^2$).  For the uniform chain ($\alpha
= 1$), the high-accuracy $\chi^*(t)$ and $C(t)$ Bethe ansatz data of Kl\"umper and
Johnston (unpublished) are shown to agree very well at low $t$ with the respective
exact theoretical low-$t$ logarithmic correction predictions of Lukyanov, Nucl.\
Phys.\ B {\bf 522}, 533 (1998).  Accurate ($\sim 10^{-7}$) independent empirical fits
to the respective data are obtained over $t$ ranges spanning 25 orders of magnitude,
$5\times 10^{-25}\leq t\leq 5$, which contain extrapolations to the respective exact
$t=0$ limits.  The infinite temperature entropy calculated using our $C(t)$ fit
function is within 8 parts in $10^8$ of the exact value $\ln 2$.  Quantum Monte Carlo
(QMC) simulations and transfer-matrix density-matrix renormalization group (TMRG)
calculations of $\chi^*(\alpha,t)$ are presented for $0.002 \leq t \leq 10$ and $0.05
\leq \alpha \leq 1$, and an accurate ($2\times 10^{-4}$) two-dimensional $(\alpha,t)$
fit to the combined data is obtained for $0.01\leq t\leq 10$ and $0\leq\alpha\leq1$. 
From the low-$t$ TMRG data, the spin gap $\Delta(\alpha)$ is extracted for
$0.8\leq\alpha\leq 0.995$ and compared with previous results, and  a fit function is
formulated for $0\leq\alpha\leq 1$ by combining these data with literature data.  We
infer from our data that the asymptotic critical regime near the uniform chain limit
is only entered for $\alpha \protect\gtrsim 0.99$.  We examine in detail the
theoretical predictions  of  Bulaevskii, Sov.\ Phys.\ Solid State {\bf 11}, 921
(1969), for $\chi^*(\alpha,t)$ and compare them with our results.  To illustrate the
application and utility of our theoretical results, we  model our experimental
$\chi(T)$ and specific heat $C_{\rm p}(T)$ data for NaV$_2$O$_5$ single crystals in
detail.  The $\chi(T)$ data above the spin dimerization temperature $T_{\rm c} \approx
34$\,K are not in quantitative agreement with the prediction for the $S = 1/2$
uniform Heisenberg chain, but can be explained if there is a moderate ferromagnetic
interchain coupling and/or if $J$ changes with $T$.  Fitting the $\chi(T)$ data using
our $\chi^*(\alpha,t)$ fit function, we obtain the sample-dependent spin gap and
range $\Delta(T = 0)/k_{\rm B} = 103(2)$\,K, alternation parameter $\delta(0)\equiv
(1-\alpha)/(1+\alpha) =0.034(6)$ and average exchange constant $J(0)/k_{\rm B} =
640(80)$\,K\@.  The $\delta(T)$ and $\Delta(T)$ are derived from the data.  A spin
pseudogap with magnitude $\approx 0.4\,\Delta(0)$ is consistently found just above
$T_{\rm c}$, which decreases with increasing temperature.  From our $C_{\rm p}(T)$
measurements on two crystals, we infer that the magnetic specific heat at low
temperatures $T\lesssim 15$\,K is too small to be resolved experimentally, and that
the spin entropy at $T_{\rm c}$ is too small to account for the entropy of the
transition.  A quantitative analysis indicates that at $T_{\rm c}$, at least 77\,\% of
the entropy change due to the transition at $T_{\rm c}$ and associated order parameter
fluctuations arise from the lattice and/or charge degrees of freedom and less
than 23\,\% from the spin degrees of freedom.
\end{abstract}
}

\section{Introduction}

An antiferromagnetic alternating-exchange Heisenberg chain is one in which
nearest-neighbor spins in the chain interact via a Heisenberg interaction, but
with two antiferromagnetic (AF) exchange constants $J_2\leq J_1,\ J_1,J_2 \geq 0$
which alternate from bond to bond along the chain; the alternation parameter is
$\alpha\equiv J_2/J_1$.  Here we will be concerned with the magnetic susceptibility
$\chi$ and specific heat $C$ versus temperature $T$ of alternating-exchange chains
consisting of spins $S = 1/2$.  The uniform AF Heisenberg chain is one limit of the
alternating chain in which the two exchange constants are equal ($\alpha = 1,\ 
J_1=J_2\equiv J$).  At the other limit is the isolated dimer in which one of the
exchange constants is zero ($\alpha = 0$).  The present work is a combined
theoretical and experimental study of $\chi(T)$ and $C(T)$ of the $S=1/2$ AF
alternating-exchange chain over the entire range $0\leq\alpha\leq 1$ of the
alternation parameter, with the emphasis on the regime $\alpha\lesssim 1$ at and
close to the uniform chain limit.  This latter regime is relevant for compounds
showing second order spin dimerization transitions with decreasing $T$.  The present
work was originally motivated by our desire to accurately extract the temperature
dependent energy gap $\Delta(T)$ for magnetic excitations, the ``spin gap'', from
experimental $\chi(T)$ data for the $S = 1/2$ chain/two-leg ladder compound ${\rm
NaV_2O_5}$ below its spin dimerization temperature $T_{\rm c}\approx 34$\,K\@.  We
found that existing theory for the alternating-exchange chain was insufficient to
accomplish this goal.  In the present work we critically examine the predictions of
previous theory, perform the required additional theoretical calculations, and then
apply the results to extract $\Delta(T)$ at $T\lesssim T_{\rm c}$ from our $\chi(T)$
data for ${\rm NaV_2O_5}$ single crystals.  We have extended the original goal
so that we also include theoretical and experimental studies of $C(T)$ and how this
quantity relates to $\chi(T)$.  In the remainder of this introduction we briefly
review the prior theoretical results pertaining to $\chi(T)$ and $C(T)$ of the
uniform and alternating-exchange chain to place our work in the proper context.  We
then review the experimental and theoretical background on ${\rm NaV_2O_5}$ and
describe the plan for the rest of the paper.

\subsection{Theory}
\vglue0.11in
The $\chi(T)$ and $C(T)$ of both limits of the $S = 1/2$ AF alternating-exchange
Heisenberg chain are known exactly.  For the dimer, the $\chi(T)$ is given by the
exact  Eq.~(\ref{EqChiDimer:a}) below and the exact $C(T)$ is also easily
calculated.  The $\chi(T)$ and $C(T)$ of the uniform chain for
$T\gtrsim 0.4 J/k_{\rm B}$ ($k_{\rm B}$ is Boltzmann's constant) were estimated from 
calculations for chains with $\leq 11$ spins by Bonner and Fisher in
1964;\cite{Bonner1964} they extended their results by extrapolating to $T = 0$, and
in the case of $\chi(T)$ to the exact $T = 0$ value.\cite{Griffiths1964}  The exact
solution for $\chi(T)$ of the uniform chain was obtained using the Bethe ansatz in
1994 by Eggert, Affleck and Takahashi, and compared with their low-$T$ results from
conformal field theory.\cite{Eggert1994}  They found, remarkably, that $\chi(T\to
0)$ has infinite slope.  Their numerical $\chi(T)$ values are up to $\sim 10$\%
larger than the Bonner-Fisher extrapolation for $T\lesssim 0.25 J/k_{\rm B}$ (for a
comparison of the two predictions, see Fig.~8.1 in Ref.~\onlinecite{Johnston1997}). 
Their conformal field theory calculations showed that the leading order correction
to the zero temperature limit is of the form $\chi(T) = \chi(0)\{1 +
1/[2\ln(T_0/T)]\}$, where the value of the scaling temperature $T_0$ is not predicted
by the field theory.  Such log terms are called ``logarithmic corrections'' in the
literature.  One of us recently presented numerical Bethe ansatz calculations of
$\chi(T)$ with a higher absolute accuracy for $\chi(T)$
estimated to be $1\times 10^{-7}$,\cite{Klumper1998} and showed that the data are
consistent with the above field theory prediction, with an additional higher order
logarithmic correction, over the temperature range $5\times 10^{-25}\leq k_{\rm
B}T/J\lesssim 10^{-3}$.  Corresponding $C(T)$ calculations were also carried
out, and logarithmic corrections were studied for this quantity as
well.\cite{Klumper1998}  Lukyanov has recently presented an exact theory for
$\chi(T)$ and $C(T)$ at low $T$, including the exact value of
$T_0$.\cite{Lukyanov1997}  In the present work, we compare the very recent numerical
Bethe ansatz results of Kl\"umper and Johnston\cite{Klumper1998} with the predictions
of Lukyanov's theory and find agreement for $\chi(T)$ to high accuracy ($\leq 1\times
10^{-6}$) over a temperature range spanning 18 orders of magnitude,
$5\times 10^{-25}\leq k_{\rm B}T/J\leq 5\times 10^{-7}$; the agreement in the lower
part of this temperatures range is much better, ${\cal O}(10^{-7})$.  For
$C(T)$, the logarithmic correction in Lukyanov's theory is insufficient to
describe the Bethe ansatz data sufficiently accurately even at very low temperatures,
so we derive the next two logarithmic corrections from the Bethe ansatz $C(T)$ data. 
For various applications, it would be desirable to have fits to the  $\chi(T)$ and
$C(T)$ Bethe ansatz data which extend to higher temperatures.  We describe the
formulation and implementation of fit functions, incorporating the influence of the
logarithmic correction terms, which yield extremely precise fits to the data for both
quantities over the entire 25 decades in temperature of the calculations,
$5\times 10^{-25}\leq k_{\rm B}T/J\leq 5$.

The $\chi(T)$ in the intermediate regime $0<\alpha<1$ has been investigated
analytically in the Hartree-Fock approximation\cite{Bulaevskii1969} and using
numerical techniques.\cite{Bonner1979,Barnes1994}  Of particular interest here is
the regime $\alpha \lesssim 1$, close to the uniform limit, which is the regime
relevant to materials exhibiting a dimerization transition with decreasing
$T$ such as occurs in materials exhibiting a spin-Peierls transition.  There are
no accurate theoretical predictions available for $\chi(T)$ of the
alternating-exchange Heisenberg chain in this regime, which is the property usually
used to initially characterize the occurrence of such a transition experimentally. 
To address this deficiency and to also cover a more extended
$\alpha$ range, we carried out extensive quantum Monte Carlo (QMC) simulations and
transfer-matrix density-matrix renormalization group (TMRG)
calculations\cite{Wang1997,Xiang1999} of $\chi(T)$ for $0.05\leq\alpha\leq 1$ over
the temperature range $0.002\leq k_{\rm B}T/J_1\leq 10$.

An interesting issue is how the spin gap $\Delta$ evolves with alternation parameter
$\alpha$ as the uniform limit is approached, $\alpha\to 1$.  Because the uniform
chain is a gapless quantum-critical system, the introduction of alternating exchange
along the chain has been theoretically predicted to yield a nonanalytic
$\Delta(\alpha)$ behavior for $\alpha\to 1$.  We derive $\Delta(\alpha)$ by fitting
our low-$t$ TMRG $\chi(T)$ data by an expression which we formulated.  The
$\Delta(\alpha)$ results are compared with those of previous numerical calculations
and with the theoretical prediction.  We infer from our data that the asymptotic
critical regime is only entered for $\alpha\gtrsim 0.99$.

In order to be optimally useful for accurately modeling experimental $\chi(T)$
data for alternating-exchange chain compounds, our QMC and TMRG $\chi(\alpha,T)$
results must first be accurately fitted by a continuous function of both $\alpha$
and $T$.  We will introduce a general fit function which eventually proves capable of
fitting these combined data for the  alternating-exchange Heisenberg chain very
accurately.  We first fit the $\chi(T)$ of the uniform chain and isolated dimer
using this function and then use the obtained fitting parameters as end-point
parameters in the fit to our combined QMC and TMRG data for intermediate values of
$\alpha$.  The final fit function is a single two-dimensional function of
$\alpha$ and $T$ for $0 \leq\alpha\leq 1$ which can be used to extract the (possibly
temperature-dependent) alternation parameter, exchange constants and spin gap from
experimental $\chi(T)$ data for compounds for which the $S = 1/2$ AF 
alternating-exchange Heisenberg chain Hamiltonian is appropriate.  Our fit function 
will also be useful as a reference for $\chi(T)$ calculated from other related $S =
1/2$ Hamiltonians such as that incorporating the spin-phonon interaction for
spin-Peierls systems.

\subsection{NaV$_2$O$_5$}
\vglue0.11in
Vanadium oxides show a remarkable variety of electronic behaviors.  For example,
the metallic fcc normal-spinel structure compound LiV$_2$O$_4$ shows local
momentlike behaviors above $\sim 50$\,K, crossing over to heavy fermion behaviors
below $\sim 10$\,K.\cite{Kondo1997}  On the other hand, the $d^1$ compound
CaV$_2$O$_5$ has a two-leg trellis-ladder-layer structure\cite{Onoda1996} in which
all of the V atoms are crystallographically equivalent and is a Mott-Hubbard
insulator.  The $\chi(T)$ shows a spin-gap $\Delta/k_{\rm B}\approx 660$\,K
arising from strong coupling of the V $S = 1/2$ spins across a
rung.\cite{Onoda1996}  Modeling of $\chi(T)$ by QMC simulations confirmed that
this compound consists magnetically of V$_2$ dimers, with an intradimer AF
exchange constant $J/k_{\rm B}\approx 665$\,K and with very weak interdimer
interactions.\cite{Miyahara1998}

The compound NaV$_2$O$_5$ can also be formed.  The crystal structure was initially
reported in 1975 to consist of two-leg ladders as in
CaV$_2$O$_5$, but in a non-centrosymmetric (acentric) structure (space group {\it
P2$_1$mn}) in which charge segregation occurs such that one leg of each ladder
consists of V$^{+4}$ and the other of crystallographically inequivalent V$^{+5}$
ions.\cite{Carpy1975}  However, recently five different crystal structure
investigations showed that the structure is actually centrosymmetric (space group
{\it Pmmn}), with all V atoms crystallographically equivalent at room
temperature,\cite{vonSchnering1998,Smolinski1998,Meetsma1998,Ludecke1999,Onoda1999}
so that (static) charge segregation between the V atoms does not, in fact, occur. 
This result is consistent with $^{51}$V NMR investigations which showed the
presence of only one type of V atom at room
temperature.\cite{Ohama1997b,Ohama1999}  This compound is thus formally a
mixed-valent 
$d^{0.5}$ system, which has been considered in a one-electon-band picture to be a
quarter-filled ladder compound.\cite{Smolinski1998,Nishimoto1998}  We note that
from modeling optical excitations in the energy range 4~meV--4~eV, Damascelli and
coworkers initially concluded that the room-temperature structure of NaV$_2$O$_5$
is acentric;\cite{Damascelli1998} their analysis was consistent with the V atoms
on a rung of a ladder having oxidation states of 4.1 and 4.9, respectively. 
However, this group subsequently explained that length- and/or
time-scale-of-measurement issues may be involved in their interpretation, such that
charge disproportionation between V atoms may only occur locally and possibly
dynamically, which could then be consistent with the (average long-range) crystal
structure refinements and NMR measurements.\cite{Damascelli1999}  Theoretical support
for this scenario was provided by Nishimoto and Ohta.\cite{Nishimoto1998b}  Factor
group analyses of the possible IR- and Raman-active phonon modes and comparisons with
experimental observations at room temperature are consistent with the centrosymmetric
space group for the
compound.\cite{Damascelli1999,Lemmens1998,Popovic1998,Popova1998}  A first-principles
electronic structure study based on the density functional method within the
generalized gradient approximation showed that the total energy of the centric
structure is about 1.0 eV/(formula unit) lower than that of the acentric
structure,\cite{Katoh1999} consistent with the recent structural studies.

One might expect that the hole-doping which occurs upon replacing Ca in
CaV$_2$O$_5$ by Na would result in metallic properties for NaV$_2$O$_5$, because
of the nonintegral oxidation state of the V cations and of the crystallographic
equivalence of these atoms.  However, NaV$_2$O$_5$ is a
semiconductor.\cite{Isobe1997a}  This has been explained by the formation of $d^1$
V-O-V molecular clusters on the rungs of the two-leg ladders, again resulting in a
Mott-Hubbard insulator due to the on-site Coulomb
repulsion,\cite{Smolinski1998,Horsch1998} where in this case a ``site'' is a V-O-V
molecular cluster.  Thus a nonintegral oxidation state and crystallographic
equivalence of transition metal atoms in a compound are not sufficient to
guarantee metallic character simply by symmetry; all nearest-neighbor pairs,
triplets, ..., of transition metal atoms must also be crystallographically
equivalent, which is not the case in NaV$_2$O$_5$, since a V$_2$ pair on a rung is
not crystallographically equivalent to one on a leg in the two-leg ladders.  In
contrast, all V atoms and pairs of V atoms in mixed-valent fcc LiV$_2$O$_4$ are
respectively crystallographically equivalent, resulting in metallic character as
demanded by symmetry.

The V-O-V rung molecular clusters which are coupled along the ladder direction in
NaV$_2$O$_5$ may be considered to form an effective $S = 1/2$ one-dimensional (1D) 
chain.\cite{Smolinski1998,Nishimoto1998,Horsch1998}  Experimental support for this
picture, often quoted in the literature, is that the magnetic
susceptibility (above $T_{\rm c}$, see below) is in agreement with
the Bonner-Fisher prediction for the $S = 1/2$ Heisenberg chain, as reported by Isobe
and Ueda.\cite{Isobe1996}  Angle-resolved photoemission spectroscopy (ARPES)
measurements on NaV$_2$O$_5$ by Kobayashi {\it et al.}\cite{Kobayashi1998} showed
that the electronic structure is essentially 1D, despite the ostensibly 2D
nature of the trellis layer, with dispersion in the oxygen and copper bands (below
the Fermi energy) occurring only in the ladder direction ($b$-axis). 
Interestingly, the dispersion in the lowest binding energy part of the occupied Cu
Hubbard band showed a lattice periodicity of 2$b$, which may reflect dynamical
short-range AF and/or crystallographic ordering in the ladder direction. 
Temperature-dependent ARPES measurements on Na$_{0.96}$V$_2$O$_5$ by the same
group from 120 to 300\,K showed evidence for the predicted spin-charge
separation in 1D magnetic systems.\cite{Kobayashi1999}

A phase transition occurs in NaV$_2$O$_5$ at a critical temperature
$T_{\rm c}\approx 33$--36\,K, below which the spin susceptibility
$\chi^{\rm spin}\to 0$ as $T\to 0$ and a lattice distortion
occurs.\cite{Isobe1996,Fujii1997,Weiden1997}  The lattice distortion
results in a supercell with lattice parameters $2a \times 2b \times
4c$.\cite{Fujii1997}  Therefore the transition was initially characterized as a
possible spin-Peierls transition, which by definition is driven by magnetoelastic
(spin-phonon) coupling, and in which there is no change in the charge/spin
distribution within the rungs/V-O-V molecular clusters.  The superstructure in
the $a$ and $c$ directions, perpendicular to the V chains which run in the
$b$ direction, would be a result of the phasing of the distortions in adjacent
chains/ladders.  In this interpretation, and within the adiabatic approximation
(discussed later), one would expect that the magnetic properties above $T_{\rm c}$
should be close to those of the $S = 1/2$ Heisenberg uniform chain, and of an $S =
1/2$  alternating-exchange Heisenberg chain below $T_{\rm c}$.  

It has become clear, however, that the phase transition occurring at
$T_{\rm c}$ in NaV$_2$O$_5$ is accompanied by charge ordering, in contrast to a
classic spin-Peierls transition.  Therefore, the magnetoelastic coupling may only
play a secondary role, and the spin gap may be a secondary order parameter. In
particular, $^{51}$V NMR experiments showed the presence of (inequivalent) V$^{+4}$
and V$^{+5}$ below $T_{\rm c}$, whereas only one V species was present above
$T_{\rm c}$.\cite{Ohama1999}  This result is consistent with the solution of the
superstructure below $T_{\rm c}$ by L\"udecke and co-workers\cite{Ludecke1999}
using synchrotron x-ray diffraction.  L\"udecke {\it et al.}\ found that there are
modulated and unmodulated chains of V atoms below $T_{\rm c}$, tentatively assigned
to magnetic and nonmagnetic chains.  One interpretation of the results is that the
$d^1$ V$^{+4}$ cations segregate into alternate two-leg ladders which are isolated
from each other within the ${\rm V_2O_3}$ trellis layer by intervening two-leg ladders
containing only nonmagnetic V$^{+5}$.\cite{Ludecke1999} The anomalous strong increase
in the thermal conductivity below $T_{\rm c}$ may also be due to charge
ordering.\cite{Vasilev1998}  From ultrasonic measurements of shear and longitudinal
elastic constants, Schwenk and co-workers have suggested that the charge ordering is
of the zig-zag type within each ladder.\cite{Schwenk1999}  In each of these scenarios
for charge ordering, static charge disproportionation occurs such that 1/2 of the V
atoms have oxidation state +4 and the other half +5, consistent with the average
formal oxidation state of +4.5 in the compound.

K\"oppen {\it et al.}\cite{Koppen1998} have concluded from thermal expansion measurements
that the phase transition at $T_{\rm c}$ actually consists intrinsically of two
closely-spaced phase transitions separated by $\lesssim 1$\,K, where the upper
transition is thermodynamically of second order whereas the lower one is first
order.  However, a double transition was not found in their specific heat
measurements on the same crystal, which they attributed to the 50\,mK temperature
oscillations required by their ac measurement technique which were thought to broaden
the two transitions and render them indistinguishable.

The nature of the possible charge ordering pattern has been studied theoretically
by several groups.  Seo and Fukuyama\cite{Seo1998} predicted (at $T = 0$) a static
zig-zag chain of V$^{+4}$ ions on each two-leg ladder, with an
interpenetrating zig-zag chain of V$^{+5}$ ions.  They proposed that pairs of
V$^{+4}$ spins, one each on adjacent ladders, would form spin singlets, resulting
in the observed spin gap.  A similar zig-zag charge configuration in each ladder
was inferred by Mostovoy and Khomskii,\cite{Mostovoy1998} with subsequent
experimental support by Smirnov {\it et al.},\cite{Smirnov1998b} and by Gros and
Valenti.\cite{Gros1999}  Motivated in part by the above thermal expansion
measurement results of K\"oppen {\it et al.},\cite{Koppen1998} Thalmeier and
Fulde\cite{Thalmeier1998} proposed that the charge ordering transition would result
in one linear chain of V$^{+4}$ and one linear chain of V$^{+5}$ on each two-leg
ladder, thereby then allowing a conventional spin-Peierls transition to occur at a
slightly lower temperature, resulting in a double transition as reported by
K\"oppen {\it et al.}\cite{Koppen1998}  A similar picture was put forward by Nishimoto and
Ohta.\cite{Nishimoto1998}  Thalmeier and Yaresko\cite{Thalmeier1999} have
extensively discussed the linear-chain and zig-zag scenarios for charge ordering,
and in addition have considered the alternating  two-leg ladder charge ordering
pattern of the type suggested by L\"udecke {\it et al.}\cite{Ludecke1999}  They point
out that in both the linear and zig-zag patterns, a secondary spin-Peierls
dimerization or spin exchange anisotropy (in spin space) may be necessary to give a
spin gap, whereas the two-leg ladder ordering has a spin gap even with no lattice
distortion.  Thalmeier and Yaresko describe the characteristic signatures of each of
the charge-ordered models to be compared with experimental inelastic neutron
scattering measurements.  Riera and Poilblanc have discussed the influence of
electron-phonon coupling on the derived charge- and spin-order phase
diagrams.\cite{Riera1999}

We have carried out $\chi(T)$ measurements from 2 to 750\,K on single crystals
of NaV$_2$O$_5$ along the ladder ($b$ axis) direction to further characterize
and clarify the nature of the magnetic interactions and ordering below and above
$T_{\rm c}$.  We find that the magnetic properties above $T_{\rm c}$ are not
accurately described by the $S = 1/2$ Heisenberg uniform chain prediction with a
$T$-independent $J$, although a mean-field ferromagnetic interchain coupling can
explain these data.  Using  our theoretical $\chi(\alpha,T)$ fit function for the
AF alternating-exchange chain below $T_{\rm c}$, we find that $\delta(0)
\equiv(1-\alpha)/(1+\alpha) = 0.034(6)$ and that the zero-temperature spin-gap of
NaV$_2$O$_5$ is $\Delta(0)/k_{\rm B} = 103(2)$\,K\@.  The $\delta(T)$ and
$\Delta(T)$ below $T_{\rm c}$ are extracted.  A spin pseudogap is
found to occur above $T_{\rm c}$ with a rather large magnitude.  From our specific
heat measurements on two crystals, we find that the magnetic specific heat at low
temperatures $T\lesssim 15$\,K is too small to be resolved experimentally, and that
the spin entropy at $T_{\rm c}$ is too small to account for the entropy of the
transition.  A quantitative analysis shows that at least 77\,\% of the entropy change
at $T_{\rm c}$ due to the transition(s) and associated order parameter
fluctuations must arise from the lattice and/or charge degrees of freedom and less
than 23\,\% from the spin degrees of freedom.

\subsection{Plan of the paper}

The rest of the paper is organized as follows.  Our notation for the Heisenberg spin
Hamiltonian and for the reduced susceptibility, temperature and spin gap are given
immediately in Sec.~\ref{SecThyIntro}.  Some general features of the high-temperature
series expansion (HTSE) for $\chi(T)$ and $C(T)$ of $S = 1/2$ Heisenberg spin
lattices and  the low-temperature limits of these quantities for one-dimensional (1D)
systems with a spin gap are then given.  We then specialize to the $S = 1/2$ AF
alternating-exchange Heisenberg chain in Sec.~\ref{SecAltChnIntro}, where we discuss
the HTSEs, the spin gap and the one-magnon dispersion relations $E(\Delta,k)$.  In
the latter subsection, we derive a one-parameter approximation for $E(\Delta,k)$
which correctly extrapolates to the $\alpha\to 0$ limit and which we will need in
order to later fit the TMRG $\chi(T)$ data to extract $\Delta(\alpha)$.  We also show
that the expressions for the low-$T$ limits of both $\chi(T)$ and $C(T)$ depend only
on the spin gap (in addition to $T$).  In Sec.~\ref{SecUniformChnThy}, we discuss
overall features of the $\chi(T)$ and $C(T)$ for the uniform chain and then focus on
the low-$T$ behavior.  The explicit forms of the logarithmic corrections previously
found for $\chi(T)$ are first discussed.  We show that a low-$T$ expansion of the
theory of Lukyanov\cite{Lukyanov1997} gives the same first two corrections, and in
addition gives the next higher-order term.  We then compare the Bethe ansatz
$\chi(T)$ results\cite{Klumper1998} directly with the theory with no adjustable
parameters or approximations.  Logarithmic corrections are also found to be
important to accurately describe the Bethe ansatz data\cite{Klumper1998} for
$C(T)$. We show that the lowest order correction is not sufficient to fit the data,
and we derive the next two higher-order corrections by fitting the data at very low
temperatures.

General features of our scheme to fit numerical $\chi(T)$ data are described in
Sec.~\ref{SecGenFit}, followed by a fit to the exact $\chi(T)$ for the
antiferromagnetic Heisenberg dimer and two fits to the numerical
$\chi(T)$ data\cite{Klumper1998} for the uniform chain.  Due to the special
requirements of, and constraints on, the two-dimensional fit function necessary to
accurately fit $\chi(\alpha,T)$ data for the alternating-exchange chain over large
ranges of both $\alpha$ and $T$, a separate section, Sec.~\ref{SecAltChnFitFcn}, is
devoted to formulating and discussing this fit function.  Using a fit function
similar to that used to fit numerical $\chi(T)$ data, in the next section an
extremely accurate and precise fit is obtained over 25 decades in temperature to the
Bethe ansatz $C(T)$ data.\cite{Klumper1998}  Our QMC and TMRG
$\chi(T)$ data for the alternating-exchange chain are presented and fitted in
Sec.~\ref{SecQMCSim}, using as end-point parameters those determined for the uniform
chain and the dimer, respectively.  The spin gap $\Delta(\alpha)$ is extracted for
$0.8\leq\alpha\leq0.995$ by fitting the TMRG $\chi(\alpha,T)$ data at low
temperatures in Sec.~\ref{SecTDMRGgaps}.  Section~\ref{SecCompPrevWork} contains a
comparison of our numerical results with previous work.  The $\Delta(\alpha)$ values
are compared with previous numerical results and with the theoretical prediction for
the asymptotic critical behavior in Sec.~\ref{SecCompareSpinGap}. Our $\chi(T)$
calculations are shown in Sec.~\ref{SecBarnesRiera} to be in good agreement with the
previous numerical results of Barnes and Riera\cite{Barnes1994} for
$0.2\leq\alpha\leq 0.8$.  The numerical calculations of
Bulaevskii\cite{Bulaevskii1969} have been extensively used in the past by
experimentalists to fit the $\chi(T)$ of spin-Peierls compounds, but up to now a
detailed analysis of the predictions of this theory has not been given.  We present
such an analysis in Sec.~\ref{SecBulaevskii} and compare our results with these
predictions.

We begin the experimental part of the paper by studying the anisotropic magnetic 
susceptibility of a high quality NaV$_2$O$_5$ single crystal in
Sec.~\ref{SecMagAnis}, where literature data on the anisotropy of the $g$ factor
and Van Vleck susceptibility are compared with our results.  In the following
sections we illustrate the utility and application of many of the theoretical
results derived and presented previously in the paper.  In
Sec.~\ref{SecNaV2O5Modeling} we present experimental $\chi(T)$ data for single
crystals of NaV$_2$O$_5$ and model these data in detail in Sec.~\ref{SecNaV2O5Fits}
using our QMC and TMRG $\chi(T)$ data fit function for the AF alternating-exchange
Heisenberg chain.  We show that qualitatively and quantitatively new information
about the temperature dependences of the spin dimerization parameter and spin gap
below $T_{\rm c}$ can be obtained from our modeling.  This analysis also shows that
spin dimerization fluctuations and a spin pseudogap are present above $T_{\rm c}$,
and we quantitatively determine their magnitudes.  Our specific heat
measurements of  NaV$_2$O$_5$ single crystals and our extensive modeling of these
data are presented in Sec.~\ref{SecCpDatModeling}, where we obtain quantitative
limits on the relative contributions of the lattice, spin and charge degrees of
freedom to the change in the entropy due to the transition at $T_{\rm c}$ and to
associated order parameter fluctuations.  A summary and concluding discussion are
given in Sec.~\ref{SecSummary}.

\section{Theory}
\label{SecThyIntro}

In this paper we will only be concerned with the spin $S = 1/2$
antiferromagnetic (AF) Heisenberg Hamiltonian
\begin{equation}
{\cal H} = \sum_{<ij>} J_{ij} \bbox{S}_i\cdot\bbox{S}_j~,
\label{EqHeisChn}
\end{equation}
where $J_{ij}$ is the Heisenberg exchange interaction between spins
$\bbox{S}_i$ and $\bbox{S}_j$ and the sum is over unique exchange bonds.  A $J_{ij} >
0$ corresponds to AF coupling, whereas $J_{ij} < 0$ refers to ferromagnetic coupling. 
Note that magnetic nearest neighbors $\bbox{S}_j$ of a given spin $\bbox{S}_i$ in
Eq.~(\ref{EqHeisChn}) need not be crystallographic nearest neighbors.  A magnetic
nearest neighbor of a given spin is any other spin with which the given spin has an
exchange interaction.

For notational convenience, we define the reduced spin susceptibilities $\chi^*$
and $\overline{\chi^*}$, reduced temperatures $t$ and $\overline{t}$ and reduced spin
gaps $\Delta^*$ and $\overline{\Delta^*}$ as
\begin{equation}
\chi^* \equiv \frac{\chi J^{\rm max}}{Ng^2\mu_{\rm B}^2}~,~~~\overline{\chi^*} \equiv
\frac{\chi J}{Ng^2\mu_{\rm B}^2}~~,\nonumber
\end{equation}
\begin{equation}
t \equiv {k_{\rm B}T\over J^{\rm max}}~,~~~\overline{t} \equiv {k_{\rm B}T\over
J}~,\label{EqChiStar}
\end{equation}
\begin{equation}
\Delta^* \equiv \frac{\Delta}{J^{\rm max}}~,~~~\overline{\Delta^*} \equiv
\frac{\Delta}{J}~,\nonumber
\end{equation}
where $J^{\rm max}$ and $J$ are, respectively, the largest and average exchange
constants in the system, $N$ is the number of spins, $g$ is the spectroscopic
splitting factor appropriate to the direction of the applied magnetic field relative
to the crystallographic axes, and $\mu_{\rm B}$ is the Bohr magneton.   

\subsection{High-temperature series expansions for the spin susceptibility and
magnetic specific heat}

For any Heisenberg spin lattice (in any dimension) in which the spins are
magnetically equivalent, i.e.\ where each spin has identical magnetic coordination
spheres, the first three to four terms of the exact quantum mechanical high
temperature series expansion of $\chi^*(t)$ have the same form, with a particularly
simple form if the series is inverted.\cite{Johnston1997}  For $S = 1/2$, one
obtains\cite{Johnston1997,Johnston1999,Rushbrooke1958}
\begin{mathletters}
\label{EqHTSGen:all}
\begin{equation}
{1\over 4\chi^*t} = \sum_{n = 0}^\infty {d_n\over t^n}~,
\label{EqHTSGen:a}
\end{equation}
\begin{equation}
d_0 = 1,~~~d_1 = {1\over 4 J^{\rm max}}\,\sum_j J_{ij},~~~d_2 = {1\over 8 {J^{\rm
max}}^2}\,\sum_j J_{ij}^2~,
\label{EqHTSGen:b}
\end{equation}
\begin{equation}
d_3 = {1\over 24 {J^{\rm max}}^3}\,\sum_j J_{ij}^3~.
\label{EqHTSGen:c}
\end{equation}
\end{mathletters}
Equation~(\ref{EqHTSGen:b}) is universal, but Eq.~(\ref{EqHTSGen:c}) holds only for
spin lattices which are not geometrically frustrated for AF ordering and in which
the magnetic and crystallographic nearest neighbors of a given spin are the same. 
Geometrically frustrated lattices typically contain closed triangular exchange paths
within the spin lattice structure, such as in the 2D triangular lattice or in the 3D
$B$ sublattices of the fcc $AB_2{\rm O}_4$ oxide normal-spinel and $A_2B_2{\rm O}_7$
oxide pyrochlore structures.  The uniform and alternating-exchange chains considered
in this paper are not geometrically frustrated, and the magnetic and
crystallographic nearest neighbors of a given spin are the same.  It has been
found\cite{Johnston1997} that the terms to
${\cal O}(1/t^3)$ on the right-hand-side of Eq.~(\ref{EqHTSGen:a}) are sufficient to
quite accurately describe the susceptibilities of a variety of nonfrustrated
\mbox{zero-,} one-, and two-dimensional $S = 1/2$ AF Heisenberg spin lattices to
surprisingly low temperatures $t \lesssim 1$.  Higher order
$d_n/t^n$ terms with $n \geq 4$ are dependent on the structure and dimensionality of
the spin lattice.  The Weiss temperature $\theta$ in the Curie-Weiss law
$\chi(T) = C/(T - \theta)$ is given by the universal expression $\theta = -d_1 J^{\rm
max}/k_{\rm B}$.

Because the spin susceptibility and the magnetic contribution $C(T)$ to the specific
heat can both be expressed, via the fluctuation-dissipation theorem and
the Heisenberg Hamiltonian, respectively, in terms of the spin-spin correlation
functions, there is a close relationship between these two
quantities.\cite{Fisher1962}  In particular, just as there is a universal expression
for the first three to four HTSE terms for $\chi(T)$ of a Heisenberg spin lattice as
discussed above, a universal expression for the first one to two HTSE terms for
$C(T)$ of such a spin lattice exists and is given for $S = 1/2$
by\cite{Johnston1997,Johnston1999,Rushbrooke1958}
\begin{equation}
{C(t)\over N k_{\rm B}} = {3\over 32}\bigg[\frac{\sum_j J_{ij}^2}{t^2 {J^{\rm
max}}^2} + \frac{\sum_j J_{ij}^3}{2 t^3 {J^{\rm
max}}^3} + {\cal O}\Big({1\over t^4}\Big)\bigg]~.
\label{EqCmHTSE}
\end{equation}
The sums are over all magnetic nearest-neighbor bonds of any given spin
$\bbox{S}_i$.  The first term is universal but the second term holds only for
geometrically nonfrustrated spin lattices in which the crystallographic and magnetic
nearest-neighbors of any given spin are the same.  Higher order terms all depend on
the structure and dimensionality of the spin lattice.

A common misconception is that $C = 0$ if the  magnetic susceptibility of a
local-moment system obeys the Curie-Weiss law.  This is only true classically.  For
Heisenberg spin lattices, one can easily show that the Weiss temperature $\theta$ in
the Curie-Weiss law arises from the first HTSE term [${\cal O}(1/t)$] of the
magnetic nearest-neighbor spin-spin correlation function, which is the same quantity
that the first HTSE term of $C(t)$ arises from.\cite{Johnston1997}  Thus,
e.g., for $S = 1/2$ Heisenberg spin lattices at temperatures $t\gg1$ at which the
Curie-Weiss law holds, the magnetic specific heat is given by the universal first
term of Eq.~(\ref{EqCmHTSE}).

\subsection{Low-temperature limit of the spin susceptibility and specific heat of 1D
systems with a spin gap}

\paragraph*{Magnetic susceptibility.}For one-dimensional (1D) $S = 1/2$ Heisenberg
spin systems with a spin gap such as the $S = 1/2$ two-leg ladder (and the
alternating-exchange chain), Troyer, Tsunetsugu, and W\"{u}rtz\cite{Troyer1994}
derived a general expression for
$\chi^*(t)$ which approximately takes into account kinematic magnon interactions,
given by
\begin{mathletters}
\label{EqTroyerGen:all}
\begin{equation}
\chi^*(t) = {1\over t}\ {z(t)\over 1+3 z(t)}~,\label{EqTroyerGen:a}
\end{equation}
\begin{equation}
z(t) = {1\over \pi}\int_0^\pi {\rm
e}^{-\varepsilon_k/t}\,d(ka)~,\label{EqTroyerGen:b}
\end{equation}
\end{mathletters}
where $\varepsilon_k\equiv E(k)/J^{\rm max}$, $E(k)$ is the nondegenerate
one-magnon (triplet) dispersion relation (the Zeeman degeneracy is already
accounted for) and $a$ is the (average) distance between spins.  This expression is
exact in both the low- and high-temperature limits.  For the isolated dimer, for which
$\varepsilon_k =
\Delta^* = 1$, Eq.~(\ref{EqTroyerGen:a}) is exact at all temperatures.  Inserting
$z(t) = {\rm e}^{-1/t}$ for the dimer into Eq.~(\ref{EqTroyerGen:a}) yields the
correct result
\begin{mathletters}
\label{EqChiDimer:all}
\begin{equation}
\chi^*(t) = {{\rm e}^{-1/t}\over t}\, {1\over 1 + 3\,{\rm e}^{-1/t}}~,~~~{\rm
(dimer)}
\label{EqChiDimer:a}
\end{equation}
\begin{equation}
\chi^*(t\to 0) = {{\rm e}^{-1/t}\over t}~.
\label{EqChiDimer:b}
\end{equation}
\end{mathletters}
The $\chi^*(t)$ in Eq.~(\ref{EqChiDimer:a}) for the antiferromagnetic Heisenberg
dimer is plotted in Fig.~\ref{Dimerchi55Fit,Dev}; the fit shown in the figure will
be presented and discussed later in Sec.~\ref{SecDimerFit}.

At low temperatures $t \ll \Delta^*$ and $t \ll$ one-magnon bandwidth/$J^{\rm max}$, 
and for a dispersion relation with a parabolic dependence on wave vector $k$
near the band minimum
\begin{equation}
\varepsilon_k \equiv {E(k)\over J^{\rm max}} = \Delta^* +  c^*(ka)^2~,
\label{EqEps(k)}
\end{equation}
one can replace $\varepsilon_k$ in Eq.~(\ref{EqTroyerGen:b}) by the
approximation~(\ref{EqEps(k)}) and replace the upper limit of the integral in
Eq.~(\ref{EqTroyerGen:b}) by $\infty$, yielding $z(t) = {\rm
e}^{-\Delta^*/t}\sqrt{t}/(2\sqrt{\pi c^*})$.  Substituting this result into
Eq.~(\ref{EqTroyerGen:a}) gives the low-$t$ limit\cite{Troyer1994}
\begin{mathletters}
\label{EqTroyer}
\begin{equation}
\chi^*(t\to 0) = {A\over t^\gamma}\,{\rm e}^{-\Delta^*/t}~,
\label{EqTroyer:a}
\end{equation}
\begin{equation}
A = {1\over 2\sqrt{\pi c^*}}~,~~~\gamma = {1\over 2}~.\label{EqTroyer:b}
\end{equation}
\end{mathletters}
This result is correct for any 1D $S = 1/2$ Heisenberg spin system with a spin
gap and with a nondegenerate (excluding Zeeman degeneracy) lowest-lying excited
triplet magnon band which is parabolic \,at \,the \,band \,minimum.  
\begin{figure}
\epsfxsize=3in
\centerline{\epsfbox{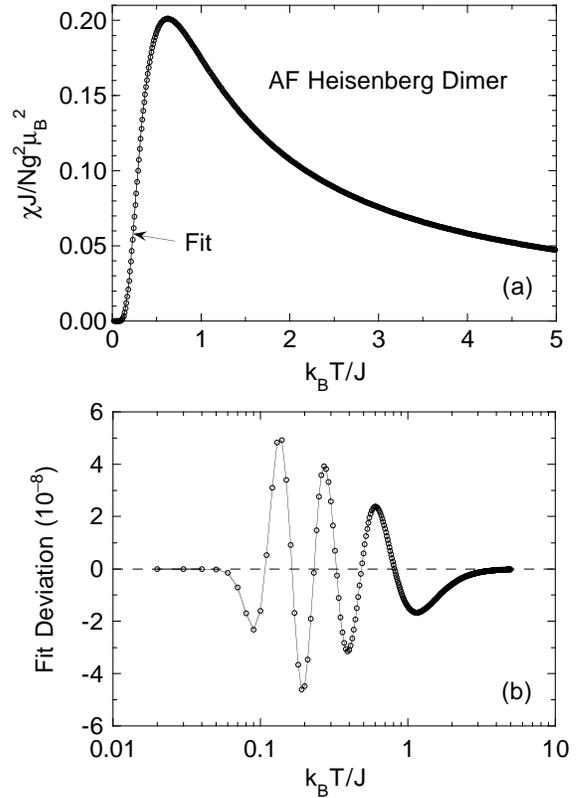}}
\vglue 0.1in
\caption{(a) Magnetic susceptibility $\chi$ ($\circ$) versus temperature $T$ for
the spin $S = 1/2$ Heisenberg dimer with antiferromagnetic exchange constant $J$. 
The fit from Sec.~\protect\ref{SecDimerFit} is shown by the solid curve.  (b)
Semilog plot of the fit deviation vs $T$.  The lines connecting the points in (b)
are guides to the eye.}
\label{Dimerchi55Fit,Dev}
\end{figure}
\noindent On the other hand,
the low-temperature limit of $\chi^*(t)$ for the isolated dimer in
Eq.~(\ref{EqChiDimer:b}) is of the same form as Eq.~(\ref{EqTroyer:a}), but with
$\gamma = 1$.  Thus, for 1D systems consisting essentially of dimers which are
weakly coupled to each other, a crossover from $\gamma = 1$ to $\gamma = 1/2$ is
expected with decreasing $t$.

The parameters $A$ and $\gamma$ can be determined if very accurate
$\chi^*(t)$ and $\Delta^*$ data are available.  Taking the logarithm
of Eq.~(\ref{EqTroyer:a}) yields the low-$t$ prediction
\begin{mathletters}
\label{EqLogChi:all}
\begin{equation}
\ln[\chi^*(t)] + {\Delta^*\over t} = \ln A - \gamma \ln t~,
\label{EqLogChi}
\end{equation}
so plotting the left-hand-side vs $\ln t$ allows these two parameters
to be determined.  Alternatively, assuming $\gamma = 1/2$, one can obtain estimates
of $A$ and $\Delta^*$ using Eq.~(\ref{EqTroyer:a}), according to
\begin{equation}
-\ln(\chi^*\sqrt{t}) = -\ln A + {\Delta^*\over t}~
\label{EqLogChi:b}
\end{equation}
and/or
\begin{equation}
-{\partial\ln(\chi^*\sqrt{t})\over \partial(1/t)} = \Delta^*.
\label{EqLogChi:c}
\end{equation}
\end{mathletters}

\paragraph*{Specific heat.}The low-$t$ limit of the magnetic contribution $C(T)$ to
the specific heat for the same model\cite{Troyer1994} is calculated to be
\begin{eqnarray}
{C(t\to 0)\over N k_{\rm B}} = {3\over 2}&&\Big({\Delta^*\over \pi
c^*}\Big)^{1/2}\Big({\Delta^*\over t}\Big)^{3/2}\nonumber\\
&&\times\bigg[1 + {t\over\Delta^*} +
{3\over 4}\Big({t\over\Delta^*}\Big)^2\bigg]{\rm e}^{-\Delta^*/t}~.
\label{EqCTroyer}
\end{eqnarray}

Note that, in addition to the ratio $t/\Delta^* = k_{\rm B}T/\Delta$ of the thermal
energy to the spin gap, the magnitude of $\chi^*$ in Eqs.~(\ref{EqTroyer}) is
determined by the actual value of the curvature $c^*$ at the triplet one-magnon band
minimum, whereas the magnitude of $C$ in Eq.~(\ref{EqCTroyer}) depends only on the
{\it ratio} of $c^*$ to  $\Delta^*$.  These formulas have been applied in the
literature to model  experimental data for alternating-exchange chain and two-leg
spin ladder compounds.  However, with one exception\cite{Johnston1996} to our
knowledge, these modeling studies have not recognized that the prefactor parameter
and the spin gap are not independently adjustable parameters.  For a given spin
lattice, they are in fact uniquely related to each other.  Their relationship for the
$S = 1/2$ two-leg Heisenberg ladder was studied in Ref.~\onlinecite{Johnston1996}. 
For the alternating-exchange chain, we estimate the relationship between $c^*$
and $\Delta^*$ below in Sec.~\ref{SecAltChnDispRlns}.

\subsection{Alternating-exchange chain}
\label{SecAltChnIntro}

The $S=1/2$ AF alternating-exchange Heisenberg chain Hamiltonian is written
in three equivalent ways as\cite{Barnes1998}
\begin{mathletters}
\label{EqAltChnHam:all}
\begin{eqnarray}
{\cal H} & = & \sum_i J_1 \roarrow{S}_{2i-1}\cdot \roarrow{S}_{2i} + J_2
\roarrow{S}_{2i}\cdot \roarrow{S}_{2i+1}\label{EqAltChnHam:a}\\
& = & \sum_i J_1 \roarrow{S}_{2i-1}\cdot \roarrow{S}_{2i} + \alpha J_1
\roarrow{S}_{2i}\cdot \roarrow{S}_{2i+1}\label{EqAltChnHam:b}\\
& = & \sum_i  J(1 + \delta) \roarrow{S}_{2i-1}\cdot \roarrow{S}_{2i} + J(1 -
\delta) \roarrow{S}_{2i}\cdot \roarrow{S}_{2i+1}~,\label{EqAltChnHam:c}
\end{eqnarray}
\end{mathletters}
where
\begin{mathletters}
\label{EqDimParams:all}
\begin{eqnarray}
J_1 & = &  J(1 + \delta) = \frac{2J}{1 + \alpha}~,\label{EqDimParams:a}\\
\nonumber\\
\alpha & = & {J_2\over J_1} = \frac{1-\delta}{1+\delta}~,\label{EqDimParams:b}\\
\nonumber\\
\delta & = & {J_1\over J} - 1 = {J_1 - J_2\over 2J} =
\frac{1-\alpha}{1+\alpha}~,\label{EqDimParams:c}\\ 
\nonumber\\
J & = & {J_1 + J_2\over 2} = J_1{1 + \alpha\over 2}~,\label{EqDimParams:d}
\end{eqnarray}
\end{mathletters}
with AF couplings $J_1 \geq J_2 \geq 0$, $0 \leq (\alpha,\,\delta) \leq 1$.  The
uniform undimerized chain corresponds to $\alpha = 1,\ \delta = 0,\ J_1 = J_2 =
J$.  The form of the Hamiltonian in Eq.~(\ref{EqAltChnHam:c}) is most appropriate
for chains showing a second-order dimerization transition at $T_{\rm c}$ with
decreasing $T$.  If the exchange modulation $\delta \ll 1$ ($\alpha \sim 1$), the
(average) $J$ below $T_{\rm c}$ can be identified with the exchange coupling in the
high-$T$ undimerized state.

In spin-Peierls systems, the spin-phonon interaction causes a lattice
dimerization to occur below the spin-Peierls transition temperature,
together with a spin-gap due to the formation of spin singlets on the dimers.  The
Hamiltonian can be mapped onto the spin  Hamiltonian~(\ref{EqAltChnHam:all})
(with renormalized exchange constants) only in the adiabatic parameter regime, in
which the relevant phonon energy is much smaller than $J$.  If such a mapping cannot
be made, dynamical phonon effects (quantum mechanical fluctuations) become important
and the $\chi(T)$ can be significantly different from that predicted from
Hamiltonian~(\ref{EqAltChnHam:all}).\cite{Sandvik1997,Wellein1998,Kuhne1999}  This
issue will be discussed further when modeling the $\chi(T)$ data for NaV$_2$O$_5$ in
Sec.~\ref{SecNaV2O5Modeling}.

\subsubsection{High-temperature series expansions}

\paragraph*{Magnetic Susceptibility.}For the alternating-exchange chain, according to
our definition one has
$J^{\rm max} = J_1$. Then using the definition for $\alpha$
in Eq.~(\ref{EqDimParams:b}), the $d_n$ HTSE coefficients in Eqs.~(\ref{EqHTSGen:b})
and~(\ref{EqHTSGen:c}) become
\begin{equation}
d_0 = 1,~d_1 = {1 + \alpha\over 4},~d_2 = {1+\alpha^2\over 8},~
d_3 = {1+\alpha^3\over 24}~.
\label{EqHTSAltChnCoeffs}
\end{equation}
One can change variables from $\alpha$ and $J_1$ in $\chi^*(\alpha,t)$ to
$\delta$ and $J$ in $\overline{\chi^*}(\delta,\overline{t})$ using
Eqs.~(\ref{EqDimParams:all}) which give
\begin{mathletters}
\label{EqBarDef:all}
\begin{equation}
t = {\overline{t}\over 1 + \delta}~,\label{EqBarDef:a}
\end{equation}
\begin{equation}
\overline{\chi^*}(\delta,\overline{t}) = {1\over 1+\delta}\ \chi^*\Big({1-\delta\over
1+\delta},{\overline{t}\over 1 + \delta}\Big)~.\label{EqBarDef:b}
\end{equation}
\end{mathletters}
We write the resulting HTSE for the inverse of
$\overline{\chi^*}(\delta,\overline{t})$ as
\begin{mathletters}
\begin{equation}
{1\over 4\overline{\chi^*}\ \overline{t}} = \sum_{n = 0}^\infty {\overline{d}_n\over
\overline{t}^n}~,
\label{EqHTSAltChnBar:a}
\end{equation}
where we find
\begin{equation}
\overline{d}_0 = 1,~\overline{d}_1 = {1\over 2},~\overline{d}_2 = {1 +
\delta^2\over 4},~\overline{d}_3 = {1 + 3\delta^2\over12}~.
\label{EqHTSAltChnBar:b}
\end{equation}
\end{mathletters}
An important feature of this HTSE of
$\overline{\chi^*}(\delta,\overline{t})$ is that it is an even (analytic) function
of $\delta$ for any finite temperature.  This constraint must be true in general and
not just for the terms listed,\cite{Klumper1998} because
$\overline{\chi^*}(\delta,\overline{t})$ cannot depend on the sign of $\delta = (J_1
- J_2)/(2J)$: the Hamiltonian in Eq.~(\ref{EqAltChnHam:c}) is invariant upon such a
sign change.  Physically, a negative $\delta$ would simply correspond to relabeling
all $\bbox{S}_i \to
\bbox{S}_{i+1}$, which cannot change the physical properties.  We will use this
constraint that $\overline{\chi^*}(\delta,\overline{t})$ must be an even function of
$\delta$ to help formulate our fitting function (after a change back in variables)
for our QMC and TMRG $\chi^*(\alpha,t)$ calculations for the alternating-exchange
chain.  This constraint is important because it allows a fit function for
$\chi^*(\alpha,t)$ to be formulated which is accurate for $\alpha\lesssim 1$
($\delta \ll 1$), a parameter regime relevant to compounds exhibiting second-order
spin-dimerization transitions with decreasing temperature.

\paragraph*{Magnetic specific heat.}Using $J^{\rm max}$ = $J_1$ and $\alpha =
J_2/J_1$, the general HTSE expression in Eq.~(\ref{EqCmHTSE}) yields the two
lowest-order HTSE terms for the magnetic specific heat $C(T)$ of the $S = 1/2$ AF
alternating-exchange Heisenberg chain as
\vglue0.2in
\begin{equation}
{C(t)\over N k_{\rm B}} = {3\over 32}\Big[{1 + \alpha^2\over t^2} + {1 +
\alpha^3\over 2t^3} + {\cal O}\Big({1\over t^4}\Big)\Big]~.
\label{EqHTSECGen2term}
\end{equation}
\vglue0.2 in

\subsubsection{Spin gap}
\vglue0.1in
The spin gap $\Delta^*(\alpha)$ of the alternating-exchange chain was determined to
high ($\leq 1\%$) accuracy for $0 \leq \alpha \leq 0.9$, in $\alpha$ increments of
0.1, using multiprecision methods by Barnes, Riera, and Tennant
(BRT).\cite{Barnes1998}  They found that their calculations could be parametrized
well by 
\vglue0.14in
\begin{mathletters}
\label{EqDimParams2:all}
\begin{equation}
\Delta^*(\alpha) \equiv \frac{\Delta(\alpha)}{J_1} \approx (1 - \alpha)^{3/4}(1 +
\alpha)^{1/4}~,\label{EqDimParams2:a}
\end{equation}
\vglue0.1in
\begin{equation}
\overline{\Delta^*}(\delta) \equiv \frac{\Delta(\delta)}{J} \approx
2\delta^{3/4}~.\label{EqDimParams2:b}
\end{equation}
\end{mathletters}
\vglue0.1in
The same $\overline{\Delta^*}(\delta)$ was found in numerical calculations
by Ladavac {\it et al.}\cite{Ladavac1998} for $0.01\leq\delta\leq 1$, whereas calculations
for $0.03\leq\delta\leq 0.06$ by Augier {\it et al.}\cite{Augier1997} yielded somewhat
smaller values of $\overline{\Delta^*}$ than predicted by
Eq.~(\ref{EqDimParams2:b}).  The asymptotic critical behavior of
$\overline{\Delta^*}$ as the uniform limit is approached ($\alpha \to 1,\ \delta \to
0$) has been given\cite{Klumper1998,Cross1979,Black1981,Affleck1989} as

\begin{equation}
\overline{\Delta^*}(\delta)\propto {\delta^{2/3}\over |\ln\delta|^{1/2}}~;
\label{EqCritGap}
\end{equation}
\vglue0.1in
\noindent
thus the parametrization in Eq.~(\ref{EqDimParams2:b}) evidently indicates that the
fitted data do not reside within the asymptotic critical regime.  Alternatively,
Barnes, Riera, and Tennant\cite{Barnes1998} suggested that Eq.~(\ref{EqCritGap}) may
not be the correct form for the asymptotic critical behavior.  On the other hand,
Uhrig {\it et al.}\cite{Uhrig1999} fitted their $T = 0$ density matrix renormalization
group (DMRG) calculations of $\overline{\Delta^*}(\delta)$ for $0.004\leq\delta\leq
0.1$ to a power-law behavior without the log correction and obtained
$\overline{\Delta^*} \approx 1.57\,\delta^{0.65}$.  We will further discuss the
above spin gap calculation results later in Sec.~\ref{SecCompareSpinGap} after
deriving our own $\Delta^*(\alpha)$ values from our TMRG $\chi^*(\alpha,t)$ data in
Sec.~\ref{SecTDMRGgaps}.

\begin{figure}
\epsfxsize=3in
\centerline{\epsfbox{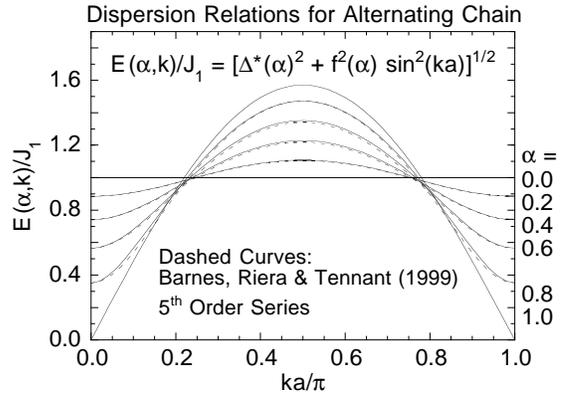}}
\vglue 0.1in
\caption{Dispersion relations $E(\alpha,k)$ for the $S=1/2$ antiferromagnetic
alternating-exchange Heisenberg chain.  The dashed curves for alternation
parameters $\alpha = 0$, 0.2, 0.4, 0.6 and 0.8 are the dimer series expansion results
of Barnes, Riera and Tennant (Ref.~\protect\onlinecite{Barnes1998}), the solid curves
for these $\alpha$ values are from our expression given in the figure and in
Eqs.~(\protect\ref{EqMyE(k)}), and the solid curve for $\alpha=1$ is the known exact
result for the uniform chain, which by construction is the same for the expression
in the figure at this $\alpha$ value.}
\label{E(ka)AltChn}
\end{figure}

\subsubsection{One-magnon dispersion relations}
\label{SecAltChnDispRlns}

Barnes, Riera, and Tennant have computed the dimer series expansion of the
dispersion relation $\varepsilon(\alpha,k)\equiv E(\alpha,k)/J_1$ for the one-magnon
($S = 1$) energy $E(\alpha,k)$ vs wave vector $k$ along the chain for the lowest-lying
one-magnon band, up to fifth order in $\alpha$,\cite{Barnes1998} which we write as
\begin{equation}
\varepsilon(\alpha,k) = \sum_{n=0}^\infty a_n(\alpha)\,\cos(2nka)~,
\label{EqBRTE(k)}
\end{equation}
where $a$ is the (average) spin-spin distance, which is 1/2 the lattice repeat
distance along the chain in the dimerized state.  Plots of
$\varepsilon(\alpha,k)$ for $\alpha = 0$, 0.2, 0.4, 0.6, and 0.8 up to fifth order in
$\alpha$, as given in Fig.~4 of Ref.~\onlinecite{Barnes1998}, are shown as the dashed
curves in Fig.~\ref{E(ka)AltChn}.  The curves are symmetric about $ka=\pi/2$, so
the same spin gap $\Delta^*(\alpha) = \sum_{n=0}^\infty a_n(\alpha)$ occurs at $ka=0$
and $\pi$.  This fifth-order approximation yields $\Delta^*(\alpha)$ values for
$\alpha\leq 0.9$ in rather close agreement with BRTs' results discussed in the
previous section.  For a dimer series expansion we expect the average energy of
the one-magnon band states to be nearly independent of $\alpha$, i.e.,
\begin{equation}
{1\over\pi}\int_0^\pi \varepsilon(\alpha,ka)\,d(ka) = 1~.
\label{EqNorm}
\end{equation}
Indeed, upon inserting BRTs' fifth order expansion coefficients
into Eq.~(\ref{EqBRTE(k)}) and the result into Eq.~(\ref{EqNorm}), we find that this
sum rule is satisfied to within 1\% for $0\leq\alpha\leq 1$.

Also shown as a solid curve in Fig.~\ref{E(ka)AltChn} is the exact result
$\varepsilon(k) = (\pi/2)|\sin(ka)|$ for the uniform chain ($\alpha =
1$).\cite{desCloizeaux1962}  This
\,$\varepsilon(k)$ \,has \,a \,cusp \,with \,infinite \,curvature \,(at \,$ka = 0$ 
\begin{figure}
\epsfxsize=3.1in
\centerline{\epsfbox{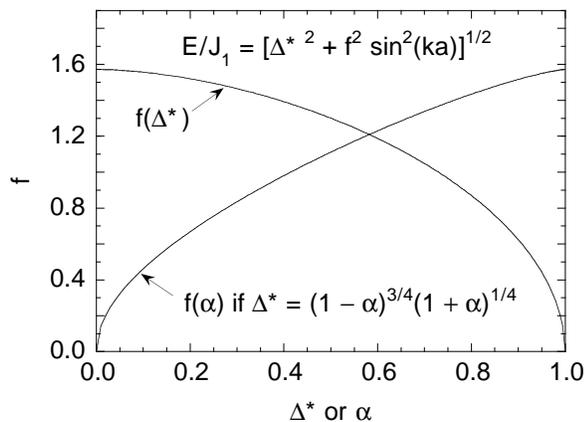}}
\vglue 0.1in
\caption{The function $f(\Delta^*)$ in the expression in the figure and in
Eqs.~(\protect\ref{EqMyE(k)}) for the one-magnon dispersion relation of the $S=1/2$
antiferromagnetic alternating-exchange Heisenberg chain, where $\Delta^*\equiv
\Delta/J_1$ is the spin gap.  The dependence $f(\alpha)$ is also plotted for the
assumed form of $\Delta^*(\alpha)$ shown.}
\label{E(ka)f(gap)Calc}
\end{figure}
\noindent and~$\pi$) which
cannot be accurately approximated by a Fourier series with a small number of terms. 
This singular behavior is evidently closely related to the critical behavior of
$\Delta^*(\alpha\to 1)$ discussed above.  In order to later model our TMRG
$\chi^*(\alpha,t)$ data close to, but not in, the low-$t$ limit, we will need an
expression for $\varepsilon(\alpha,k)$ which is correct in the limit $\alpha\to 1$
and which also reproduces reasonably well the $\varepsilon(\alpha,k)$ of BRT.  We
found that the simple one-parameter ($\Delta^*$) form suggested earlier by one of us
in the context of the $S = 1/2$ two-leg ladder\cite{Johnston1996}
\begin{mathletters}
\label{EqMyE(k)}
\begin{equation}
\varepsilon(\Delta^*,k) = [{\Delta^*}^2 + f^2(\Delta^*)\sin^2(ka)]^{1/2}~,
\label{EqMyE(k):a}
\end{equation}
is satisfactory in these regards for the AF alternating-exchange chain over the
entire range $0\leq\Delta^*\leq 1$.  The function $f(\Delta^*)$ is determined here by
the sum rule~(\ref{EqNorm}), which yields the condition
\begin{equation}
{\rm E}\bigg[-{f^2(\Delta^*)\over{\Delta^*}^2}\bigg] =
{\pi\over 2\Delta^*}~,
\label{Eqf(a)}
\end{equation}
\end{mathletters}
where E($x$) is the complete elliptic integral of the second kind.  From
Eq.~(\ref{Eqf(a)}), $f$ varies nonlinearly with $\Delta^*$ from
$f(\Delta^*=0)=\pi/2$ to $f(\Delta^* = 1)=0$, as shown in
Fig.~\ref{E(ka)f(gap)Calc}.  From an independently determined dependence of
$\Delta^*$ on $\alpha$ as in Eq.~(\ref{EqDimParams2:a}), one can then determine
$f(\alpha)$ as also shown in Fig.~\ref{E(ka)f(gap)Calc}.  Using the fifth-order
$\Delta^*(\alpha)$ values of BRT  in Fig.~\ref{E(ka)AltChn}, the resulting dispersion
relations~(\ref{EqMyE(k)}) for
$\alpha = 0,$ 0.2, 0.4, 0.6, and 0.8 were calculated and are shown as the solid
curves in Fig.~\ref{E(ka)AltChn}, where they are seen to be in close agreement with
the respective dashed curves of BRT.  An important difference for large $\alpha$,
however, is that the $\varepsilon(\Delta^*,k)$ in Eqs.~(\ref{EqMyE(k)}) properly
reduces by construction to the exact $\varepsilon(\alpha,k)$ for $\alpha\to 1$,
whereas the one in Eq.~(\ref{EqBRTE(k)}) with a finite number of terms does not.

Close to the one-magnon band minimum, the square root and the sine function in
the dispersion relation in Eq.~(\ref{EqMyE(k):a}) can be expanded, yielding
\begin{equation}
\varepsilon(\Delta^*,k\to 0) \approx \Delta^* +
{1\over 2}\,{f^2(\Delta^*)\over\Delta^*}(ka)^2~.
\label{EqE(kto0)}
\end{equation}
A comparison of Eqs.~(\ref{EqE(kto0)}) and~(\ref{EqEps(k)}) shows that the parameter
$c^*$ in the formulas for $\chi^*(t\to 0)$ [Eqs.~(\ref{EqTroyer})] and $C(t\to 0)$
[Eq.~(\ref{EqCTroyer})] is a unique function of $\Delta^*$ which in our
approximation is given by
\begin{equation}
c^*(\Delta^*) = {1\over 2}\,{f^2(\Delta^*)\over\Delta^*}~,
\end{equation}
with $f^2(\Delta^*)$ given by Eq.~(\ref{Eqf(a)}).  Thus both $\chi^*(t\to 0)$ and
$C(t\to 0)$ for the alternating-exchange chain only depend on the single
parameter $\Delta^*$ (in addition to $t$).  Explicitly, we
obtain
\begin{equation}
\chi^*(t\to 0) = {1\over
\sqrt{2\pi}\,f(\Delta^*)}\,\Big({\Delta^*\over t}\Big)^{1/2}{\rm e}^{-\Delta^*/t}~.
\label{EqMyLoTChi}
\end{equation}
As might have been anticipated, the only thermodynamic variable is
the ratio $t/\Delta^* = k_{\rm B}T/\Delta$ of the thermal energy to the spin gap. 
The numerical prefactor depends explicitly (only) on the reduced spin gap $\Delta^*
\equiv
\Delta/J_1$.  Similarly, the magnetic specific heat is obtained as
\begin{eqnarray}
{C(t\to 0)\over N k_{\rm B}} = {3\over
\sqrt{2\pi}}&&{\Delta^*\over f(\Delta^*)}\Big({
\Delta^*\over t}\Big)^{3/2}\nonumber\\ &&\times\bigg[1 + {t\over\Delta^*} + {3\over
4}\Big({t\over\Delta^*}\Big)^2\bigg]{\rm e}^{-\Delta^*/t}~,
\label{EqCTroyer2}
\end{eqnarray}
where again the same characteristics are present as just discussed for $\chi^*(t)$. 
The variations of the prefactors with $\Delta^*$ for $\chi^*(t)$ and $C(t)$ can
both be ascertained from the plot of $f(\Delta^*)$ in Fig.~\ref{E(ka)f(gap)Calc}. 
In particular, when $\alpha \sim 1$ ($\delta\ll 1$) for which $\Delta^*\ll1$, $f$ is
nearly a constant.  For our and our readers' convenience when modeling materials
showing small spin gaps, we have fitted our numerical $f(\Delta^*)$ calculations for
$0\leq\Delta^*\leq0.4$ by a third-order polynomial to within 2 parts in $10^4$,
given by
\begin{equation}
f(\Delta^*) = {\pi\over 2} -0.034289 \Delta^* - 1.18953{\Delta^*}^2
+ 0.40030{\Delta^*}^3.
\label{Eqf(Delta^*):a}
\end{equation}

By a change in variables to $(J,\delta)$ and using the 
$\Delta^*(\alpha)$ in Eq.~(\ref{EqDimParams2:all}),\cite{Barnes1998} we obtain the
following forms which are more useful for modeling materials with small spin gaps,
especially those showing second-order spin dimerization transitions with decreasing
$T$:
\vglue-0.1in
\begin{mathletters}
\label{EqsChiCfbar:all}
\begin{equation}
\overline{\chi^*}(\overline{t}\to 0) = {1\over
(1+\delta)\sqrt{2\pi}\,f(\overline{\Delta^*})}\,\Big({\overline{\Delta^*}\over
\overline{t}}\Big)^{1/2}{\rm e}^{-\overline{\Delta^*}/\overline{t}}~.
\label{EqsChiCfbar:a}
\end{equation}
\begin{eqnarray}
{\overline{C}(t\to 0)\over N k_{\rm B}} = &&{3\over (1+\delta)§
\sqrt{2\pi}}{\overline{\Delta^*}\over\overline{f}(\overline{\Delta^*})
}\Big({\overline{\Delta^*}\over\overline{t}
}\Big)^{3/2}\nonumber\\ &&\times\bigg[1 +
{\overline{t}\over\overline{\Delta^*}} + {3\over
4}\Big({\overline{t}\over\overline{\Delta^*}}\Big)^2\bigg]{\rm
e}^{-\overline{\Delta^*}/\overline{t}}~,
\label{EqsChiCfbar:b}
\end{eqnarray}
\begin{equation}
\overline{f}(\overline{\Delta^*}) = {\pi\over 2} - 0.033933 \overline{\Delta^*} -
1.19607 {\overline{\Delta^*}}^2 + 0.92430{\overline{\Delta^*}}^3.
\label{EqsChiCfbar:c}
\end{equation}
\end{mathletters}
Note that in these formulas
$\Delta^*/t=\overline{\Delta^*}/\overline{t}=\Delta/(k_{\rm B}T)$.

\section{Theory: $\bbox{S = 1/2}$ Uniform Heisenberg Chain}
\label{SecUniformChnThy}

\subsection{Magnetic spin susceptibility}

\begin{figure}
\epsfxsize=3in
\centerline{\epsfbox{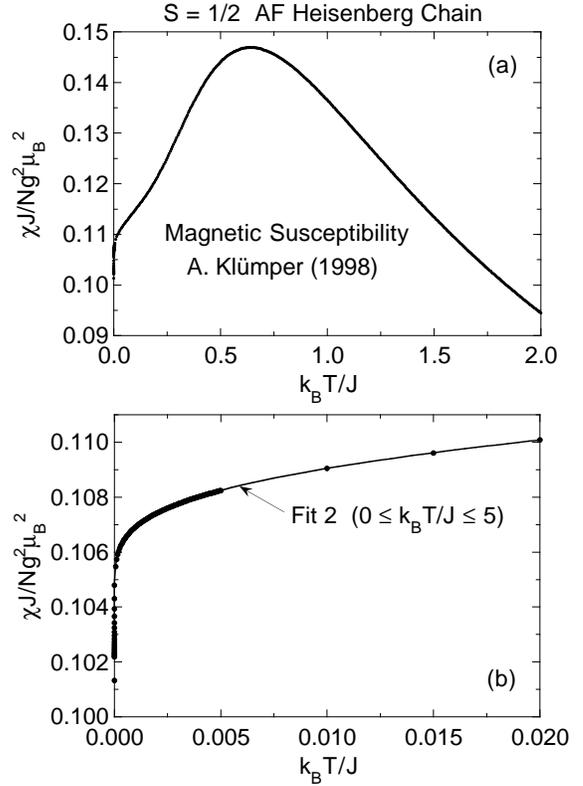}}
\vglue 0.1in
\caption{(a) Magnetic susceptibility $\chi$ versus temperature $T$ for
the spin $S = 1/2$ nearest-neighbor antiferromagnetic Heisenberg
chain (Ref.~\protect\onlinecite{Klumper1998}) ($\bullet$).  (b)  Expanded plot of the
data in (a) for $0\leq t \leq 0.02$, together with a fit (Fit~2, solid curve) obtained
in Sec.~\ref{SecChnFit} to the data of Kl\"umper and Johnston
(Ref.~\protect\onlinecite{Klumper1998}).  The fit is not shown in (a) because on the
scale of this figure the fit is indistinguishable from the data.}
\label{KlumperDatFit2}
\end{figure}

The uniform $S=1/2$ chain is one limit of the alternating-exchange chain with $J_{ij}
\equiv J,\ \alpha = 1,\ \delta = 0$, and with no spin gap [the $\chi^*(t\to0)$ and
$C(t\to 0)/t$ are finite]. The spin susceptibility was calculated accurately by
Eggert, Affleck and Takahashi in 1994,\cite{Eggert1994} and recently refined by
Kl\"umper\cite{Klumper1998} as shown in Fig.~\ref{KlumperDatFit2}(a) where only the
calculations up to $t = 2$ are shown.  An expanded plot of the data for $t\leq
0.02$, including the exact value $1/\pi^2$ at $t=0$, is shown in
Fig.~\ref{KlumperDatFit2}(b), along with a fit (Fit~2) to the data to be
derived and discussed in Sec.~\ref{SecChnFit}.  The most recent calculations of
Ref.~\onlinecite{Klumper1998} have an absolute accuracy estimated to be
$\approx 1\times 10^{-9}$ and show a broad maximum at a temperature $T^{\rm max}$ with
a value $\chi^{\rm max}$.  By fitting data points near the maximum by up to
6th order polynomials, we determined these numerical values to be given by
\begin{mathletters}
\label{EqChainPars:all}
\begin{equation}
T^{\rm max} = 0.6\,408\,510(4) J/k_{\rm B}~,\label{EqChainPars:a}
\end{equation}
\begin{equation}
\frac{\chi^{\rm max}J}{Ng^2\mu_{\rm B}^2} = 0.146\,926\,279(1)~,\label{EqChainPars:b}
\end{equation}
\begin{equation}
\chi^{\rm max}T^{\rm max} = 0.0\,941\,579(1) \frac{Ng^2\mu_{\rm B}^2}{k_{\rm
B}}~.\label{EqChainPars:c}
\end{equation}
\end{mathletters}
These values are consistent within the errors
with those found by Eggert {\it et al.},\cite{Eggert1994} but are much more accurate. 
For one mole of spins, setting $N = N_{\rm A}$ (Avogadro's number) in
Eq.~(\ref{EqChainPars:c}) yields
\begin{equation}
\chi^{\rm max}T^{\rm max} = 0.0\,353\,229(3)\, g^2\, {\rm \frac{cm^3\,K}{mol}}~.
\label{EqChimaxTmax}
\end{equation}
Note that the product $\chi^{\rm max}T^{\rm max}$ in
Eqs.~(\ref{EqChainPars:c}) and~(\ref{EqChimaxTmax}) is independent of $J$, and hence
is a good initial test of whether the $S = 1/2$ AF uniform Heisenberg chain model
might be applicable to a particular compound.  

\subsubsection{High-temperature series expansions}

The coefficients $c_n$ of the HTSE for $\chi^*(t)$, 
\begin{mathletters}
\label{EqHTS1:all}
\begin{equation}
4\chi^*t  = \sum_{n = 0}^\infty {c_n\over t^n}~,
\label{EqHTS1:a}
\end{equation}
are given up to ${\cal O}(1/t^7)$ by\cite{Obokata1967}
\begin{eqnarray}
c_0 & =	& 1~,~~c_1 =	-{1\over 2}~,~~c_2 =	0~,~~c_3 =	{1\over 24}~,~~c_4 =	{5\over
384}~,\nonumber\\
c_5 & =	& -{7\over 1280}~,~~c_6 =	-{133\over 30720}~,~~
c_7 =	{1\over 4032}~. \label{EqHTS1:b}
\end{eqnarray}
\end{mathletters}

\noindent Inverting the series, we obtain the corresponding $d_n$ coefficients in
Eq.~(\ref{EqHTSGen:a}) as
\begin{eqnarray}
d_0 & =	& 1~,~~d_1 =	{1\over 2}~,~~d_2 =	{1\over 4}~,~~d_3 =	{1\over 12}~,~~d_4 =
{1\over 128}~, \nonumber\\ 
d_5 & =	& -{29\over 3840}~,~~d_6 = -{317\over 92160}~,~~d_7 =	{11\over
71680}~. \label{EqHTS2}
\end{eqnarray}
The $d_n$ coefficients with $n = 0,$ 1, 2, and~3 are of course in agreement with
Eq.~(\ref{EqHTSAltChnCoeffs}) for $\alpha = 1$.

\subsubsection{Logarithmic corrections at low temperatures}
\label{SecLogCorr}

At low temperatures a simple expansion of thermodynamic properties in for instance
the variable $t$ is not possible. Such a nonanalyticity in $t$ can
be viewed as due to the strong correlation of the quasiparticles, i.e., the elementary
excitations of the system are not strictly free; they show rather nontrivial
scattering processes.  Spinons with low energies $\epsilon_1$ and $\epsilon_2$ have a
scattering phase $\phi(\epsilon_1,\epsilon_2)\simeq
\phi_0+\hbox{const}/|\log(\epsilon_1\epsilon_2)|$. From this property it is
clear\cite{McRae1998} that an expansion in the single variable $t$ is not possible,
but has to be supplemented by a term $1/\log(t)$. Although being very intuitive, this
physical picture on the basis of scattering processes of spinons has not played any
important role in the investigation of logarithmic corrections until
recently.\cite{Klumper1998}

Logarithmic dependencies of physical quantities have been known for the spin-1/2
Heisenberg chain for a rather long time.  Usually, a quantum chain with critical
couplings leads to critical correlations only in the thermodynamic limit $1/L=0$
and at $T=0$, where $L$ is the length of the chain. If one of these conditions is not
met the physical properties receive nonanalytic contributions in terms of $1/L$ or
$T$.  From the renormalization group point of view the existence of logarithmic
corrections is reflected by the perturbation of the (critical) fixed point
Hamiltonian by some marginal operator.  Such operators usually exist only for
isotropic systems.

The investigation of the size dependence of energy levels of
critical quantum chains was started more than a decade ago.  For the isotropic
spin-1/2 Heisenberg chain, expansions in $1/L$ and additional logarithmic
corrections ($1/L\log L$ etc.) were found in lattice approaches (Bethe
ansatz\cite{WoynE87,Nomura93,Karbach95}) as well as in field theory [RG study of the
Wess-Zumino-Witten (WZW) model with topological term
$k=1$ (Refs.~\onlinecite{Affleck1989,Cardy,LudwigCardy})].

Many of these earlier results are still relevant for the issues discussed in this
section due to an equivalence of many-particle systems at $T=0$, $1/L>0$
(groundstate properties of finite chains) and those at $T>0$, $1/L=0$
(thermodynamics of the bulk).  This leads to asymptotic series where $T$ and $1/L$
play very similar roles.  To our knowledge the first explicit report on $\log(T)$
corrections in the magnetic susceptibility resulting in an infinite slope at $T=0$
was given in Ref.~\onlinecite{Eggert1994}. Including higher order terms, the
asymptotic expansion $\chi_{\rm lt}^*(t)$ for $\chi^*(t)$
is\cite{Eggert1994,Klumper1998,McRae1998,Kim1998}
\begin{mathletters}
\label{EqChLoT:all}
\begin{eqnarray}
\chi_{\rm lt}^*(t) &=& {1\over \pi^2}\bigg[1 + {1\over 2{\cal L}} -
\frac{\ln\big({\cal L} + {1\over 2}\big)}{(2{\cal L})^2} + \cdots \bigg]~,
\label{EqChLoT:a}\\
&=& {1\over \pi^2}\bigg[1 + {1\over 2{\cal L}} -
\frac{\ln{\cal L}}{(2{\cal L})^2} - \frac{1}{(2{\cal L})^3}
+ \cdots \bigg]~,
\label{EqChLoT:b}
\end{eqnarray}
\begin{equation}
{\cal L} \equiv \ln(t_0/t)~,
\label{EqChLoT:c}
\end{equation}
\end{mathletters}
\vglue0.07in
\noindent where $t_0$ is a nonuniversal (undetermined) parameter.   In
Ref.~\onlinecite{Eggert1994} the field theoretical prediction on the basis of the
WZW model was compared with the results of thermodynamic Bethe ansatz calculations
and showed convincing agreement in an intermediate temperature regime.  Using up to
the first logarithmic correction term in Eq.~(\ref{EqChLoT:a}), Eggert, Affleck, and
Takahashi estimated $t_0 \approx 7.7$,\cite{Eggert1994} so at low temperatures $t
\lesssim 0.01$ the parameter ${\cal L} \gg 1$.

A general feature of field theoretical and lattice approaches is their restriction to
``low'' and ``high'' temperatures, respectively. Field theoretical studies suffer at
high temperatures from the neglect of (infinitely many) irrelevant operators. Lattice
studies show convergence problems at low temperatures as increasingly larger systems
have to be studied in order to avoid finite-size effects. In addition, the
comparison of field theory and lattice results can only verify or falsify the
universal aspects of an asymptotic expansion.  Non-universal quantities like $t_0$
which derive from some coupling constant of a marginal or irrelevant operator
(undetermined within the field theory) can at best be fitted as done
in Ref.~\onlinecite{Eggert1994}.

The latter problem of determining the coupling constants in an effective field theory
was solved by Lukyanov\cite{Lukyanov1997} who used a bosonic representation of the
Heisenberg chain. The coupling constants were fixed by a  comparison of the
susceptibility data $\chi(T=0,h)$ obtained by him with Bethe ansatz calculations for
magnetic field $h$ at $T=0$.  Eventually, the $\chi(T>0,h=0)$ data could be calculated
without any need of a fit parameter.

Lukyanov\cite{Lukyanov1997} obtained the following analytical low-temperature
expansion of $\chi^*(t)$,
\begin{mathletters}
\label{EqsLukyanov:all}
\begin{eqnarray}
\chi_{\rm lt,g}^*(t) = {1\over\pi^2}\bigg\{1 &+& {g\over 2} + {3 g^3\over 32} + {\cal
O}(g^4)\nonumber\\
&+& {\sqrt{3}\over \pi} t^2 [1 + {\cal O}(g)]\bigg\}~,\label{EqsLukyanov:a}
\end{eqnarray}
where $g$ obeys the transcendental equation
\begin{equation}
 {1\over 2}{\rm ln}g + {1\over g} = {\cal L}~\label{EqsLukyanov:b}
\end{equation}
or equivalently
\begin{equation}
\sqrt{g}\,{\rm e}^{1/g} = {t_0\over t}~,\label{EqsLukyanov:c}
\end{equation}
with a unique value of $t_0$ given by
\begin{equation}
t_0 = \sqrt{\pi\over 2}\ {\rm e}^{\gamma+(1/4)}\approx
2.866\,257\,058~,\label{EqsLukyanov:d}
\end{equation}
\end{mathletters}
where $\gamma \approx 0.577\,215\,665$ is Euler's constant.  Lukyanov showed that his
$\chi_{\rm lt,g}^*(t)$ is in agreement with the numerical data of Eggert, Affleck
and Takahashi\cite{Eggert1994} at low temperatures $t\geq 0.003$.

In the following, we will compare high-accuracy numerical Bethe ansatz
calculations carried out to much lower temperatures by Kl\"umper and
Johnston\cite{Klumper1998} with this theory\cite{Lukyanov1997} in some detail because
this theory is exact at low temperatures with no adjustable parameters.  The
calculations of Ref.~\onlinecite{Klumper1998} are based on lattice studies, however
without suffering from the usual shortcomings.  By means of a lattice path integral
representation of the finite temperature Heisenberg chain and the formulation of a
suitable quantum transfer matrix (both  quite analogous to the numerical TMRG
calculations presented later in this paper) a set of numerically well-posed
expressions for the free energy was derived. In more physical terms the method can
be understood as an application of the familiar though often rather vague concept of
quasiparticles to a quantitative description of the many particle system valid for
all temperatures $T$ and magnetic field values $h$.\cite{Klumper1998}  The
work can be understood as an evaluation of the full scattering
theory of spinons and antispinons.\cite{Klumper1998}

\begin{figure}
\epsfxsize=3in
\centerline{\epsfbox{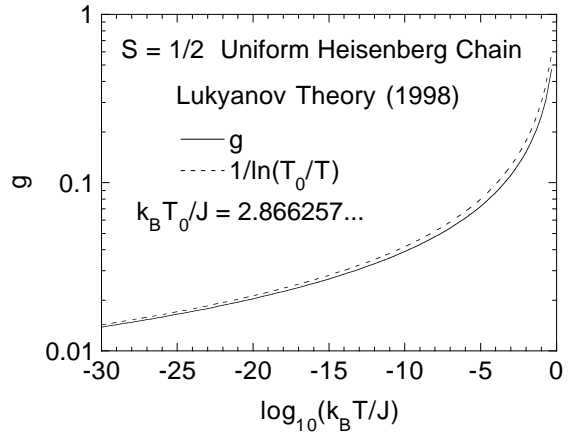}}
\vglue 0.1in
\caption{Log-log plot vs temperature $T$ of the function $g(T)$ (solid curve) and
its lowest-order approximation $g(T) = 1/\ln(T_0/T)$ (dashed curve) in Lukyanov's
theory (Ref.~\protect\onlinecite{Lukyanov1997}) for the $S=1/2$ antiferromagnetic
uniform Heisenberg chain over the temperature range $10^{-30} \leq t \leq 0.5$.}
\label{Lukyanov_g}
\end{figure}

Our iterative solution of Eq.~(\ref{EqsLukyanov:b}) yields the expansion
\begin{equation}
g = {1\over {\cal L}}\bigg\{1 - {\ln{\cal L}\over 2{\cal L}} + {(\ln{\cal
L})^2 - \ln{\cal L}\over (2{\cal L})^2} + {\cal O}\bigg[{1\over (2{\cal
L})^3}\bigg]\bigg\}~.
\label{EqgSoln}
\end{equation}
A log-log plot of $g$ vs $t$ obtained by numerically solving
Eq.~(\ref{EqsLukyanov:c}) is shown in Fig.~\ref{Lukyanov_g} (solid curve), along
with its lowest-order approximation $g(t) \approx 1/{\cal L} = 1/\ln(t_0/t)$ (dashed
curve).  This approximation is 1.1\% larger than the exact result at
$t=10^{-30}$, with the discrepancy increasing steadily to 5.8\% at $t = 10^{-15}$
and 8.5\% at $t = 10^{-7}$.  Substituting Eq.~(\ref{EqgSoln})
into~(\ref{EqsLukyanov:a}) gives
\begin{eqnarray}
\chi_{\rm lt,log}^*(t) = {1\over \pi^2}\bigg\{1 &+& {1\over 2{\cal L}} -
\frac{\ln{\cal L}}{(2{\cal L})^2}\nonumber\\
&+& \frac{(\ln{\cal L})^2-\ln{\cal L}+(3/4)}{(2{\cal L})^3}
+{\cal O}\bigg[{1\over (2{\cal L})^4}\bigg]\nonumber\\ 
&+& {\sqrt{3}\over \pi}\,t^2 \Big[1 + {\cal O}\Big({1\over 2{\cal L}}\Big)\Big]
\bigg\}~.
\label{EqLukLog}
\end{eqnarray}
The first three terms are identical with those in Eq.~(\ref{EqChLoT:b}), but the
constant term in the numerator of the fourth term is not the same as in
Eq.~(\ref{EqChLoT:b}), indicating that Eq.~(\ref{EqChLoT:b}) is not accurate to
order higher than ${\cal O}[1/(2{\cal L})^2]$.

\begin{figure}
\epsfxsize=3in
\centerline{\epsfbox{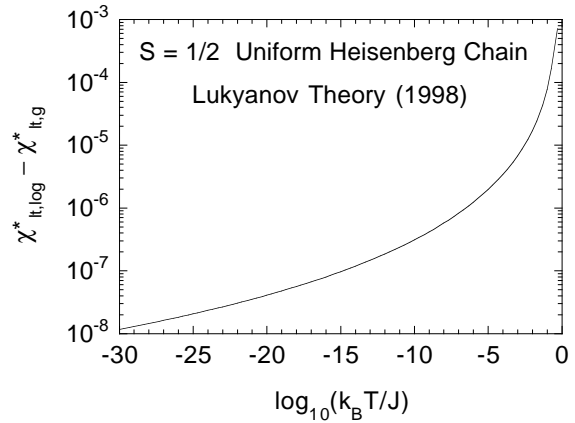}}
\vglue 0.1in
\caption{Log-log plot vs temperature $T$ of the difference between our approximate
logarithmic expansion $\chi^*_{\rm lt,log}$ of Lukyanov's
theory (Ref.~\protect\onlinecite{Lukyanov1997}) and his exact prediction $\chi^*_{\rm
lt,g}$ for the low-temperature limit of the magnetic susceptibility of the spin $S =
1/2$ antiferromagnetic uniform Heisenberg chain, for the temperature range $10^{-30}
\leq t \leq 0.5$.}
\label{LukChilog-Chig}
\end{figure}

An important issue is the accuracy to which the log expansion approximation
$\chi^*_{\rm lt,log}(t)$ in Eq.~(\ref{EqLukLog}) reproduces the $\chi^*_{\rm
lt,g}(t)$  prediction of the original Eqs.~(\ref{EqsLukyanov:all}).  We have
calculated both quantities to high accuracy and plot the difference vs $t$, for the
range $10^{-30}\leq t \leq 0.5$, in Fig.~\ref{LukChilog-Chig}.  The $\chi^*_{\rm
lt,log}(t)$ is seen to increasingly diverge from $\chi^*_{\rm lt,g}(t)$ with
increasing $t$.

When comparing the predictions of Lukyanov's theory with numerical results such
as obtained from the Bethe ansatz, it is important to know at what temperature the
low temperature expansion in Eqs.~(\ref{EqsLukyanov:all}) ceases to be accurate
(``accurate'' must be defined) with increasing temperature.  There are three aspects
of this issue that need to be addressed.  The first and second aspects concern the
temperatures at which the unknown ${\cal O}(g^4)$ and ${\cal O}(g)$ terms in
Eq.~(\ref{EqsLukyanov:a}) become significant, respectively; we will return to these
two issues shortly.  The third aspect is whether the log expansion approximation
$\chi^*_{\rm lt,log}(t)$ in Eq.~(\ref{EqLukLog}) can be used in this comparison. 
The absolute accuracy of the most recent Bethe ansatz calculations\cite{Klumper1998}
is estimated to be $\approx 1\times 10^{-9}$.  From Fig.~\ref{LukChilog-Chig}, we see
that $\chi^*_{\rm lt,log}(t)$ approximates $\chi^*_{\rm lt,g}(t)$ to this degree only
for temperatures $t\lesssim 10^{-30}$ [we infer that the previous
Eqs.~(\ref{EqChLoT:all}) only apply to this accuracy at similarly very low
temperatures].  Therefore, to avoid this unnecessary approximation as a source of
error at higher temperatures, we will henceforth compare the numerical Bethe ansatz
calculations with $\chi^*_{\rm lt,g}(t)$ calculated from Lukyanov's original
Eqs.~(\ref{EqsLukyanov:all}).

\begin{figure}
\epsfxsize=3in
\centerline{\epsfbox{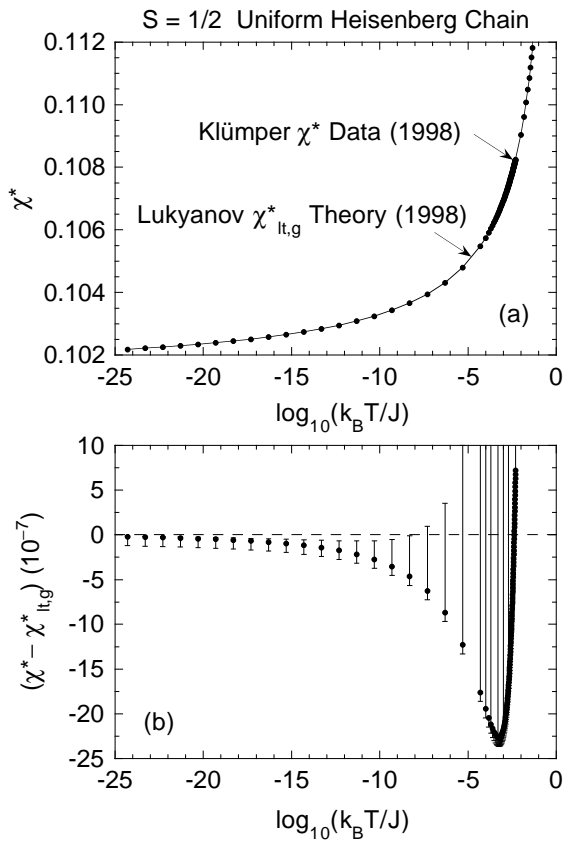}}
\vglue 0.1in
\caption{Semilog plots vs temperature $T$ at low $T$ of (a) 
numerical Bethe ansatz magnetic susceptibility ($\chi^*$) data for the $S=1/2$
uniform Heisenberg chain (Ref.~\protect\onlinecite{Klumper1998}) ($\bullet$) and the
prediction $\chi^*_{\rm lt,g}$ (solid curve) of Lukyanov's
theory (Ref.~\protect\onlinecite{Lukyanov1997}), and (b) the difference between these
two results ($\bullet$).  In (b), the upper error bar is the estimated uncertainty in
$\chi^*_{\rm lt,g}$ (see text).}
\label{KlumperLukyanov}
\end{figure}

\vglue0.08in
A comparison of the low-temperature Bethe ansatz $\chi^*(t)$
calculations\cite{Klumper1998} and Lukyanov's theoretical $\chi^*_{\rm lt,g}$
prediction is shown in Fig.~\ref{KlumperLukyanov}(a).  On the scale of this figure,
the two results are identical.  The (small) quantitative differences between them
are shown as the filled circles in Fig.~\ref{KlumperLukyanov}(b).  The lower error
bar on each data point in Fig.~\ref{KlumperLukyanov}(b) is $1\times 10^{-7}$ to
indicate the scale.  The upper error bar is the estimated uncertainty in $\chi^*_{\rm
lt,g}$ arising from the presence of the unknown ${\cal O}(g^4)$ and higher-order
terms in Eq.~(\ref{EqsLukyanov:a}), which was set to $g^4(t)/\pi^2$; the uncertainty
in the $t^2$ contribution, $\sim\sqrt{3}t^2g(t)/\pi^3$, is negligible at low $t$
compared to this.  At the lower temperatures, the data agree extremely well with the
prediction of Lukyanov's theory.  At the highest temperatures $t\gtrsim 10^{-3}$,
higher or-

\noindent der $t^n$ terms also become important, as inferred from our empirical fits
(Fits~1 and~2) below to the numerical data.

Irrespective of the uncertainties in the theoretical prediction at high temperatures
just discussed, we can safely conclude directly from Fig.~\ref{KlumperLukyanov}(b)
that the Bethe ansatz $\chi^*(t)$ data\cite{Klumper1998} are in agreement with the
exact theory of Lukyanov\cite{Lukyanov1997} to within an absolute accuracy of
$1\times 10^{-6}$ (relative accuracy $\approx 10$ ppm) over a temperature range
spanning 18 orders of magnitude from $t = 5\times 10^{-25}$ to $t = 5\times
10^{-7}$.  The agreement is much better than this at the lower temperatures.

\subsection{Magnetic specific heat}

The magnetic specific heat $C$ of the $S = 1/2$ AF uniform Heisenberg chain was
recently calculated to high accuracy by Kl\"umper and Johnston over the temperature
range $5\times 10^{-25} \leq k_{\rm B}T/J \leq 5$.\cite{Klumper1998}  The accuracy is
estimated to be $3\times 10^{-10}C(t)$.  The results for $T\leq 2 J/k_{\rm B}$ are
shown in Fig.~\ref{KlumperCp}(a).  The initial $T$ dependence is approximately (see
below) \,linear, and \,is \,given \,exactly \,in

\begin{figure}
\epsfxsize=3in
\centerline{\epsfbox{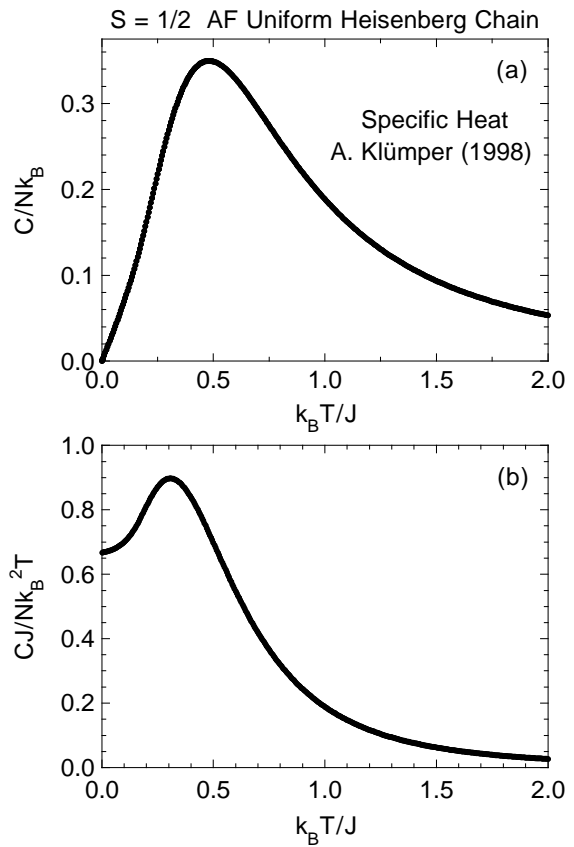}}
\vglue 0.1in
\caption{(a) Specific heat $C$ vs temperature $T$ ($\bullet$) for the $S = 1/2$
antiferromagnetic uniform Heisenberg chain (Ref.~\protect\onlinecite{Klumper1998}).
(b)~Specific heat coefficient $C/T$ vs $T$ from the data in~(a).  The area under the
curve in~(b) from $T = 0$ to $T = 5 J/k_{\rm B}$ is 99.4\% of $\ln 2$.}
\label{KlumperCp}
\end{figure}

\noindent the $t = 0$ limit by
\begin{equation}
{C(t\to 0)\over N k_{\rm B}} = {2\over 3}\,t~.
\label{EqC(tto0)}
\end{equation}
The data show a maximum with a value $C^{\rm max}$ at a temperature $T_C^{\rm
max}$.  By fitting 3--7 data points in the vicinity of the maximum by up to 6th order
polynomials, these values were found to be
\begin{eqnarray}
{k_{\rm B}T_C^{\rm max}\over J} &=& 0.48\,028\,487(1)~,\nonumber\\
\\
{C^{\rm max}\over Nk_{\rm B}} &=& 0.3\,497\,121\,235(2)~.\nonumber
\end{eqnarray}

The electronic specific heat coefficient $C(T)/T$ is plotted vs temperature in
Fig.~\ref{KlumperCp}(b).  As expected from Eq.~(\ref{EqC(tto0)}), the data approach
the value $(2/3)Nk_{\rm B}^2/J$ for $t\to 0$.  The initial deviation from this
constant value is positive and approximately (see below) quadratic in $t$.  The data
exhibit a smooth maximum with a value $(C/T)^{\rm max}$ at a temperature $T_{\rm
C/T}^{\rm max}$, values which we determined by fitting polynomials to the data in the
vicinity of the peak to be
\begin{figure}
\epsfxsize=3.4in
\centerline{\epsfbox{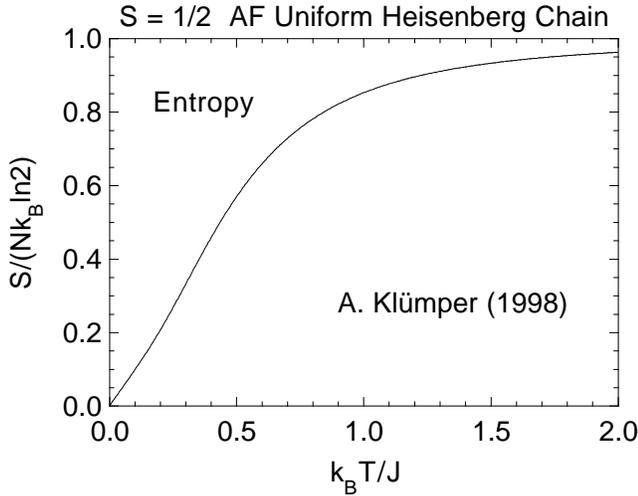}}
\vglue 0.1in
\caption{Entropy $S$ vs temperature $T$ for the $S = 1/2$ antiferromagnetic uniform
Heisenberg chain, obtained from the data in Fig.~\protect\ref{KlumperCp}(b).  The
entropy is normalized by $S(T = \infty) = N k_{\rm B}\ln\,2$.}
\label{KlumperS(T)}
\end{figure}

\begin{eqnarray}
{k_{\rm B}T_{C/T}^{\rm max}\over J} &=& 0.30\,716\,996(2)~,\nonumber\\
\\
{(C/T)^{\rm max}J\over Nk_{\rm B}^2} &=& 0.8\,973\,651\,576(5)~.\nonumber
\end{eqnarray}

The magnetic entropy $S(T)$ is determined by integrating the $C(T)/T$ data in
Fig.~\ref{KlumperCp}(b) vs $T$ and the result, normalized by $S(T\to\infty) = N
k_{\rm B}\ln2$, is plotted vs $T$ in Fig.~\ref{KlumperS(T)}.  This figure allows
one to estimate the maximum magnetic entropy that can be associated with a
dimerization transition or any other magnetic transition involving $S = 1/2$
Heisenberg chains which are weakly coupled to each other [assuming that the
(average) $J$ does not change at the transition].  For example, for
${\rm NaV_2O_5}$ where $k_{\rm B}T_{\rm c}/J\approx 0.057$, one can estimate from
Fig.~\ref{KlumperS(T)} that the magnetic entropy at $T_{\rm c}$ cannot exceed $0.056
R\ln 2 = 0.32$\,J/mol\,K, where $R$ is the molar gas constant.  The reason this value
is the upper limit is that magnetic critical fluctuations will increase the specific
heat, and hence the entropy, above $T_{\rm c}$ and thus reduce it at (and below)
$T_{\rm c}$, by conservation of magnetic entropy, compared to the values for the
isolated chain at the same reduced temperatures.  Similarly, the $C(T)$ data in
Fig.~\ref{KlumperCp}(a) allow one to estimate the minimum lattice specific heat
contribution $C^{\rm lat}(T)$ above $T_{\rm c}$ if the $C^{\rm lat}(T)$ has not been
determined previously from experiments and/or theory directly.

At low temperatures, the electronic specific heat coefficient $C(T)/T$ becomes
independent of temperature (apart from logarithmic corrections, see below), as does
the spin susceptibility $\chi^*(t)$, just as in a metal (Fermi liquid).  Therefore
it is of interest to compute a normalized ratio of these two quantities.  For a
metal, the relevant \ \,quantity \ \,is \ \,the \ \,Wilson-Sommerfeld \ \,ratio, \
\,which \ \,for 
\begin{figure}
\epsfxsize=3.5in
\centerline{\epsfbox{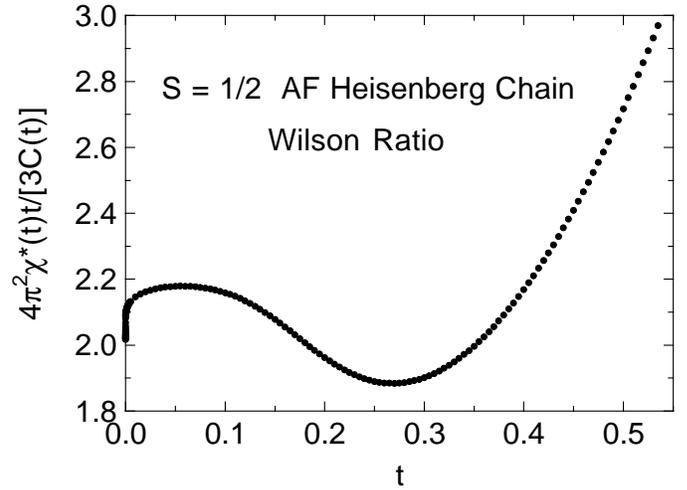}}
\vglue 0.1in
\caption{\mbox{The Wilson-Sommerfeld ratio} $R_{\rm W} = 4\pi^2\chi^*(t)t/$ $[3 C(t)]$
between the spin susceptibility and electronic specific heat coefficient for the $S
= 1/2$ AF Heisenberg chain vs reduced temperature $t = k_{\rm B}T/J$.  In the Wilson
ratio, we have set $k_{\rm B} = 1$.}
\label{KlumperWilsonRatio}
\end{figure}
\noindent $S = 1/2$
quasiparticles reads, in the notation of this paper and with $k_{\rm B}$ set to 1,
\begin{equation}
R_{\rm W}(t) = {4\pi^2\chi^*(t)t\over 3 C(t)}~.
\label{EqWilRat}
\end{equation}
\vglue0.1in
\noindent In a degenerate free electron gas, $R_{\rm W} = 1$ and is independent of
$t$.  For exchange-enhanced metals $1 < R_{\rm W} \lesssim 10$, for $S = 1/2$ Kondo
impurities in a metal the Wilson ratio associated with the impurities is $R_{\rm W}
= 2$, and for many heavy fermion metals $R_{\rm W} \sim 2$.\cite{Johnston1999a} 
Plotted in Fig.~\ref{KlumperWilsonRatio} is $R_{\rm W}(t)$ for the $S = 1/2$ AF
Heisenberg chain, where $C(t)/t$ and $\chi^*(t)$ were given above in
Figs.~\ref{KlumperCp}(b) and~\ref{KlumperDatFit2}, respectively.  For $t\to 0$,
the Wilson ratio for the $S = 1/2$ Heisenberg chain is exactly 2.  With
increasing $t$, $R_{\rm W}$ is seen to be nearly independent of $t$ to within
$\pm 10$\% up to $t \approx 0.4$, but the influence of the logarithmic corrections to
both $\chi(T)$ and $C(T)$ are quantitatively important.  Although the logarithmic
corrections for $\chi(T)$ and $C(T)$ oppose each other in their ratio in $R_{\rm
W}(t)$, the logarithmic corrections for $\chi(t)$ win out, giving a net $\sim 10$\%
increase in $R_{\rm W}(t)$ with increasing $t$ at low $t$.  At higher $t$, the system
crosses over to the expected local moment Heisenberg behavior where $R_{\rm
W}\propto t^2$.  Thus as far as the thermodynamics is concerned, the uniform
Heisenberg chain behaves at low temperatures as expected for a Fermi liquid, apart
from the influence of the logarithmic corrections.  This quasi-Fermi liquid behavior
arises because the elementary excitations at low temperatures are
$S = 1/2$ spinons which are fermions with a Fermi surface (i.e., Fermi points in
one dimension).  Since the spinons carry no charge, the chain is an insulator. 
The deviation of the Wilson ratio from unity and the logarithmic corrections are due
to spinon interactions.

\subsubsection{High-temperature series expansions}

The HTSE for the specific heat of a spin $S$ AF uniform Heisenberg chain
is\cite{Rushbrooke1958}
\begin{mathletters}
\label{EqCHTSEGen0:all}
\begin{equation}
{C(T)\over Nk_{\rm B}} = {x^2\over 3 t^2}\bigg[1 + \sum_{n=1}^\infty {c_n(x)\over
t^n}\bigg]~,
\end{equation}
\begin{equation}
x = S(S+1)~,~~~t = {k_{\rm B}T\over J}~,
\end{equation}
\[
c_1 = {1\over 2}~,~~c_2 = {1\over 15}(3 - 8 x - 3 x^2)~,
\]
\[
c_3 = {1\over36}(3 - 16 x - 4 x^2)~,
\]
\[
c_4 = {1\over5040}(192 - 1432 x + 1123 x^2 + 800 x^3 + 160 x^4)~,
\]
\begin{equation}
c_5 = {1\over21600}(414 - 3768 x + 6635 x^2 + 2624 x^3 + 480 x^4)~.
\end{equation}
\end{mathletters}
Specializing Eqs.~(\ref{EqCHTSEGen0:all}) to $S = 1/2$ $(x = 3/4$) then gives
\begin{mathletters}
\label{EqCHTSEGen:all}
\begin{equation}
{C(T)\over Nk_{\rm B}} = {3\over 16 t^2}\bigg[1 + \sum_{n=1}^\infty {c_n\over
t^n}\bigg]~,
\label{EqCHTSEGen:a}
\end{equation}
\begin{equation}
c_1 = {1\over 2}~,~~c_2 = c_3 = -{5\over 16}~,~~c_4 = {7\over256}~,~~c_5 =
{917\over7680}~.
\label{EqCHTSEGen:b}
\end{equation}
\end{mathletters}
The two $C(T)$ HTSE terms of order $1/t^2$ and $1/t^3$ in
Eqs.~(\ref{EqCHTSEGen:all}) are in agreement with the general expression for the two
lowest-order HTSE expansion terms for $C(T)$ of the $S = 1/2$ alternating-exchange
Heisenberg chain in Eq.~(\ref{EqHTSECGen2term}) with alternation parameter $\alpha =
1$.

In a later section, the Bethe ansatz $C(T)$ data\cite{Klumper1998} will be fitted to
obtain a function accurately representing the $C(T)$ of the $S = 1/2$ AF uniform
Heisenberg chain.  In order that we are not required to change our fitting equations
from those we use for fitting magnetic susceptibility data, the coefficients
for the series inverted from that in Eq.~(\ref{EqCHTSEGen:a}) are required.  We
obtain
\begin{mathletters}
\label{EqCHTSEGen2:all}
\begin{equation}
{C(T)\over Nk_{\rm B}} = {3\over 16 t^2}\bigg[1 + \sum_{n=1}^\infty {d_n\over
t^n}\bigg]^{-1}~,
\label{EqCHTSEGen2:a}
\end{equation}
\begin{equation}
d_1 = -{1\over 2},~~d_2 = {9\over 16},~~d_3 = -{1\over 8},~~d_4 =
{7\over128},~~d_5 = {7\over1920}~.
\label{EqCHTSEGen2:b}
\end{equation}
\end{mathletters}
\begin{figure}
\epsfxsize=3.1in
\centerline{\epsfbox{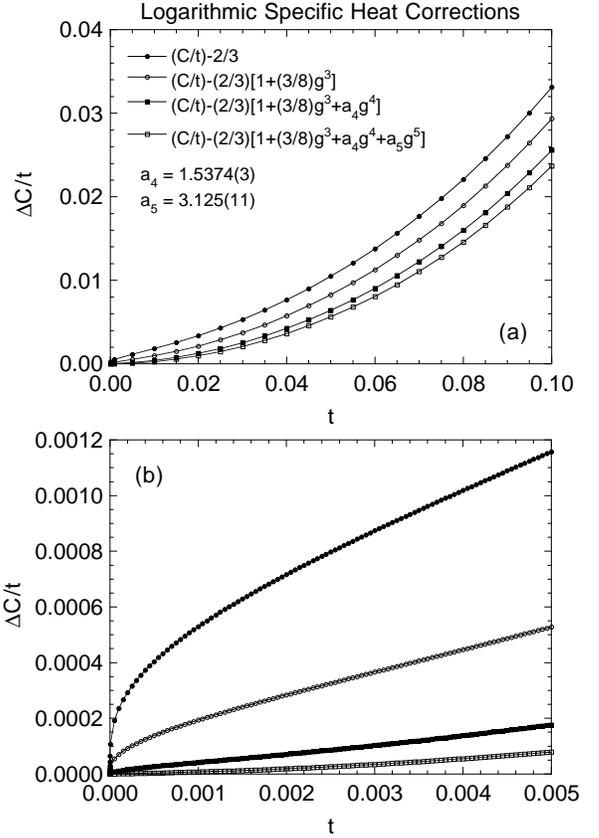}}
\vglue 0.1in
\caption{(a) Difference $\Delta C/t$ between the electronic specific heat coefficient
from the Bethe ansatz data (Ref.~\protect\onlinecite{Klumper1998}) and the nominal
coefficient of 2/3 (top data set), plotted vs reduced temperature $t$.  Moving down
the figure, successive data sets show the influence of correcting for cumulative
logarithmic correction terms.  (b) Expanded plots at low temperatures.  The
reduced temperature is
$t = k_{\rm B}T/J$ and we have set $N = k_{\rm B} = 1$.  In both (a) and (b), the
lines connecting the data points are guides to the eye.}
\label{KlumperLukCgCorr/T}
\end{figure}

\subsubsection{Low-temperature logarithmic corrections}
\label{SecCLogCorr}

At first sight, from Fig.~\ref{KlumperCp} there appear to be no singularities in
the temperature dependence of the specific heat for the $S = 1/2$ AF uniform
Heisenberg chain.  However, if the electronic specific heat coefficient $C(T)/T$ is
examined in detail, one sees anomalous behavior at low temperatures.  Shown as the
top curve in Fig.~\ref{KlumperLukCgCorr/T}(a) is a plot of the difference between
the electronic specific heat coefficient and its zero-temperature value, $\Delta
C(t)/Nk_{\rm B}t\equiv [C(t) - (2/3)t]/(Nk_{\rm B}t)$ for $0\leq t\leq 0.1$ [compare
with Fig.~\ref{KlumperCp}(b)].  From this figure, there is still nothing
particularly strange about the data.  However, upon further expanding the plot to
study the range $0\leq t\leq 0.005$ as shown in Fig.~\ref{KlumperLukCgCorr/T}(b), we
see that $\Delta C/Nk_{\rm B}t$ is developing an infinite slope as $t \to 0$.  This
is the signature of the existence of logarithmic corrections to the specific heat at
temperatures $t\ll1$, just as it was for the magnetic susceptibility.

\begin{figure}
\epsfxsize=3in
\centerline{\epsfbox{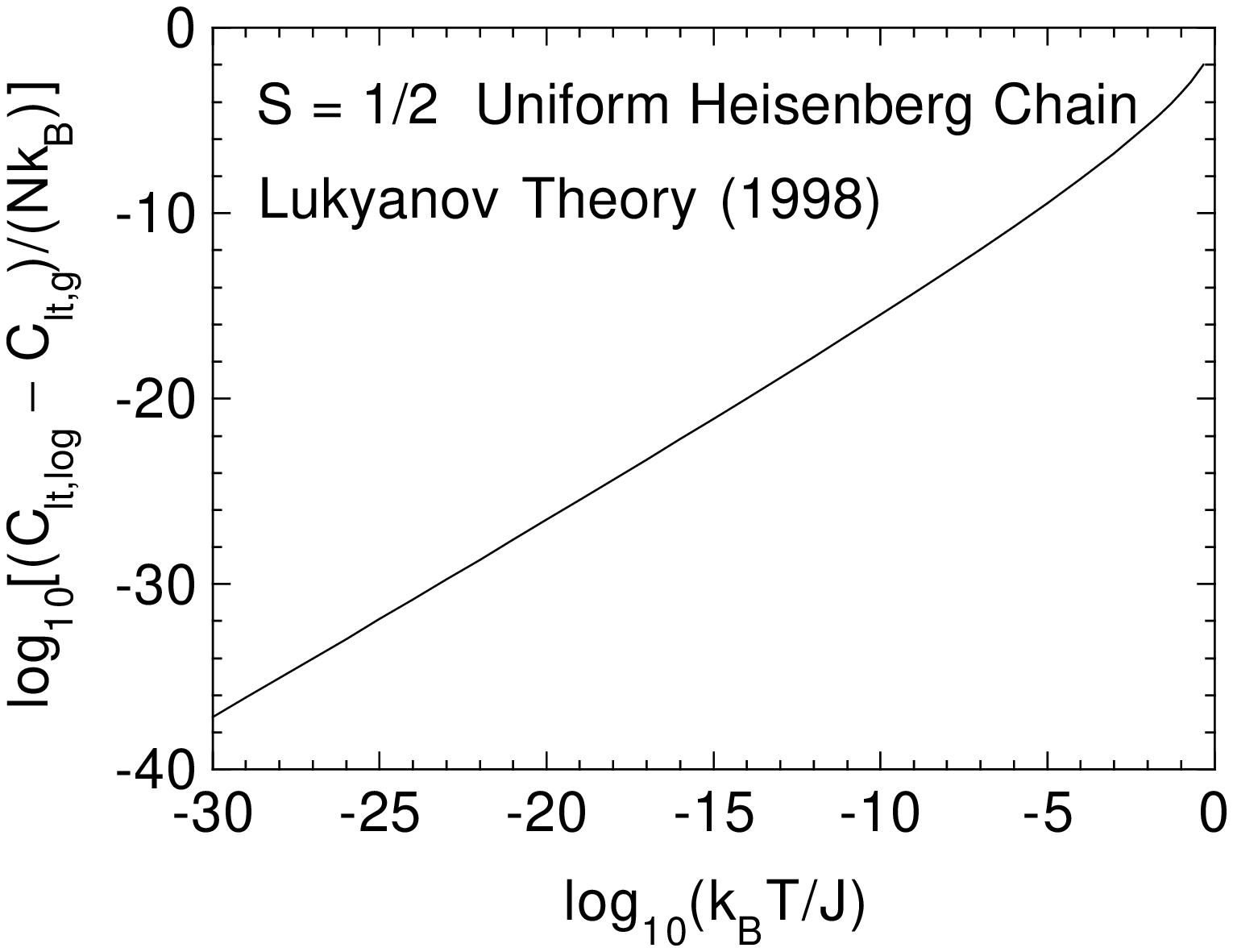}}
\vglue 0.1in
\caption{Log-log plot vs temperature $T$ of the difference between our approximate
logarithmic expansion $C_{\rm lt,log}/(N k_{\rm B})$ of Lukyanov's
theory (Ref.~\protect\onlinecite{Lukyanov1997}) and his exact prediction $C_{\rm
lt,g}/(N k_{\rm B})$ for the low-temperature limit of the magnetic specific heat of
the spin $S = 1/2$ antiferromagnetic uniform Heisenberg chain, for the temperature
range $10^{-30} \leq t \leq 0.5$.}
\label{LukClog-Cg}
\end{figure}

Kl\"umper,\cite{Klumper1998} Lukyanov,\cite{Lukyanov1997} and others have
found a logarithmic correction to the low-$t$ limit in Eq.~(\ref{EqC(tto0)}).  
Lukyanov's exact asymptotic expansion for the free energy per spin in zero
magnetic field is
\begin{eqnarray}
f = -J\ln 2 &-& {(k_{\rm B}T)^2\over 3 J}\Big[1 + {3\over 8}\,g^3 + {\cal
O}(g^4)\Big]\nonumber\\
\nonumber\\
&-& {3^{3/2}(k_{\rm B}T)^4\over 10\pi J^3}[1 + {\cal O}(g)]~,
\end{eqnarray}
where $g(t/t_0)$ and $t_0$ are the same as given in Eqs.~(\ref{EqsLukyanov:b})
and~(\ref{EqsLukyanov:d}), \,respectively, and \,where \,$g(t)$ \,was \,plotted \,in
Fig.~\ref{Lukyanov_g}.  The specific heat at constant volume is calculated
using
$C = -T\partial^2f/\partial T^2$, yielding
\begin{eqnarray}
{C_{\rm lt,g}(T)\over N k_{\rm B}} = {2 k_{\rm B}T\over 3 J}\Big[1 &+& {3\over
8}\,g^3 + {\cal O}(g^4)\Big]\nonumber\\
\nonumber\\
&+& {2(3^{5/2})\over 5\pi}\Big({k_{\rm B}T\over J}\Big)^3[1 + {\cal O}(g)]~.
\label{EqCLusnikov}
\end{eqnarray}
This formula shows that the electronic specific heat coefficient $C(T)/T$
increases quadradically with $T$ at low $T$ (after subtracting the logarithmic 
corrections).  The numerical prefactor of the $t^3$ term is $1.98478\cdots$.  If
the approximate expansion for $g({\cal L})$ in Eq.~(\ref{EqgSoln}) is substituted
into Eq.~(\ref{EqCLusnikov}), one obtains
\begin{eqnarray}
{C_{\rm lt,log}(T)\over N k_{\rm B}}\ \ = &\mbox{}&{2 k_{\rm B}T\over 3 J}\bigg\{1 +
{3\over (2{\cal L})^3} + {\cal O}\Big[{1\over (2{\cal L})^4}\Big]\bigg\}\nonumber\\
\nonumber\\
&+& {2(3^{5/2})\over 5\pi}\Big({k_{\rm B}T\over J}\Big)^3\Big[1 + {\cal
O}\Big({1\over 2{\cal L}}\Big)\Big]~,
\end{eqnarray}
\vglue-0.06in
\noindent where the prefactor 3/8 in the logarithmic correction term was
found independently  by Kl\"umper,\cite{Klumper1998} confirming
\begin{figure}
\epsfxsize=3.02in
\centerline{\epsfbox{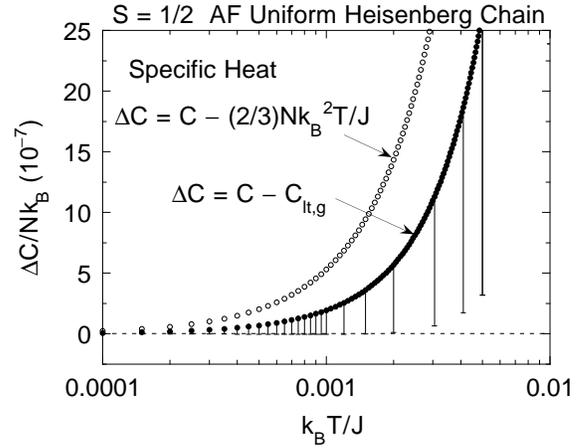}}
\vglue 0.1in
\caption{Semilog plot vs temperature $T$ of the difference $\Delta C = C -C_{\rm
lt,g}\ (\bullet)$ between the Bethe ansatz numerical specific heat $C(T)$
data (Ref.~\protect\onlinecite{Klumper1998}) and Lukyanov's
theory (Ref.~\protect\onlinecite{Lukyanov1997}) [$C_{\rm lt,g}(T)$] for the spin $S =
1/2$ antiferromagnetic uniform Heisenberg chain at low temperatures.  The error bar on
each data point is an estimated uncertainty in the theory due to higher order
correction terms that were not calculated.  Also shown is the deviation $\Delta C = C
- (2/3)N k_{\rm B}^2T/J\ (\circ)$ of the numerical data from the $T \to 0$ limit of
$C(T)$.}
\label{KlumperLukyanovCDev}
\end{figure}
\noindent Refs.~\onlinecite{Affleck1989} and~\onlinecite{Karbach95}.  The difference
between
$C_{\rm lt,log}(T)$ and $C_{\rm lt,g}(T)$ is plotted vs temperature in
Fig.~\ref{LukClog-Cg}, where the difference becomes $> 10^{-10}$ only for
$t\gtrsim 10^{-5}$.

Shown in Fig.~\ref{KlumperLukyanovCDev} is the deviation $\Delta C/N k_{\rm B}$
($\bullet$) of the Bethe ansatz data\cite{Klumper1998} from Lukyanov's theoretical
prediction in Eq.~(\ref{EqCLusnikov}).  For temperatures $t\lesssim 10^{-4}$, the
agreement is better than $10^{-8}$.  At higher temperatures, the uncertainty in the
theoretical prediction due to the unknown ${\cal O}(g^4)$ and higher order correction
terms becomes an important factor in the comparison.  The length of the error bar on
each data point in Fig.~\ref{KlumperLukyanovCDev} has arbitrarily been set to $(4/3)t
g^4(t)$ [cf. Eq.~(\ref{EqCLusnikov})]; the ${\cal O}(g)$ uncertainty in the $T^3$
term is negligible compared to this.  Also plotted in Fig.~\ref{KlumperLukyanovCDev}
is the deviation of the numerical data from the extrapolated linear low-$T$ behavior
($\circ$).  A comparison of the two data sets indicates that the ${\cal
O}(g^3)$ logarithmic correction term is responsible for at least most of this latter
difference for temperatures $t\lesssim 0.001$.

\begin{figure}
\epsfxsize=3in
\centerline{\epsfbox{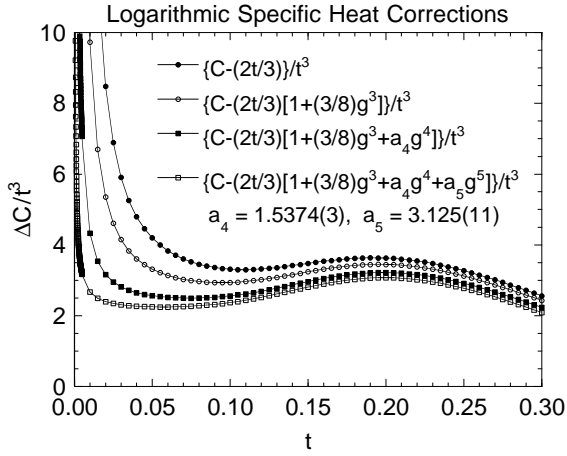}}
\vglue 0.1in
\caption{Coefficient $\Delta C/t^3$ of the expected $t^2$ dependence of the
electronic specific heat coefficient at low temperatures (top data set). 
Moving down the figure, successive data sets show the influence of
correcting for cumulative logarithmic correction terms.  If all logarithmic
corrections were accounted for, the data would become independent of $t$ for
$t\to 0$.  Here, the reduced temperature is $t = k_{\rm B}T/J$ and we have set
$N = k_{\rm B} = 1$.  The lines connecting the data points are guides to the eye.}
\label{KlumperLukCgCorr}
\end{figure}

A more rigorous evaluation of the influence of the above logarithmic correction term
is obtained by correcting for it in the plot of $\Delta C/t$ vs $t$, as shown by the
second curve from the top  in each of Figs.~\ref{KlumperLukCgCorr/T}(a)
and~\ref{KlumperLukCgCorr/T}(b).  From the latter figure, we infer that although
subtracting this correction term from the data helps to remove the zero-temperature
singularity, a singularity is still present but with reduced amplitude.  This means
that additional logarithmic correction terms are important, within the accuracy and
precision of the data.  Another indication of this is shown in
Fig.~\ref{KlumperLukCgCorr}, where we have plotted $\Delta C/t^3$ vs $t$.  According
to Eq.~(\ref{EqCLusnikov}), after accounting for the logarithmic correction
term(s), the result should be independent of $t$ at low $t$.  Instead, both before 
and after accounting for the log correction term, there \,is \,a strong 
\noindent upturn at low
temperatures although the strength of the upturn is smaller after subtracting the
influence of the log correction term.

The numerical Bethe ansatz specific heat data\cite{Klumper1998} are sufficiently
accurate and precise that we can estimate the coefficients of the next two
logarithmic correction ($g^4,\ g^5$) terms in Eq.~(\ref{EqCLusnikov}) from these data
as follows.  From Eq.~(\ref{EqCLusnikov}), if we plot the numerical data as
$[C(t)/(Nk_{\rm B}t) - (2/3)(1+3g^3/8)]/g^4$ vs $g$ at low temperatures, where the
$t^3$ term can be neglected, and fit the lowest $t$ data by a straight line, the $y$
intercept for $g\to 0$ gives the coefficient of the $g^4$ term and the slope gives
the coefficient of the $g^5$ term.  This plot is given in Fig.~\ref{Cg4coeff}.  This
type of plot places extreme demands on the accuracy of the data.  Even so, we see
that the data follow the required linear behavior even at the lowest temperatures.
We fitted a straight line to the data from $t= 5\times10^{-25}$ up to a maximum
temperature $t^{\rm max}$.  The fit parameters and rms deviation held nearly
constant for $t^{\rm max} = 5\times10^{-15}$ (11 data points) up to $t^{\rm max} =
5\times10^{-8}$ (18 data points), but both quantities changed rapidly upon
further increasing $t^{\rm max}$.  The fit for $t^{\rm max} =
5\times10^{-8}$ is shown as the straight line in Fig.~\ref{Cg4coeff}, along with the
fit parameters.  From the parameters of the fit [after accounting for the prefactor
of 2/3 in Eq.~(\ref{EqCLusnikov})], we include our estimated coefficients in
Eq.~(\ref{EqCLusnikov}) explicitly as
\begin{mathletters}
\label{EqCLusnikov+Us}
\begin{eqnarray}
{C_{\rm lt,g}(T)\over N k_{\rm B}} = {2 k_{\rm B}T\over 3 J}\Big[1 &+& {3\over
8}\,g^3 + a_4g^4 + a_5 g^5 + {\cal O}(g^6)\Big]\nonumber\\
\nonumber\\
&+& {2(3^{5/2})\over 5\pi}\Big({k_{\rm B}T\over J}\Big)^3[1 + {\cal O}(g)]~,
\label{EqCLusnikov+Us:a}
\end{eqnarray}
\begin{figure}
\epsfxsize=3in
\centerline{\epsfbox{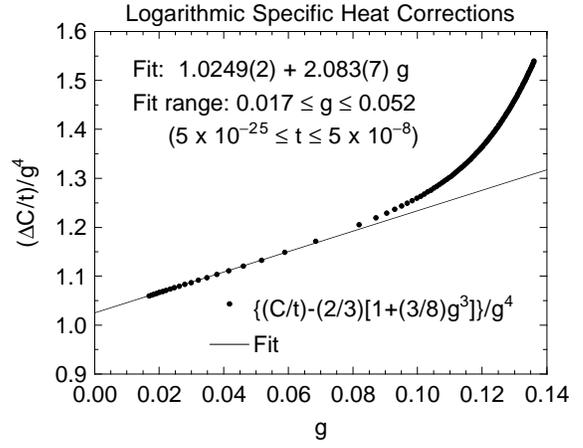}}
\vglue 0.1in
\caption{Plot showing the estimation of the coefficients of the ${\cal O}(g^4)$
and ${\cal O}(g^5)$ logarithmic correction terms in Eq.~(\protect\ref{EqCLusnikov})
for the magnetic specific heat of the $S = 1/2$ AF uniform Heisenberg chain.  The
reduced temperature is $t = k_{\rm B}T/J$ and we have set $N = k_{\rm B} = 1$.}
\label{Cg4coeff}
\end{figure}
\begin{equation}
a_4 = 1.5374(3)~,~~~a_5 = 3.125(11)~.
\end{equation}
\end{mathletters}

The influences of these $g^4$ and $g^5$ logarithmic correction terms on the data in
Figs.~\ref{KlumperLukCgCorr/T} and~\ref{KlumperLukCgCorr} are shown as the two
additional data sets in each figure, where accounting for these two terms is
seen to largely remove the remaining singular behavior as $t\to 0$.  From
Fig.~\ref{KlumperLukCgCorr}, we can now estimate that the coefficient of the
$t^3$ term in Eq.~(\ref{EqCLusnikov+Us}) is a little larger than 2, contrary to the
exact value 1.98478$\cdots$.  The magnitude of this difference is about as expected
from the ${\cal O}(g)$ logarithmic correction to the $t^3$ term, since $g(t\sim
0.1)\sim 0.1$.  The remaining upturn at low temperatures in
Fig.~\ref{KlumperLukCgCorr} is due to residual logarithmic corrections which are not
accounted for.

If the approximate expansion for $g({\cal L})$ in Eq.~(\ref{EqgSoln}) is inserted 
into Eq.~(\ref{EqCLusnikov+Us:a}), one obtains
\begin{eqnarray}
{C_{\rm lt,g}\over N k_{\rm B}} &=& {2 k_{\rm B}T\over 3 J}\bigg\{1 + 
{3\over (2{\cal L})^3} - {9 \ln({\cal L}) - 16 a_4\over (2{\cal L})^4}\nonumber\\
&+& {\ln{\cal L}[18\ln({\cal L}) - 64 a_4 -9] + 32 a_5\over (2{\cal L})^5}
 + {\cal O}\Big[{1\over (2{\cal L})^6}\Big]\bigg\}\nonumber\\
\nonumber\\
&+& {2(3^{5/2})\over 5\pi}\Big({k_{\rm B}T\over J}\Big)^3\Big[1 + {\cal
O}\Big({1\over 2{\cal L}}\Big)\Big]~.
\label{EqCLusnikov+UsLog}
\end{eqnarray}

\section{Fits to $\bbox{\chi^*(\lowercase{t})}$ and $\bbox{C(\lowercase{t})}$ of
Heisenberg Spin Lattices}

\subsection{General $\bbox{\chi^*(t)}$ fit considerations}
\label{SecGenFit}

The general expression we use to fit theoretical numerical $\chi^*(t)$ data for
$S = 1/2$ Heisenberg spin lattices is
\begin{mathletters}
\label{EqChi*Gen:all}
\begin{equation}
\chi^*(t) = {{\rm e}^{-\Delta^*_{\rm fit}/t}\over 4t}{\cal
P}^{(q)}_{(r)}(t)~,\label{EqChi*Gen:a}
\end{equation}
\begin{equation}
{\cal P}^{(q)}_{(r)}(t) = \frac{1 + \sum_{n=1}^q N_n/t^n}{1 + \sum_{n=1}^r D_n/t^n}~,
\label{EqChi*Gen:b}
\end{equation}
\end{mathletters}
where the orders $q$ and $r$ of the Pad\'e approximant ${\cal P}^{(q)}_{(r)}$
are often constrained by the behavior of $\chi^*(t)$ at low $t$, and the fitted
gap $\Delta^*_{\rm fit}$ is not necessarily the same as the true gap.  At high
$t$, $\chi^*(t)$ in Eqs.~(\ref{EqChi*Gen:all}) approaches the Curie law $1/(4t)$ as
required [for a general spin $S$ lattice, the numerical prefactor 1/4 in
Eq.~(\ref{EqChi*Gen:a}) would be replaced by $S(S+1)/3$].

The $N_n$ and $D_n$ parameters in Eq.~(\ref{EqChi*Gen:b}) are not in general
independent if one or more of the HTSE conditions in Eqs.~(\ref{EqHTSGen:b})
and~(\ref{EqHTSGen:c}) are invoked.  For example, for $n = 1$--3 one finds
\vglue0.05in
\begin{mathletters}
\label{EqDconstraints:all}
\begin{equation}
D_1 = (d_1 + N_1) - \Delta^*_{\rm fit}~,\label{EqDconstraints:a}
\end{equation}
\begin{equation}
D_2 = (d_2 + d_1 N_1 + N_2) - \Delta^*_{\rm fit}(d_1 + N_1) + {{\Delta^*_{\rm
fit}}^2\over 2}~,~\label{EqDconstraints:b}
\end{equation}
\begin{eqnarray}
D_3 = (d_3 &+& d_2 N_1 + d_1 N_2 + N_3)  - \Delta^*_{\rm fit}(d_2 + d_1 N_1 +
N_2)\nonumber\\
 &+& {{\Delta^*_{\rm fit}}^2\over 2}(d_1 + N_1) - {{\Delta^*_{\rm fit}}^3\over
6}~.
\label{EqDconstraints:c}
\end{eqnarray}
\end{mathletters}
\vglue0.1in
\noindent
In general, one has
\begin{equation}
D_n = \sum_{p=0}^n\ \sum_{m=0}^{n-p} \frac{(-{\Delta^*_{\rm fit}})^p}{p!}\,d_m
N_{n-p-m}~.
\label{EqDGeneral}
\end{equation}
\noindent
A fit of experimental or theoretical
$\chi^*(t)$ data by Eqs.~(\ref{EqChi*Gen:all}) can be constrained by inserting one or
more of Eqs.~(\ref{EqDconstraints:all}) and~(\ref{EqDGeneral}) into
Eq.~(\ref{EqChi*Gen:b}).  These constraints are especially useful for high-$t$
extrapolations when $\chi^*(t)$ data are not available for high temperatures $t\gg
1$, and/or to reduce the number of fitting parameters required to obtain a fit of
specified precision.  In  the following fits to the numerical $\chi^*(t)$ data for
the dimer, the uniform chain, and finally our QMC and TMRG data for the
alternating-exchange chain, the three constraints in
Eqs.~(\ref{EqDconstraints:all}) on $D_1,\ D_2$, and $D_3$, respectively, are enforced
in each case, where $d_1,\ d_2$, and $d_3$ for the alternating-exchange chain are
given in Eq.~(\ref{EqHTSAltChnCoeffs}).

All of the fits reported in this paper were carried out on a 400\,MHz
\bbox{Macintosh} G3 (B\&W) computer with 1GB of RAM\@.  Most fits were implemented
using the program \bbox{Mathematica~3.0}, although a few of the simpler ones (fits to
experimental data) were done using \bbox{KaleidaGraph~3.08c.}  The fits using
Mathematica sometimes required prodigious amounts of memory, e.g., 930~MB for the
28-parameter fit to the combined 2551 data point QMC and TMRG $\chi^*(\alpha,t)$
data set for the alternating-exchange chain in Sec.~\ref{SecQMCSim} below.

\subsection{Fit to $\bbox{\chi^*(t)}$ of the $\bbox{S = 1/2}$ antiferromagnetic
Heisenberg dimer}
\label{SecDimerFit}
\vglue-0.1in
The spin gap of the $S = 1/2$ Heisenberg dimer is $\Delta = J$, where $J$ is the
antiferromagnetic exchange constant within the dimer.  The spin susceptibility and
its low-temperature limit are given by Eqs.~(\ref{EqChiDimer:all}).  The $\chi^*(t)$
is plotted in Fig.~\ref{Dimerchi55Fit,Dev}(a) for $0.02\leq t\leq 4.99$.  In order
to later obtain a continuous fit function for
$\chi^*(\alpha,t)$ for the entire range $0\leq\alpha\leq 1$ of the
alternating-exchange chain, it is necessary to first obtain a high accuracy fit to
the exact expression~(\ref{EqChiDimer:a}) for the dimer by our general fitting
function in Eqs.~(\ref{EqChi*Gen:all}), in addition to Fit~1 obtained for the uniform
chain below.  The form of our fit function in Eqs.~(\ref{EqChi*Gen:all}) allows
both the low- and high-$t$ limiting forms of $\chi^*(t)$ for the dimer to be exactly
reproduced.  The low-$t$ limit in Eq.~(\ref{EqChiDimer:b}) requires that $r = q$ and
that $D_q = N_q/4$ in the Pad\'e approximant ${\cal P}^{(q)}_{(r)}$; we also take
$\Delta_{\rm fit} = \Delta$, so the total number of fitting parameters is $2q - 4$.

We fitted the 498-point double-precision representation of $\chi^*(t)$ in
Fig.~\ref{Dimerchi55Fit,Dev}(a) from $t = 0.02$ to $t = 4.99$ by
Eqs.~(\ref{EqChi*Gen:all}) using the above constraints.  The variances of the four
fits for $q = r = 4,$ 5, 6 and 7 were $2.5\times 10^{-13}$, $1.17\times 10^{-16}$,
$5.3\times 10^{-17}$, and $5.6\times 10^{-19}$, respectively, showing that
Eqs.~(\ref{EqChi*Gen:all}) have the potential for very high accuracy fits with a
relatively small number of fitting parameters.  The six $N_n$ ($n = 1$--5) and
$D_4$ parameters of the fit for $q,r = 5$ are given in
Table~\ref{TableDimChnPars}, along with $D_1,\ D_2$, and $D_3$ computed from
Eqs.~(\ref{EqDconstraints:all}) and $D_5 = N_5/4$.  The Pad\'e approximant ${\cal
P}^{(5)}_{(5)}$ in the fit function has no poles or zeros on the positive $t$
axis.  The fit is shown by the solid curve in Fig.~\ref{Dimerchi55Fit,Dev}(a), and
the deviation of the fit from the exact susceptibility in Eq.~(\ref{EqChiDimer:a}) is
plotted versus $t$ in Fig.~\ref{Dimerchi55Fit,Dev}(b).

\subsection{Fits to $\bbox{\chi^*(\lowercase{t})}$ of the $\bbox{S = 1/2}$
Antiferromagnetic Uniform Heisenberg Chain}
\label{SecChnFit}
\vglue-0.1in

\paragraph*{Fit~1: $0.01 \leq t \leq 5$.}
Fits to the uniform chain $\chi^*(t)$ calculated by Eggert, Affleck and
Takahashi\cite{Eggert1994} for limited temperature regions were obtained
previously.\cite{Johnston1997}  Here we obtain a fit (Fit~1) to the higher accuracy
data of Kl\"umper and Johnston\cite{Klumper1998} for the temperature region
$0.01\leq t\leq 5$ (999 data points) using Eqs.~(\ref{EqChi*Gen:all}), the results
of which will be utilized later in the fit function for $t\geq 0.01$ for our QMC and
TMRG alternating-exchange chain $\chi^*(\alpha,t)$ data.  This uniform chain fit can
be accurately extrapolated to arbitrarily high $t$.

The requirement that $\chi^*(t\to 0)$ is a finite non-zero value requires
$\Delta_{\rm fit}^* = 0$ and $r = q +1$ in Eqs.~(\ref{EqChi*Gen:all}).  We found that 
using $q=5$ and $r = 6$ produces a fit sufficiently accurate for use in the fit
function for our QMC and TMRG calculations for the alternating chain.  The seven
$N_n$ ($n = 1$--5) and $D_n$ ($n = 4$--6) parameters obtained for the fit with $q =
5,\ r = 6$ are given in the column \ labeled \ ``Fit 1'' \ in
\ Table~\ref{TableDimChnPars}, \ along with $D_1$, $D_2$, 
\begin{figure}
\epsfxsize=2.8in
\centerline{\epsfbox{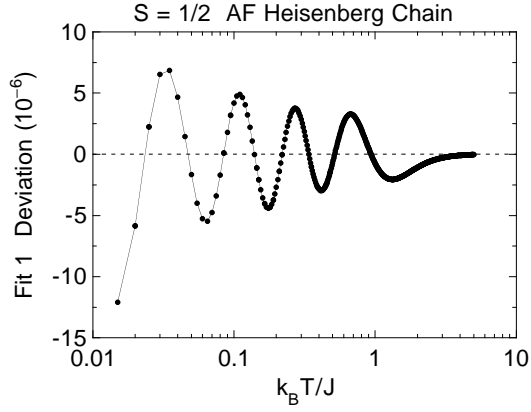}}
\vglue 0.1in
\caption{Semilog plot vs temperature $T$ of the deviation of Fit~1 from the
magnetic susceptibility calculations of Kl\"umper and Johnston
(Ref.~\protect\onlinecite{Klumper1998}) for the $S=1/2$ uniform antiferromagnetic
Heisenberg chain.}
\label{KlumperFit1Dev}
\end{figure}
\noindent and $D_3$ computed from Eqs.~(\ref{EqDconstraints:all}).  The
Pad\'e approximant ${\cal P}^{(5)}_{(6)}$ in the fit function has no poles or zeros
on the positive $t$ axis.  The deviation of the fit from the data is plotted in
Fig.~\ref{KlumperFit1Dev}.  The variance of the fit is $2.97\times 10^{-12}$, and
the relative rms deviation of the fit from the data in the fitted $t$ region is
14.5\,ppm.  Extrapolation of the fit to higher temperatures is very accurate.

\widetext
\begin{table}
\caption{Fitted parameters for $\chi^*(t)$ of the $S = 1/2$
antiferromagnetically coupled Heisenberg dimer ($\alpha = 0$)
[Eqs.~(\protect\ref{EqChi*Gen:all}) with $\Delta^*_{\rm fit} = 1$] and
$\chi^*(t)$ and $C(t)$ [Eqs.~(\protect\ref{EqCFit:all})] for the uniform chain
($\alpha = 1$).  $\chi^*(t)$ Fit~1 for the uniform chain ($0.01\leq t\leq 5$)
[Eqs.~(\protect\ref{EqChi*Gen:all}) with $\Delta^*_{\rm fit} = 0$] uses powers of
$1/t$ only, whereas $\chi^*(t)$ Fit~2 ($0\leq t\leq 5$)
[Eqs.~(\protect\ref{EqChiFit2:all})] also incorporates logarithmic correction terms.}
\begin{tabular}{lllld}
parameter & $\chi^*(\alpha = 0)$ & $\chi^*(\alpha=1)$ Fit 1 &
$\chi^*(\alpha=1)$ Fit 2 & $C(\alpha = 1)$\\
\hline
$N_1$ & 0.6342798982    & $-$0.053837836 & $-$0.240262331211&$-$0.018890951426    \\
$N_2$ & 0.1877696166    & 0.097401365    & 0.451187371598     & 0.024709724025    \\
$N_3$ & 0.03360361730   & 0.014467437    & 0.0125889356883  &$-$0.0037086264240   \\
$N_4$ & 0.003861106893  & 0.0013925193   & 0.0357903808997   &  0.0030159975962   \\
$N_5$ & 0.0002733142974 & 0.00011393434  & 0.00801840643283 &$-$0.00037946929995 \\
$N_6$ &                 &                & 0.00182319434072  &  0.000042683893992 \\
$N_7$ &                 &                & 0.0000533189078137 &  \\
$N_8$ &                 &                & 0.000184003448334  &  \\
$N_{81}$&               &                & 1.423476309767     &  \\
$N_{82}$&               &                & 0.341607132329     &  \\
$t_1$ &                 &                & 5.696020642244      &  \\
$D_1$ & $-$0.1157201018 & 0.44616216     & 0.259737668789    &$-$0.51889095143     \\
$D_2$ & 0.08705969295   & 0.32048245     & 0.581056205993     &  0.59657583453     \\
$D_3$ & 0.005631366688  & 0.13304199     & 0.261450372018    &$-$0.15117343936     \\
$D_4$ & 0.001040886574  & 0.037184126    & 0.142680453011     &  0.074445241148    \\
$D_5$ & 0.00006832857434 & 0.0028136088  & 0.0572246926066   &$-$0.0024804135233   \\
$D_6$ &                 & 0.00026467628  & 0.0176410851952   &$-$0.00053631174698  \\
$D_7$ &                 &                & 0.00390435823809   &  0.00082005310111  \\
$D_8$ &                 &                & 0.000119767935536 &$-$0.00010820401214 \\
$D_9$ &                 &                &                    &  0.000011991365422 \\
$a_1$ &                 &                &                   &$-$0.000015933393 \\
$a_2$ &                 &                &                    &  0.013021564       \\
$a_3$ &                 &                &                    &  0.0043275575      \\
$a_4$ &                 &                &                    & 49.422168          \\
$a_5$ &                 &                &                    &  0.00040160786     \\
$a_6$ &                 &                &                    &325.22706        \\
\end{tabular}
\label{TableDimChnPars}
\end{table}
\narrowtext

The quality of Fit~1 does not approach the limitation imposed by the absolute
accuracy of the data ($1 \times 10^{-9}$).  For an ideal fit, the variance
is expected to be $\sim 10^{-18}$ and the relative rms deviation $\sim
0.01$~ppm.  As can be inferred from Fit~2 in the following section, the reason that
Fit~1 cannot be optimized to this extent is due to the $t=0$ critical point and
associated logarithmic divergence in the slope of $\chi^*(t)$ as $t\to 0$; this
divergence cannot be fitted accurately by a finite polynomial or Pad\'e
approximant.    We  attempted to improve the accuracy of the fit over the same
temperature range
$0.01\leq t\leq 5$ by replacing the Pad\'e approximant ${\cal P}^{(5)}_{(6)}$ in the
fit function by ${\cal P}^{(6)}_{(7)}$, which incorporates two additional fitting
parameters. The variance improved somewhat to $2.18\times 10^{-12}$ and the
relative rms deviation improved slightly to 12.2 ppm, but the Pad\'e approximant
developed a pole at $1/t = 129.23$, and hence this fit was discarded.  Although the
temperature at which this pole occurs is below the fitted temperature range, as a
general rule we cannot allow poles in the fit function at low temperatures because
of problems that can occur when using the fit function to model experimental data
which include data at temperatures lying below the fitted temperature range of the
fit function.  In fact, we will encounter this situation frequently in
modeling experimental data later.  For exam-

\newpage
\noindent
ple, for $\rm NaV_2O_5$, $t=0.01$ corresponds to
an absolute temperature $T\approx 7$\,K, whereas the experimental data and modeling
extend down to $\approx 2$\,K\@.

\paragraph*{Fit~2: $0 \leq t \leq 5$.}

We can greatly improve the accuracy of the fit compared to that of Fit~1, and extend
the fit to $t=0$, by restricting the high-temperature limit of the fit and using in
the fit function one or more low-temperature logarithmic correction terms discussed
in Sec.~\ref{SecLogCorr}.  In particular, in this section we obtain a very high
precision fit (Fit~2) to the exact $t = 0$ value and to the calculations of
Kl\"umper and Johnston\cite{Klumper1998} over the entire temperature range $5\times
10^{-25} \leq t \leq 5$ of the calculations.  We do not use this fit in our
formulation of the fit function for the alternating-exchange chain.  However, Fit~2
will be generally useful for evaluating the accuracy of other theoretical
calculations of $\chi^*(t)$ for the uniform chain, such as our TMRG calculations to
be presented below, and for modeling appropriate experimental $\chi(T)$ data whose
scaled upper temperature limit is below $t = 5$.

We initially formulated a fit function utilizing a modified Pad\'e approximant in
which the last term of the numerator and/or denominator contained the $\chi^*_{\rm
lt,log}$ expansion in Eq.~(\ref{EqLukLog}), such that the low-temperature expansion
of the fit function yielded $\chi^*_{\rm lt,log}$ to lowest orders in $t$.  The
best fit to the data from $t=5\times 10^{-25}$ to~2.5 (777 data points) was
unsatisfactory, with a variance $v = 2.4\times 10^{-11}$ and a relative rms
deviation $\sigma_{\rm rms} = 45$\,ppm.  Allowing an arbitrary $t^2$ coefficient in
place of the exact value $\sqrt{3}/\pi^3$ yielded an improved fit  with $v = 1.1\times
10^{-12}$ and $\sigma_{\rm rms} = 9.6$\,ppm.  However, this fit was still
unsatisfactory, given the high absolute accuracy ($1\times 10^{-9}$) of the data. 
From these results it became clear that a fit function which can fit the data to much
higher accuracy over such a large temperature range would indeed have to include an
expression $\chi^*_{\rm log}(t)$ containing logarithmic correction terms, but where
the form and/or coefficient of one or more of these terms would have to be empirically
determined by trial and error.  This process yielded the formulation we now describe.

The $\chi^*_{\rm log}(t)$ function is incorporated into our fit function in
Eqs.~(\ref{EqChi*Gen:all}) as follows.  As in Fit~1, the finite value of $\chi^*(0)$
requires $\Delta_{\rm fit}=0$ in Eq.~(\ref{EqChi*Gen:a}) and $r = q+1$ in the Pad\'e
approximant ${\cal P}^{(q)}_{(r)}(t)$ in Eq.~(\ref{EqChi*Gen:b}).  Since the two
terms highest order in $1/t$ in ${\cal P}^{(q)}_{(r)}(t)$ (one each in the numerator
and denominator) dominate the fit as $t\to 0$ and become small for
$t\gtrsim 1$, relative to the other terms in the numerator and denominator,
respectively, we incorporate $\chi^*_{\rm log}(t)$ into the last term in the
numerator of a modified ${\cal P}^{(q)}_{(r)}(t)$.  Trial fits showed that to obtain
the optimum accuracy of the fit required $q=8$ and
$r=9$.

Our final fit function for Fit~2 is
\vglue-0.1in
\begin{mathletters}
\label{EqChiFit2:all}
\begin{equation}
\chi^*(t) = \Big(\frac{1}{4t}\Big)\,\frac{1 +
\big[\sum_{n = 1}^{7} N_n/t^n\big] + 4 N_8\chi_{\rm log}^*(t)/t^8} {1 +
\big[\sum_{n = 1}^8 D_n/t^n\big] + N_8/t^{9}} ~,
\label{EqChiFit2:a}
\end{equation}
\begin{figure}
\epsfxsize=3in
\centerline{\epsfbox{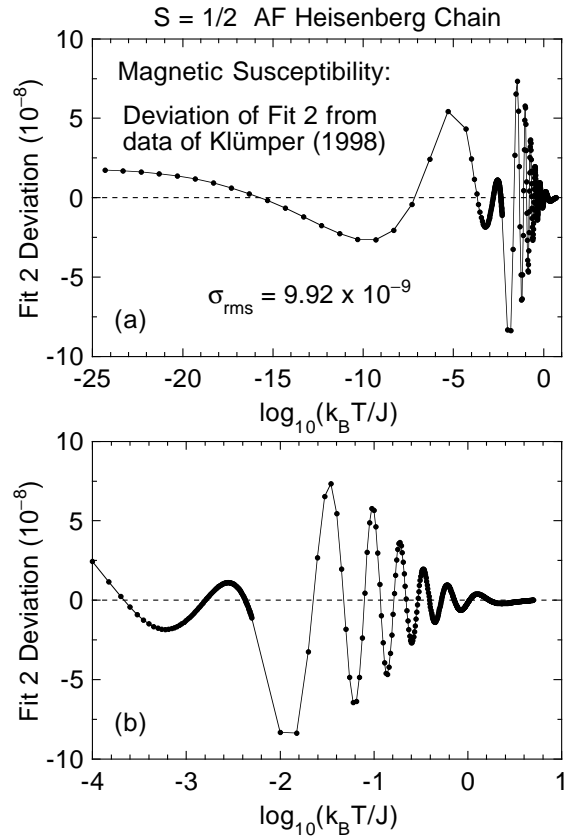}}
\vglue 0.1in
\caption{(a) Semilog plot of the absolute deviation of Fit~2 ($0\leq t\leq 4$) from
the magnetic susceptibility calculations of Kl\"umper and Johnston 
(Ref.~\protect\onlinecite{Klumper1998}) versus temperature $T$.  (b) Expanded plot of
the data in (a) at the higher temperatures.}
\label{KlumperFit893DevLog6}
\end{figure}
\begin{equation}
\chi_{\rm log}^*(t) = {1\over \pi^2}\bigg[1 + {1\over 2{\cal L}} -
\frac{\ln\big({\cal L} + {1\over 2}\big)-N_{81}}{(2{\cal L})^2} +
\frac{N_{82}}{(2{\cal L})^3} \bigg]~,
\label{EqChiFit2:b}
\end{equation}
\begin{equation}
{\cal L} \equiv \ln(t_1/t)~,
\label{EqChiFit2:c}
\end{equation}
\end{mathletters}
subject to the three constraints on $D_1,\ D_2$, and $D_3$ 
in Eqs.~(\ref{EqDconstraints:all}) which are required by the HTSE.  Two of the four
logarithmic correction terms in Eq.~(\ref{EqChiFit2:b}) are identical to the first
two such terms in Eq.~(\ref{EqChLoT:a}).  By construction, the exact
$\chi^*(0)=1/\pi^2$ is fitted exactly.  

We fitted all of the numerical $\chi^*(t)$ data,\cite{Klumper1998} calculated 
over the range $5\times 10^{-25}\leq t\leq 5$ (1119 data points), by
Eqs.~(\ref{EqChiFit2:all}).  The 19 fitting parameters of the fit
function~(\ref{EqChiFit2:all}), which are $N_n$ ($n = 1$--8), $D_n$ ($n = 4$--8),
$N_{81},\ N_{82}$ and $t_1$, are given in the column labeled ``Fit~2'' in
Table~\ref{TableDimChnPars}, along with $D_1,\ D_2$, and $D_3$ computed from
Eqs.~(\ref{EqDconstraints:all}).  The data to parameter ratio is 59.  The denominator
of the modified Pad\'e approximant in Eq.~(\ref{EqChiFit2:a}) has no zeros for any
real positive $t$.  The fit is shown in the low-temperature region
$0\leq t\leq 0.02$ in Fig.~\ref{KlumperDatFit2}(b) [over the larger $t$ range
plotted in Fig.~\ref{KlumperDatFit2}(a), the fit is indistinguishable from the data
and is therefore not plotted there].

The deviation of Fit~2 from the numerical data for $10^{-25}\leq t\leq 5$ is plotted
vs $\log_{10}t$ in Fig.~\ref{KlumperFit893DevLog6}(a), and an expanded plot at the
higher temperatures is shown in Fig.~\ref{KlumperFit893DevLog6}(b).  Due to a
logarithmic divergence in $\chi_{\rm log}^*(t)$ at $t = t_1 = 5.696$, Fit~2 should
not be used (e.g., for modeling experimental data) at temperatures $t \gtrsim 5$. 
The variance of the fit is $9.8 \times 10^{-17}$, and the relative rms deviation is
$\sigma_{\rm rms} = 0.087$\,ppm.  These values are both much smaller than for Fit~1
above.  The relatively large number of fitting parameters in Fit~2 is justified {\it a
posteriori} by the extremely high quality of the fit over a temperature range
spanning 25 orders of magnitude.

\subsection{Fit to $\bbox{C(t)}$ for the $\bbox{S = 1/2}$ antiferromagnetic uniform
Heisenberg chain}
\label{SecChainCFit}
\vglue0.27in
The logarithmic corrections to the magnetic specific heat $C(t)$ at low
temperatures, discussed above in Sec.~\ref{SecCLogCorr}, do not pose as serious a 
problem for fitting the data as for $\chi^*(t)$, because the strength of these log
corrections is much smaller for $C(t)$ than for $\chi^*(t)$.  In addition, since here
we fit $C(t)$, and not the electronic specific heat coefficient $C(t)/t$, the
influence of the log corrections is ameliorated  by the multiplicative leading order
$t^1$ dependence of $C(t)$.  Even so, in order to obtain the optimum fit to the
highly accurate Bethe ansatz $C(t)$ data,\cite{Klumper1998} we found it necessary to
take the influence of the logarithmic corrections into account.

Our fit to the Bethe ansatz $C(t)$ data,\cite{Klumper1998} some of which were shown
previously in Fig.~\ref{KlumperCp}(a), was carried out in two stages.  First, the
data from $t = 0.01$ to the maximum temperature $t = 5$ of the calculations were
fitted by the  Pad\'e approximant ${\cal P}^{(q)}_{(r)}$ in Eq.~(\ref{EqChi*Gen:b})
with a prefactor $3/(16 t^2)$ to satisfy the HTSE in Eqs.~(\ref{EqCHTSEGen2:all}) to
lowest order in $1/t$.  The orders $q$ and $r$ of ${\cal P}^{(q)}_{(r)}$ were chosen
to satisfy $r = q+3$ so that $C(t\to 0) \propto t$.  To obtain
a fit of the required accuracy (see the fit deviations given below) we found that
$q=6$ and $r = 9$ are of sufficiently high order.   Due to the presence of
the log corrections at very low $t$, we did not require the parameters $N_6$ 
and $D_9$ to yield the exact coefficient $\gamma = 2/3$ in the expression
$C(t)/Nk_{\rm B} = \gamma t$, in a low-$t$ expansion of the fit function.  We also
found that to obtain the best fit, only the one additional HTSE constraint (on
$D_1$) in Eq.~(\ref{EqDconstraints:a}) (with
$\Delta^*_{\rm fit} = 0$) could be used.  It was quite difficult to find the
region in parameter space in which the absolute minimum in the variance of the
fit resided; the initial starting parameters usually flowed to regions
with local variance minima in them with much larger values (by two to four orders of
magnitude) \ than \ the \ smallest \ variance we ultimately \,found.  Then \,the
\,deviation \,of \,the fit from all the data for $5\times 10^{-25}\leq t\leq 5$ was computed.  The fit
deviations for $t\geq 0.01$ were very small [${\cal O}(10^{-8})$], but the log
corrections which become \,most \ important \,at \,lower 
\begin{figure}
\epsfxsize=3.4in
\centerline{\epsfbox{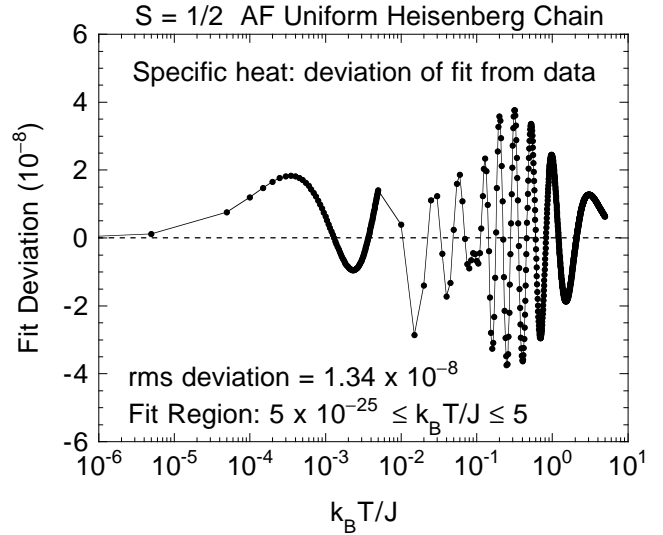}}
\vglue 0.1in
\caption{Semilog plot of the absolute deviation of the fit ($10^{-25}\leq k_{\rm
B}T/J \leq 5$) from the specific heat calculations of Kl\"umper and Johnston
(Ref.~\protect\onlinecite{Klumper1998}) in Fig.~\protect\ref{KlumperCp}(a) versus
temperature $T$.  The fit deviation is negligible at low temperatures $T \leq
10^{-6}J/k_{\rm B}$.  The lines connecting the data points are guides to the eye.}
\label{Klumper0to5CFit}
\end{figure}

\noindent temperatures resulted in fit
deviations at $t < 0.01$ an order of magnitude larger.  We therefore fitted the fit
deviation versus $t$ for
$0 < t\leq 0.1$ by a separate empirically determined function $F(t)$, so the net fit
function consists of the Pad\'e approximant fit function minus the fit function to
the low-$t$ fit deviations.  In the final fitting cycles the two
functions were refined simultaneously.

Our final fit function for $C(t)$ in the range $0\leq t\leq 5$ is
\begin{mathletters}
\label{EqCFit:all}
\begin{equation}
{C(t)\over N k_{\rm B}} = \frac{3}{16t^2}\,{\cal P}^{(6)}_{(9)}(t) - F(t)~,
\label{EqCFit:a}
\end{equation}
\begin{equation}
{\cal P}^{(6)}_{(9)}(t) = 
\frac{1 + \big[\sum_{n = 1}^{6} N_n/t^n\big]} {1 +
\big[\sum_{n = 1}^9 D_n/t^n\big]} ~,
\label{EqCFit:b}
\end{equation}
\begin{equation}
F(t) = a_1 t\sin\Big({2\pi\over a_2 + a_3 t}\Big)\,{\rm e}^{-a_4 t} + a_5 t\,{\rm
e}^{-a_6 t}~,
\label{EqCFit:c}
\end{equation}
\end{mathletters}
\vglue0.19in
\noindent
subject to the constraint on $D_1$ in Eq.~(\ref{EqDconstraints:a}) which is
required by the HTSE.  By construction, the exact $C(0) = 0$ is fitted exactly.  The
20 fitting parameters, $N_n\ (n = 1-6),\ D_n\ (n = 2-9)$ and $a_n\ (n = 1-6)$, are
given in Table~\ref{TableDimChnPars}, together with the constrained parameter $D_1$
computed from Eq.~(\ref{EqDconstraints:a}) with $\Delta^*_{\rm fit} = 0$ and $d_1$
given in Eq.~(\ref{EqCHTSEGen2:b}).  The deviation of the fit from the data is shown
in a semilog plot vs temperature in Fig.~\ref{Klumper0to5CFit}.  The maximum
deviations of $\approx\pm 4\times 10^{-8}$ occur at $t\approx 0.3$.  The absolute
rms deviation of the fit from all the data (1119 data points), which extend over the
temperature range $5\times 10^{-25} \leq t \leq 5$, is $1.34\times 10^{-8}$, and the
relative rms deviation for $0.01 \leq t \leq 5$ (999 data points) is 0.50\,ppm.

At high temperatures, our $C(t)$ fit function reduces by construction
to the lowest order $1/t^2$ and $1/t^3$ terms of the HTSE of $C(t)$ in
Eqs.~(\ref{EqCHTSEGen2:all}), so extrapolation of our $C(t)$ fit function to
arbitrarily higher temperatures should be very accurate (see
Fig.~\ref{Klumper0to5CFit}).  In particular, even though our fit was to $C(t)$
and hence not optimized as a fit to the electronic specific heat coefficient
$C(t)/t$, the magnetic entropy $S$ at $t = \infty$ computed from our $C(t)$ fit
function is
\begin{equation}
{S(t = \infty)\over N k_{\rm B}} \equiv \int_0^\infty {C(t)\over N k_{\rm B}\,t}\,dt
= 0.693\,147\,235~,
\end{equation}
which is the same as the exact value $\ln 2 = 0.693\,147\,181$ to within 8
parts in $10^8$.  This agreement reflects well on our fit function, and of course
also strongly confirms the high accuracy of the Bethe ansatz $C(t)$
data.\cite{Klumper1998}

\subsection{Fit function for the $\bbox{S = 1/2}$ AF alternating-exchange Heisenberg 
chain $\bbox{\chi^*(\alpha,t)}$}
\label{SecAltChnFitFcn}
\vglue0.08in
Here we formulate a single two-dimensional ($\alpha,t$) function to
accurately fit numerical calculations of $\chi^*(\alpha,t)$ for the $S = 1/2$
alternating-exchange Heisenberg chain for the
entire range $0 \leq
\alpha \leq 1$, and for the entire temperature range $t\geq0.01$ over
which our Fit~1 for $\chi^*(t)$ of the uniform chain is most accurate, subject to
four general requirements as follows.  (i) The HTSE of the $\chi^*(\alpha,t)$
fit function must give the correct result to ${\cal O}(1/t^4)$, as satisfied by the
fit functions for the dimer and uniform chain (Fit~1) susceptibilities above, so that
the fit can be accurately extrapolated to higher temperatures.  (ii)  We require
the $\chi^*(\alpha,t)$ fit function to become identical with those found above for
the isolated dimer and for the uniform chain (Fit 1) when $\alpha = 0$ and $\alpha =
1$, respectively.  As discussed above in Sec.~\ref{SecAltChnIntro}, at any finite
temperature, $\overline{\chi^*}(\delta,\overline{t})$ in the variables
$\delta$ and $\overline{t}$ is an {\it even} ({\it analytic}) function of
$\delta$.  Therefore, as a minimum accommodation of this fact, (iii) we require that
the fit function for $\chi^*(\alpha,t)$, when transformed to the form
$\overline{\chi^*}(\delta,\overline{t})$, must have the  property
$\partial\overline{\chi^*}(\delta,\overline{t})/\partial\delta|_{\delta=0} = 0$ at
all finite temperatures.  This requirement is clearly the minimum necessary in order
to accurately interpolate the fit vs $\alpha$ for $\alpha\to 1$ at each $t$, and to
thereby accurately model the susceptibility of materials which are in or near this
limit.  Finally, the QMC and TMRG calculations of $\chi^*(\alpha,t)$ to be presented
below are sufficiently accurate and cover sufficiently large ranges of $\alpha$ and
$t$ with sufficient resolution that (iv) we require the {\it nonanalytic} energy gap
$\Delta(\alpha)$ [see Eqs.~(\ref{EqDimParams2:all}) and~(\ref{EqCritGap})] to be
included in the fit function in order to fit the data for $\alpha \lesssim 1$ at
$t\ll 1$, so as to avoid the alternate necessity of including high-order power series
in $\alpha$ and $t$ in the fit function.  We note that according to
Eq.~(\ref{EqDimParams2:b}) or~(\ref{EqCritGap}),
$\partial\overline{\Delta^*}(\delta)/\partial\delta|_{\delta=0} = \infty$.  The
major obstacle we faced in formulating the fit function for $\chi^*(\alpha,t)$ was to
simultaneously satisfy both requirements (iii) and (iv), which at first sight seem to
require mutually exclusive forms for the fit function. 

We found that these four requirements can all be satisfied by an
extension of the form of the fit function in Eqs.~(\ref{EqChi*Gen:all}) which was
used above for the isolated dimer and for the uniform chain Fit~1.  This extension
consists of using a modified Pad\'e approximant ${{\cal P}_{\rm m}}^{(7)}_{(8)}$ in
the fit function in place of the former ${\cal P}^{(q)}_{(r)}$.  The fit function is
\begin{mathletters}
\label{EqChi*AltChn:all}
\begin{equation}
\chi^*(\alpha,t) = {{\rm e}^{-\Delta^*_{\rm fit}(\alpha)/t}\over 4t}\,{{\cal
P}_{\rm m}}^{(7)}_{(8)}(\alpha,t)~,\label{EqChi*AltChn:a}
\end{equation}
\FL
\[
{{\cal P}_{\rm m}}^{(7)}_{(8)}(\alpha,t) = 
\]
\begin{equation}
\frac{[\sum_{n=0}^6
N_n/t^n] + (N_{71} \alpha + N_{72}
\alpha^2)(\Delta^*_0/t)^y/t^7}{[\sum_{n=0}^7 D_n/
t^n]+(D_{81}\alpha+D_{82}\alpha^2)(\Delta^*_0/t)^z
{\rm e}^{(\Delta^*_0-\Delta^*_{\rm fit})/t}/t^8},
\label{EqChi*AltChn:b}
\end{equation}
\begin{equation}
\Delta^*_{\rm fit}(\alpha) = 1 - {1\over 2}\alpha - 2 \alpha^2 + {3\over 2}\alpha^3~,
\label{EqChi*AltChn:c}
\end{equation}
\begin{eqnarray}
\Delta^*_0(\alpha)=(1&-&\alpha)^{3/4}(1+\alpha)^{1/4}\nonumber\\
&+&g_1 \alpha(1-\alpha)+g_2 \alpha^2(1-\alpha)^2~,
\label{EqChi*AltChn:d}
\end{eqnarray}
\begin{equation}
N_0 = D_0 = 1~,\label{EqChi*AltChn:e}
\end{equation}
\begin{equation}
N_n(\alpha) = \sum_{m=0}^4 N_{nm}\alpha^m~~~~(n = 1-6)~,
\label{EqChi*AltChn:f}
\end{equation}
\begin{equation}
D_n(\alpha) = \sum_{m=0}^4 D_{nm}\alpha^m~~~~(n = 1-7)~.
\label{EqChi*AltChn:g}
\end{equation}
\end{mathletters}

To satisfy requirement (i), $D_1(\alpha),\ D_2(\alpha)$, and
$D_3(\alpha)$ are determined from the $N_1(\alpha),\ N_2(\alpha)$, and
$N_3(\alpha)$ fitting parameters according to the three constraints  in
Eqs.~(\ref{EqDconstraints:all}) demanded by the HTSE.  In order to satisfy
requirement (ii), the $\{N_{n0},D_{n0}\}$ parameters are set to be
identical with those determined above for the dimer, and we require
$\{N_n(1),D_n(1)\}$ to be identical with the corresponding fit parameters determined
above in Fit~1 for the uniform chain.  In order to satisfy requirement~(iii), the 
$N_{nm}$ and $D_{nm}$ coefficients must satisfy
\begin{equation}
\sum_{m=0}^4 (n-2 m)(N_{nm}\ {\rm or}\ D_{nm}) = 0~,
\end{equation}
so that no $\delta^1$ term appears in the Taylor series expansions in $\delta$ of the
transformed  $\{\overline{N}_n(\delta),\overline{D}_n(\delta)\}$. 
These various constraints on the $\{N_{nm},D_{nm}\}$ parameters reduce
the number of independent fitting parameters within this set from 50 to 20. 
Together with the parameters $N_{71},\ N_{72},\ N_{81},\ N_{82},\ y,\ z$ in
Eq.~(\ref{EqChi*AltChn:b}) and $g_1,\ g_2$ in Eq.~(\ref{EqChi*AltChn:d}), the total
number of independent fitting parameters in the fit function is 28.

The quantity $\Delta^*_{\rm fit}(\alpha)$ in the exponential prefactor to 
${{\cal P}_{\rm m}}^{(7)}_{(8)}$ in Eq.~(\ref{EqChi*AltChn:a}) cannot be set
equal to the true nonanalytic gap $\Delta^*(\alpha)$, because this  prefactor affects
the fit at all $t$, and would not allow requirement (iii) above to be fulfilled.  In
addition, the nonanalytic critical behavior of $\Delta^*(\alpha\to 1)$ in practice
only becomes manifest in $\chi^*(\alpha,t)$ at low temperatures $t\ll 1$. 
Therefore, we separated the spin gap into an analytic part $\Delta^*_{\rm
fit}(\alpha)$ which goes into the argument of the exponential prefactor in
Eq.~(\ref{EqChi*AltChn:a}), and a nonanalytic part $\Delta^*_0(\alpha)$ [satisfying
requirement (iv)] which is placed into the argument of the exponential in the last
term of the denominator of ${{\cal P}_{\rm m}}^{(7)}_{(8)}$ in
Eq.~(\ref{EqChi*AltChn:b}) and which therefore only becomes important at low
temperatures.  The first two terms of $\Delta^*_{\rm fit}(\alpha)$ (to order
$\alpha^1$) in Eq.~(\ref{EqChi*AltChn:c}) are the first two terms of the exact dimer
series expansion up to ${\cal O}(\alpha^9)$ given by Barnes, Riera and
Tennant\cite{Barnes1998} for the AF alternating-exchange chain, and the last two are
included so that $\partial\overline{\Delta^*_{\rm
fit}}(\delta)/\partial\delta|_{\delta=0}=0$, in accordance with requirement (iii). 
The nonanalytic $\Delta^*_0(\alpha)$ in Eq.~(\ref{EqChi*AltChn:d}) contains the
behavior in Eq.~(\ref{EqDimParams2:a}) proposed by Barnes, Riera, and
Tennant,\cite{Barnes1998} plus two analytic terms which are included to adjust the
$\alpha$ dependence for $\alpha\to 1$ but which make no contribution at $\alpha=0$
or $\alpha=1$.  Provided that the inequality
$y,z>4/3$ is satisfied by the powers $y$ and $z$ in Eq.~(\ref{EqChi*AltChn:b}), the
last term in each of  the numerator and denominator of ${{\cal P}_{\rm
m}}^{(7)}_{(8)}(\alpha,t)$, when transformed to the variables
$(\delta,\overline{t})$, has a partial derivative with respect to $\delta$ which is
zero at $\delta = 0$.

We have now shown that at $\delta = 0\ (\alpha = 1)$, the partial derivative of each
part of $\overline{\chi^*}(\delta,\overline{t})$ with respect to $\delta$ is zero
(if $y,z > 4/3$, which is confirmed in the actual fit later).   Hence, the entire fit
function has the property
$\partial\overline{\chi^*}(\delta,\overline{t})/\partial\delta|_{\delta=0}=0$ at all
finite temperatures, thus satisfying requirement (iii), despite the fact that the
fit function contains the nonanalytic $\Delta_0^*(\alpha)$ as required by
requirement~(iv).

At the lowest temperatures, the last term in each of the numerator and denominator of
${{\cal P}_{\rm m}}^{(7)}_{(8)}$ in Eq.~(\ref{EqChi*AltChn:b}) should dominate the
fit, together with the exponential prefactor to ${{\cal P}_{\rm m}}^{(7)}_{(8)}$ in
Eq.~(\ref{EqChi*AltChn:a}), so in this limit our fit function for $0 < \alpha < 1$
becomes 
\begin{equation}
\chi^*(\alpha,t\to 0) = \frac{N_{71}\alpha + N_{72}\alpha^2}{4(D_{81}
\alpha+D_{82}\alpha^2)}\Big[{\Delta^*_0(\alpha)\over t}\Big]^{y-z}{\rm
e}^{-\Delta_0^*(\alpha)/t}~.
\label{EqAltChnLoT}
\end{equation}
This expression has the form of Eq.~(\ref{EqTroyer:a}) (with $\gamma = y-z$) as
required in the low-$t$ limit.  In fact, the forms of the last term in each of the
numerator and denominator of ${{\cal P}_{\rm m}}^{(7)}_{(8)}$ were designed to result
in the form of Eq.~(\ref{EqTroyer:a}) in the low-$t$ limit, with $\Delta^*_0$ and
$t$ entering the prefactor only as their ratio as in Eq.~(\ref{EqMyLoTChi}), in
addition to being consistent with requirements (iii) and (iv).  One might expect the
fitted $y$ and $z$ powers to satisfy $y-z=\gamma = 1/2$ as
in Eq.~(\ref{EqTroyer:b}).  However, if a fit of
$\chi^*(t)$ data by Eq.~(\ref{EqTroyer:a}) is not carried out completely within the
low-$t$ limit, an effective exponent $\gamma\sim 1$ is often inferred [see, e.g., 
Eq.~(\ref{EqBul}) and subsequent discussion, and Fig.~\ref{BulLog[ChiExp]2} below]. 
Similarly, since many of our calculated $\chi^*(\alpha,t)$ data sets for different
$\alpha$ in the fitted temperature range $t\geq 0.01$ are not, or do not contain
extensive data, in the low-$t$ limit, we did not impose the constraint $y - z =
1/2$.  On the basis of the above discussion we expect the actual fitted values of
$y$ and $z$ to yield $y - z\sim 1$.  In fact, as will be seen in the next section,
our fitted parameters $y$ and $z$ give $y - z = 1.14$.

\section{QMC and TMRG $\bbox{\chi^*(\alpha,t)}$ Calculations and Fit for the
$\bbox{S = 1/2}$ AF Alternating-Exchange Heisenberg Chain}
\label{SecQMCSim}

QMC simulations of $\chi^*(\alpha,t)$ were carried out on $S = 1/2$
alternating-exchange chains containing 100 spins for $\alpha = 0.05,\ 0.1,\ 0.15,\
\ldots,$ 0.9, 0.92, 0.94, 0.96, 0.97, 0.98, and 0.99 in various temperature ranges
spanned by $0.01\leq t\leq 4$.

Complementary TMRG calculations of $\chi^*(\alpha,t)$ of $S = 1/2$
alternating-exchange chains were carried out for $\alpha =
0.80$, 0.82, ..., 0.96, 0.97, 0.98, 0.99, 0.995 and 1, where the number of states
kept was $m = 150$ or 256.  The calculations were carried out for temperatures
given by $1/t = 0.1,\ 0.2,\ ...,\ (1/t)^{\rm max}$, with $(1/t)^{\rm max} \lesssim
500$ increasing with increasing $\alpha$, and comprised a total of 22\,370
($\alpha,t$) parameter combinations.  The details of the calculational method are
given in Refs.~\onlinecite{Wang1997} and~\onlinecite{Xiang1999}.  It should be noted
that the TMRG calculations by their nature are explicitly in the thermodynamic limit.

\begin{figure}
\epsfxsize=3in
\centerline{\epsfbox{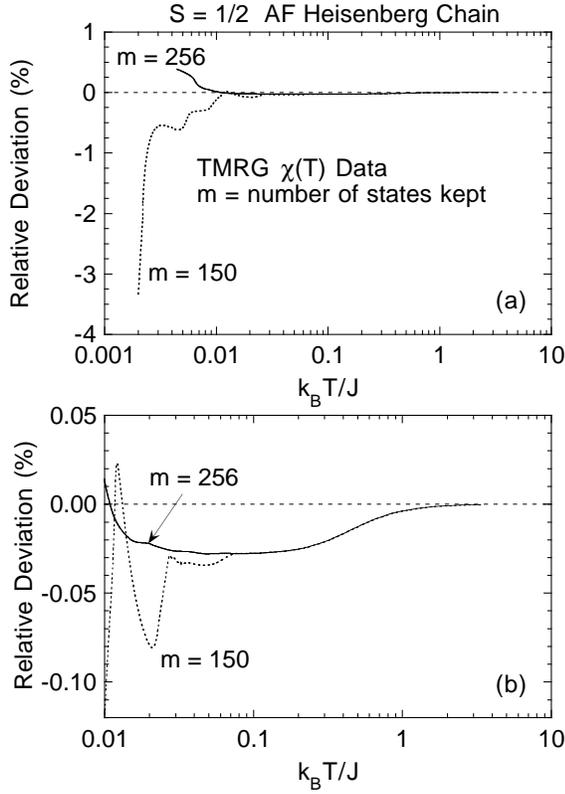}}
\vglue 0.1in
\caption{(a) Semilog plot vs temperature $T$ of the relative deviation of the
magnetic susceptibility $\chi$ of the $S = 1/2$ antiferromagnetic uniform spin
Heisenberg chain calculated with the TMRG technique from that
calculated (Ref.~\protect\onlinecite{Klumper1998}) using the Bethe ansatz. 
(b)~Expanded plot of the data in~(a) at the higher temperatures.}
\label{Alpha=1TDMRGDev}
\end{figure}

The reason for doing TMRG calculations for the uniform chain ($\alpha = 1$) was to
enable comparison of the results with the values\cite{Klumper1998} computed with the
Bethe ansatz which have a high absolute accuracy of $1\times 10^{-9}$.  This
comparison was done using the above very accurate and precise Fit~2 for the Bethe
ansatz data.  The relative deviation of the TMRG data from Fit~2 is shown in
Fig.~\ref{Alpha=1TDMRGDev}(a), and an expanded plot for the higher temperature
region $t\geq 0.01$ is shown in Fig.~\ref{Alpha=1TDMRGDev}(b).  This comparison
indicates that the accuracy of the TMRG calculations for both $m = 150$ and~256 in
the range $t\geq 0.01$ is better than 0.1\,\%, which is the same as the
estimate\cite{Wang1997} made previously for $m = 80$.  However, the accuracy of
these calculations deteriorates rapidly at lower $t$, to about 3\,\% at the lowest
temperatures $t\approx 0.002$ for $m = 150$.

Since the TMRG calculations extend close to the $t=0$ limit for
most of the above-stated $\alpha$ values, the spin gaps can be estimated from these
data.  Comparisons with previous \ work \ can \ then \ be \ made \ of
the dependence of \ the \ spin \ gap \ on \ $\alpha$.  An important question, not
yet  

\noindent  answered in previous work, is the approximate $\alpha$ value at which the
asymptotic critical region is entered upon approaching the uniform limit.  Performing
these estimates and comparisons will be postponed to the following sections.  In the
present section, we present the QMC and TMRG $\chi^*(\alpha,t)$ data and obtain a
fit to these combined data by the fit function formulated in the previous section.

\begin{figure}
\epsfxsize=3.035in
\centerline{\epsfbox{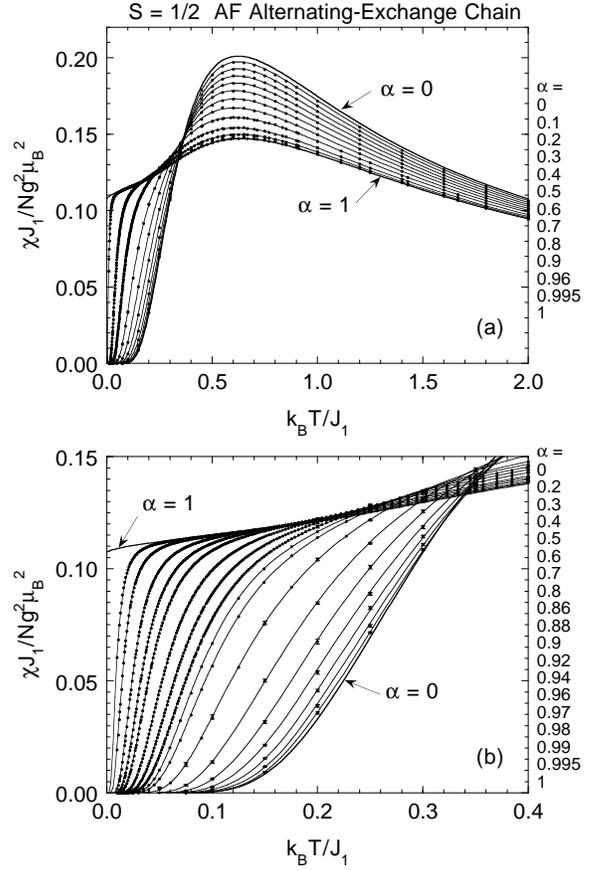}}
\vglue 0.1in
\caption{(a) Magnetic susceptibility $\chi$ versus temperature $T$ for
the spin $S = 1/2$ antiferromagnetic (AF) alternating-exchange Heisenberg chain with
alternation parameter $\alpha = J_2/J_1$ from~0 to~1, as shown.  The small 
filled circles are a selection of the calculated QMC and TMRG data, where for clarity
only a small subset of the available data are plotted.  The set of curves through
the data is obtained from the global two-dimensional ($\alpha,t$) fit function in
Eqs.~(\protect\ref{EqChi*AltChn:all}) with parameters given in
Table~\protect\ref{TableAltChnPars}.  The solid curves for
$\alpha = 0$ and~1 are plots of the fit function for the dimer and uniform chain
susceptibilities, respectively, for which no data are plotted.  The parameter $J_1$
is the larger of the two alternating exchange constants.  (b) Expanded plot of the
fit for a selection of data at low temperatures.  Error bars are plotted with the QMC
data in (b), but are not plotted in (a) because they are not visible on the scale
of this figure.}
\label{QMCDMRGFinalFit}
\end{figure}

Some of the results for $t\leq 2$ are shown as the filled symbols without error bars
in Fig.~\ref{QMCDMRGFinalFit}(a) (the error bars are smaller than the data symbols);
an expanded plot of data for $t\leq 0.4$ is shown in
Fig.~\ref{QMCDMRGFinalFit}(b).  [A log-log plot of the TMRG $\chi^*(\alpha,t)$ data
at low $t$ is shown below in Fig.~\ref{TDMRGLoTFits}.]  Also shown in both figures as
the two bounding  solid curves with no data points are the fits we obtained above to
$\chi^*(t)$ for the dimer and uniform chain (Fit~1), respectively.  The data points
plotted for a given $\alpha$ value are the subset below the upper temperature limits
of the figures, of the subset of available data points which were fitted by our fit
function as described below.

We fitted a combined QMC and TMRG $\chi^*(\alpha,t)$ data set containing 2551
selected data points over the temperature range $0.01\leq t\leq 10$.  The 802 QMC
data points covered \ the \ ranges \ $0.01\lesssim t\leq 4$ \ and
\ $0.05\leq\alpha\leq0.99$. 
\vglue-0.01in
\noindent The average estimated absolute accuracy of these QMC data is
$1.7\times10^{-4}$. The best estimated accuracy among these QMC data is
$7.7\times10^{-6}$ and the worst  is $1.5\times10^{-3}$, with the better accuracies
occurring at the highest temperatures.  The 1749 TMRG data points covered the ranges
$0.01\leq t\leq10$ and
$0.8\leq\alpha\leq 0.995$.  We did not use all 22\,370 TMRG data points in the
available data set, because this would have weighted the region $\alpha\lesssim 1$
too heavily in the fit, and in any case a large fraction of these are for
temperatures below our low-temperature fitting limit of $t = 0.01$.  We used the
low-temperature data to determine the spin gaps as described in the following
section.

We fitted this $\chi^*(\alpha,t)$ data set by Eqs.~(\ref{EqChi*AltChn:all}), with
the constraints on the parameters discussed above.  Obtaining a reliable
28-parameter two-dimensional fit to these data over the full above-cited ranges of
$t$ and $\alpha$, with no poles in the fit, posed a very difficult challenge.  The
particular choice of starting parameters and the detailed sequence of refinements
were found to be important to avoiding poles in the final fit.  Since there are a
total of 28 parameters in the fit function for 2551 data points, the data to
parameter ratio is 91.  The number of fitting parameters seems large, until it is
realized that we are simultaneously fitting $\chi^*(t)$ data for 29 different
$\alpha$ values, so on average a $\chi^*(t)$ data set for a given
$\alpha$ value is fitted by a single parameter.  A weighting function was not
included during the variance minimization, because we were interested in obtaining
a fit which treated all the data points the same on an absolute scale;  this choice
optimizes the fit for use in modeling experimental data.

The parameters of the fit are given in Table~\ref{TableAltChnPars}, where we have
also included the constrained parameters for completeness and for ease of
implementation of our fit function by the reader.  From
Eqs.~(\ref{EqDconstraints:all}), the constrained parameters $D_2$
and $D_3$ contain products of the third-order (in $\alpha$) polynomial
$\Delta^*_{\rm fit}$ with itself and/or with the fourth-order $N_1$ fitting
polynomial, so $D_2$ and $D_3$ are of seventh and tenth-order, respectively.  The
two-dimensional fit is shown as the set of solid curves
in Fig.~\ref{QMCDMRGFinalFit}.  The variance of the fit is $v =
3.77\times 10^{-8}$.  The absolute rms deviation $\sqrt{v}\approx 1.9\times 10^{-4}$
is about the same as the average estimated accuracy of the QMC data noted above,
indicating that the fit function is appropriate and that the fit is a reliable
representation of 
\widetext
\begin{table}
\caption{Parameters in the fit function [Eqs.~(\protect\ref{EqChi*AltChn:all})] for
$\chi^*(\alpha,t)$ of the $S=1/2$  antiferromagnetic alternating-exchange Heisenberg
chain.  Note that $D_2$ and $D_3$ are respectively of seventh and tenth order in
$\alpha$.}
\begin{tabular}{cccccc}
parameter  & $m = 0$ & $m = 1$ & $ m= 2$ & $m = 3$  & $m = 4$ \\
\hline
$N_{1m}$ &0.63427990 &$-$2.06777217 &$-$0.70972219  & 4.89720885
         &$-$2.80783223  \\
$N_{2m}$ &0.18776962 &$-$2.84847225 &5.96899688 &$-$3.85145137 &0.64055849 \\
$N_{3m}$ &0.033603617 &$-$0.757981757  & 4.137970390 &$-$6.100241386 &
    2.701116573 \\
$N_{4m}$ &0.0038611069 &0.5750352896 &$-$2.3359243110 &2.934083364
         &$-$1.1756629304 \\
$N_{5m}$ &0.00027331430 &$-$0.10724895512 &0.40345647304 &$-$0.48608843641 &
    0.18972153852 \\
$N_{6m}$ &0 &0.00578123759 &$-$0.02313572892 &0.02892774508 &$-$0.01157325374 
    \\
$N_{7m}$ & &2.59870347$\times 10^{-7}$ & $-$2.39236193$\times 10^{-7}$ \\
$D_{1m}$ & $-$0.11572010 &$-$1.31777217 &1.29027781 &3.39720885 &$-$2.80783223
     \\
$D_{2m}$ &0.08705969 &$-$1.44693321 & 5.09401919 & $-$10.51861382 & 8.97655318
   \\
 & 5.75312680 ($m=5$) & $-$11.83647774 ($m=6$) & 4.21174835 ($m=7$) \\
$D_{3m}$ &0.00563137 & 0.65986015 & $-$1.38069533 &$-$0.09849603 &7.54214913 
  \\
 & $-$22.31810507 ($m=5$) & 27.60773633 ($m=6$) & $-$6.39966673 ($m=7$) \\
 & $-$15.69691721 ($m=8$) & 13.37035665 ($m=9$) & $-$3.15881126 ($m=10$) \\
$D_{4m}$ &0.0010408866 &0.1008789796 &$-$0.9188446197 &1.6052570070
         &$-$0.7511481272 \\
$D_{5m}$ &0.0000683286 &$-$0.1410232710 &0.6939435034 &$-$0.9608700949 &
    0.4106951428 \\
$D_{6m}$ & 0 &0.0367159872 &$-$0.1540749976 &0.1982667100 &$-$0.0806430233 \\
$D_{7m}$ & 0 &$-$0.00314381636 &0.01140642324 &$-$0.01338139741 &0.00511879053
   \\
$D_{8m}$ & &1.25124679$\times 10^{-7}$ & $-$1.03824523$\times 10^{-7}$\\
\hline
$g_1$ & $g_2$ & $y$ & $z$\\
\hline
0.38658545 & $-$0.20727806 & 4.69918784 & 3.55692695\\
\end{tabular}
\label{TableAltChnPars}
\end{table}
\narrowtext
\begin{figure}
\epsfxsize=2.54in
\centerline{\epsfbox{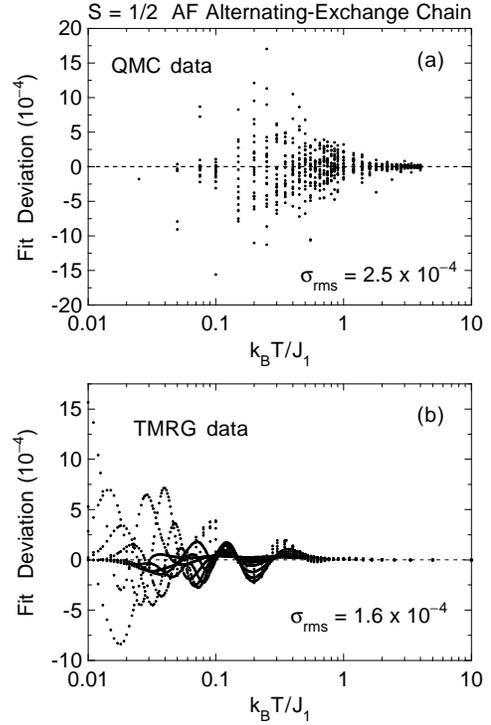}}
\vglue 0.1in
\caption{Deviation in absolute units of the fit function for the magnetic
susceptibility $\chi^*$ versus temperature $T$ for the spin $S = 1/2$
antiferromagnetic (AF) alternating-exchange Heisenberg chain, with alternation
parameter $\alpha = J_2/J_1$ from~0.05 to~0.995, from the QMC data (a) and TMRG data
(b).  The parameter $J_1$ is the larger of the two alternating exchange constants. 
The absolute rms deviations of the respective data from the fit are given in the
figures.  The fit function is given in Eqs.~(\protect\ref{EqChi*AltChn:all}) with the
parameters in Table~\protect\ref{TableAltChnPars}.}
\label{QMCDMRGFitDevs}
\end{figure}
\newpage
\begin{figure}
\epsfxsize=3.02in
\centerline{\epsfbox{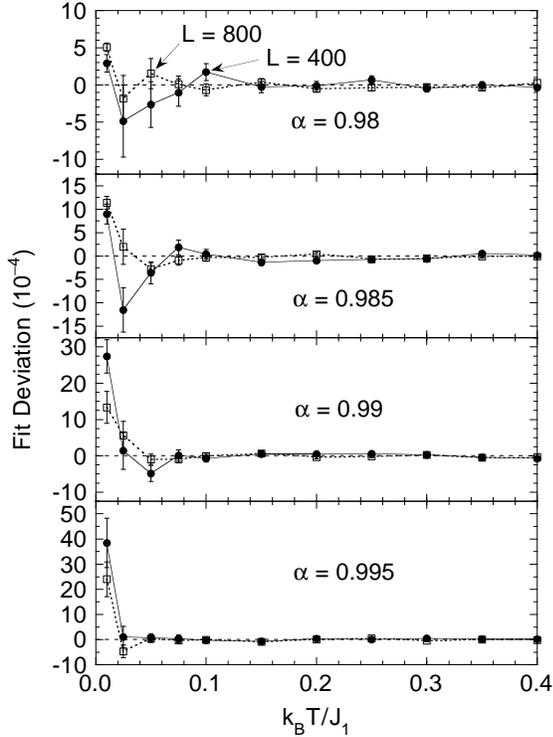}}
\vglue 0.1in
\caption{Deviation in absolute units of the fit function for the magnetic
susceptibility $\chi^*$ versus temperature $T$ for the spin $S = 1/2$
antiferromagnetic (AF) alternating-exchange Heisenberg chain, from $L = 400$
($\bullet$) and $L = 800$ (open squares) QMC data for $\alpha = 0.98$, 0.985, 0.99
and 0.995.  The only significant deviation is at the lowest temperature $T = 0.01
J/k_{\rm B}$.  The fit function is given in Eqs.~(\protect\ref{EqChi*AltChn:all})
with the parameters in Table~\protect\ref{TableAltChnPars}.}
\label{TroyerAltChnL=400}
\end{figure}
\noindent the data.  The fit deviations from the 802 QMC and 1749 TMRG
data are shown separately in Figs.~\ref{QMCDMRGFitDevs}(a)
and~\ref{QMCDMRGFitDevs}(b), respectively.  A comparison of the two figures shows
that the TMRG data are, on average, significantly more precise at a given temperature.

After the parameters in the present $\chi^*(\alpha,t)$ fit function were finalized,
as a check on the accuracy of the fit function for $\alpha$ values close to the
uniform limit, we carried out QMC $\chi^*(t)$ simulations for alternating-exchange
chains of length $L = 400$ and 800, factors of four and eight longer than the chains
for which QMC data were combined with TMRG data to determine the fit function,
respectively.  The simulations were carried out for $\alpha = 0.98, 0.985, 0.99$,
and~0.995 at temperatures $0.01\leq t\leq 4$.  Overall, the fit function was found
to be in extremely good agreement with the QMC data.  For
$0.4\leq t\leq 4$, the $\chi^*(\alpha,t)$ fit function agreed with the simulation
data to within about $\pm 5\times 10^{-5}$ or better.  The deviations of the fit
function from the data for $0.01\leq t\leq 0.4$ are shown in
Fig.~\ref{TroyerAltChnL=400}, along with the error bars on the QMC data.  As can be
seen from the figure, the only significant deviation of the fit function from the
QMC data in this $t$ range is at the lowest \ temperature \ $t = 0.01$ \ for \ each
\ of \ the \ four \ $\alpha$ 
\begin{figure}
\epsfxsize=3in
\centerline{\epsfbox{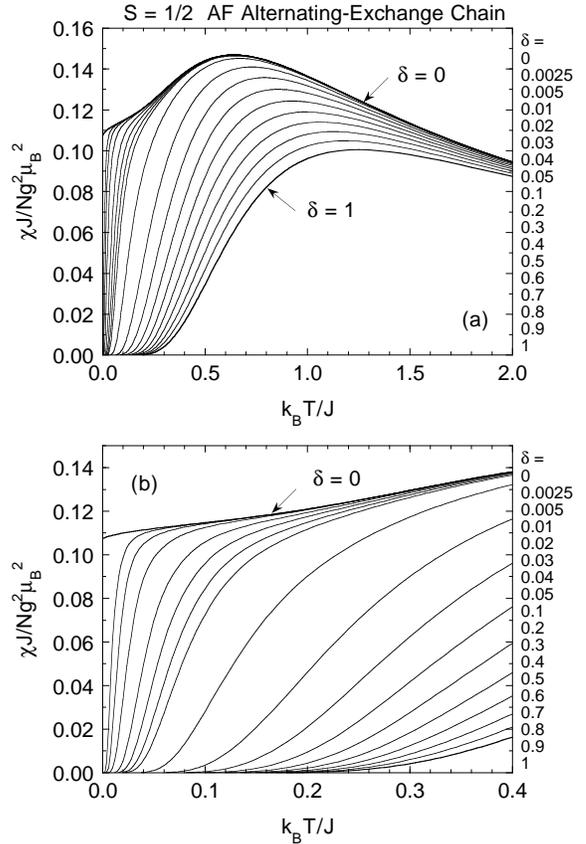}}
\vglue 0.1in
\caption{(a) Magnetic susceptibility $\chi$ versus temperature $T$ for
the spin $S = 1/2$ antiferromagnetic alternating-exchange Heisenberg chain with
values of alternation parameter $\delta$ from 0 to~1 as shown, where $\chi$ is
scaled by $1/J$ and $T$ by $J$ in contrast to Fig.~\protect\ref{QMCDMRGFinalFit}. 
The parameter $J=J_1/(1+\delta)$ is the average of the two exchange constants $J_1$
and $J_2$ alternating along the chain.  (b) Expanded plot at low
temperatures.  These $\overline{\chi^*}(\delta,\overline{t})$ plots were generated
using our two-dimensional $\chi^*(\alpha,t)$ fit function which was converted to the
variables $(\delta,\overline{t})$ using Eq.~(\protect\ref{EqBarDef:b}).}
\label{QMCDMRGFinalFitbar}
\end{figure}
\vglue-0.03in
\noindent values.  Because the fit deviations at this temperature remain upon
increasing the length of the simulated chain from $L = 400$ to $L = 800$, these fit
deviations are most likely due to inaccuracies in the fit function, as expected at
this lowest fitted temperature.

For compounds showing spin-Peierls or other types of second-order spin dimerization
transitions, it is more appropriate to scale $\chi$ by $1/J$ and $T$ by
$J$, where $J$ is the average of $J_1$ and $J_2$, in which case the appropriate
alternation parameter is $\delta$ rather than $\alpha$.  It is straightforward to
convert our $\chi^*(\alpha,t)$ fit function to the form
$\overline{\chi^*}(\delta,\overline{t})$, where
$\overline{t}\equiv k_{\rm B}T/J$, using Eq.~(\ref{EqBarDef:b}).  We have done this
and plot the $\overline{\chi^*}(\delta,\overline{t})$ fit function versus
temperature for a series of $\delta$ values in Fig.~\ref{QMCDMRGFinalFitbar}(a).  An
appealing  monotonic progression of $\overline{\chi^*}(\delta,\overline{t})$ with
increasing $\delta$ is seen in Fig.~\ref{QMCDMRGFinalFitbar}(a); an expanded plot at
lower temperatures is shown in Fig.~\ref{QMCDMRGFinalFitbar}(b).  This formulation of
the fit function allows accurate estimates to be made of the temperature-dependent
spin gap in compounds exhibiting spin-dimerization transitions, provided that the
nearest neighbor $S = 1/2$ AF alternating-exchange Heisenberg model is appropriate
to them.  An illustration of the procedure and the results to be gained will be
given later when we model the $\chi(T)$ data for NaV$_2$O$_5$.

\section{Spin Gap from TMRG $\bbox{\chi^*(\alpha,\lowercase{t})}$}
\label{SecTDMRGgaps}

According to Eq.~(\ref{EqLogChi:c}), if highly precise $\chi^*(t)$ data in the
low-$t$ limit are available, the spin gap $\Delta^*$ can in principle be computed
directly from the derivative of these data with respect to
inverse temperature.  However, in general the maximum temperature of the low-$t$
limit region is ill defined since it depends on how precise and accurate the data
are and the accuracy to which $\Delta^*$ is to be determined.  Therefore, in practice
one could define a temperature-dependent effective spin gap $\Delta^*_{\rm eff}$ from
Eq.~(\ref{EqLogChi:c}) as
\begin{equation}
\Delta^*_{\rm eff}(t) = -{\partial\ln(\chi^*\sqrt{t})\over \partial(1/t)}~,
\label{EqDeff}
\end{equation}
and then try to ascertain $\Delta^*$ from the extrapolated zero-temperature limit
$\Delta^* = \lim_{t\to 0}\Delta^*_{\rm eff}(t)$.  Using Eq.~(\ref{EqLogChi:b}) would
be less desirable and precise because a fit of this type typically averages
$\Delta^*_{\rm eff}(t)$ over a rather large temperature range.

An overview of $\Delta^*_{\rm eff}(\alpha,t)$ determined from our TMRG
$\chi^*(\alpha,t)$ data for $0.8\leq\alpha\leq 0.995$ using Eq.~(\ref{EqDeff}) is
shown in Fig.~\ref{TDMRGdln}(a).  At the lowest temperatures, and for $\alpha$ not
too close to 1, the $\Delta^*_{\rm eff}(\alpha,t)$ data do approach a constant value
with decreasing $t$, confirming the applicability of Eqs.~(\ref{EqTroyer}) and
prior assumptions and hence Eqs.~(\ref{EqLogChi:all}) and~(\ref{EqDeff}) to the
alternating chain, and the approximate values of $\Delta^*(\alpha)$ can be estimated
from the figure.  Closer inspection reveals that $\Delta^*_{\rm eff}(\alpha,t)$
shows a weak maximum before decreasing by $\approx{1\over 2}$\% to $\Delta^*$ as
$t\to 0$, as illustrated in Fig.~\ref{TDMRGdln}(b) for $\alpha=0.8$.  For this among
other reasons, we will not use Eq.~(\ref{EqDeff}) to extract the spin gaps from our
TMRG
$\chi^*(\alpha,t)$ data.  On the other hand, we need to know whether such behavior is
expected, since it could conceivably arise from systematic errors in the TMRG
calculations.  Therefore, in the next section we study the $\Delta^*_{\rm
eff}(\alpha,t)$ expected at low temperatures for the alternating-exchange chain.  As
part of this study, we formulate and discuss the fit function which we will use in
Sec.~\ref{SecTMRGLoTFit} to extract $\Delta^*(\alpha)$ from our TMRG
$\chi^*(\alpha,t)$ data at low temperatures.

\subsection{Effective spin gap $\bbox{\Delta^*_{\rm eff}(\Delta^*,t)}$ for the
alternating-exchange chain}

From our definition of $\Delta^*_{\rm eff}$ in Eq.~(\ref{EqDeff}), a discussion of
how this quantity varies with $t$ at low $t$ requires an independent estimate of
$\chi^*(\alpha,t)$ for the alternating-exchange 
\begin{figure}
\epsfxsize=3in
\centerline{\epsfbox{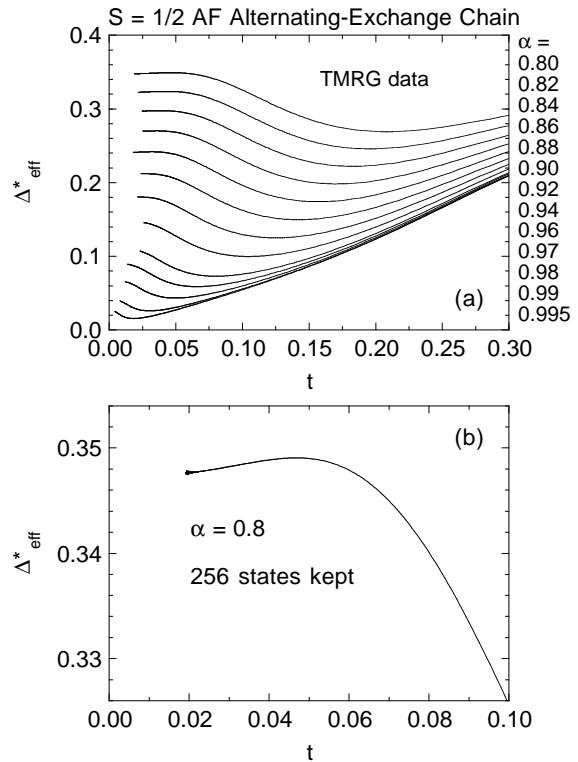}}
\vglue 0.1in
\caption{(a) Overview of the effective spin gap
$\Delta^*_{\rm eff}(\alpha,t)$ vs temperature $t$ for the $S=1/2$ AF
alternating-exchange Heisenberg chain, derived from our TMRG $\chi^*(\alpha,t)$ data
using the definition in  Eq.~(\protect\ref{EqDeff}), where $\alpha = J_2/J_1$ is the
alternation parameter.  (b) Expanded plot of $\Delta^*_{\rm eff}(t)$ for $\alpha =
0.8$ at low temperatures from (a).}
\label{TDMRGdln}
\end{figure}
\vglue-0.03in
\noindent chain, which must
include at least the leading order correction to the low-$t$ limit in
Eqs.~(\ref{EqTroyer}).  As a first attempt, we used the general expression for
$\chi^*(t)$ in Eqs.~(\ref{EqTroyerGen:all}), which requires as input the one-magnon
dispersion relation $\varepsilon(k)$ for the alternating chain.  For this we used the
explicit $\varepsilon(\Delta^*,k)$ for $0\leq\alpha\leq 1$ in Eqs.~(\ref{EqMyE(k)})
that we presented and discussed previously.  The resultant $\chi^*(t)$ is plotted for
eleven $\Delta^*$ values in Fig.~\ref{AltChnChi(Gap,T)2}(a), where the
results are designated by $\chi^{*(1)}$ in the figure.  Although
$\chi^{*(1)}(\Delta^*,t)$ is exact in both the low- and high-$t$ limits, the results
are only qualitatively correct at intermediate temperatures, as can be seen by
comparing Fig.~\ref{AltChnChi(Gap,T)2}(a) with the QMC and TMRG data and fit in
Fig.~\ref{QMCDMRGFinalFit}.  Troyer, Tsunetsugu and W\"urtz\cite{Troyer1994} obtained
a very good fit of $\chi^{*(1)}(\Delta^*,t)$ to QMC
$\chi^*(t)$ simulation data over a large temperature range for the $S=1/2$ two-leg
Heisenberg ladder with spatially isotropic exchange; however, they assumed a
$\varepsilon(\Delta^*,k)$ in the fit function which was later found to be inaccurate
over much of the Brillouin zone.

We formulated an approximation [designated as $\chi^{*(2)}$] which is more
accurate in the low-temperature range, and which we will use in the next
section as a fit function to fit our TMRG $\chi^*(t)$ data at low $t$ to extract
$\Delta^*(\alpha)$.  The function $\chi^{*(2)}$ was obtained by summing the
susceptibilities of isolated dimers with a distribution of singlet-trip-
\begin{figure}
\epsfxsize=3.03in
\centerline{\epsfbox{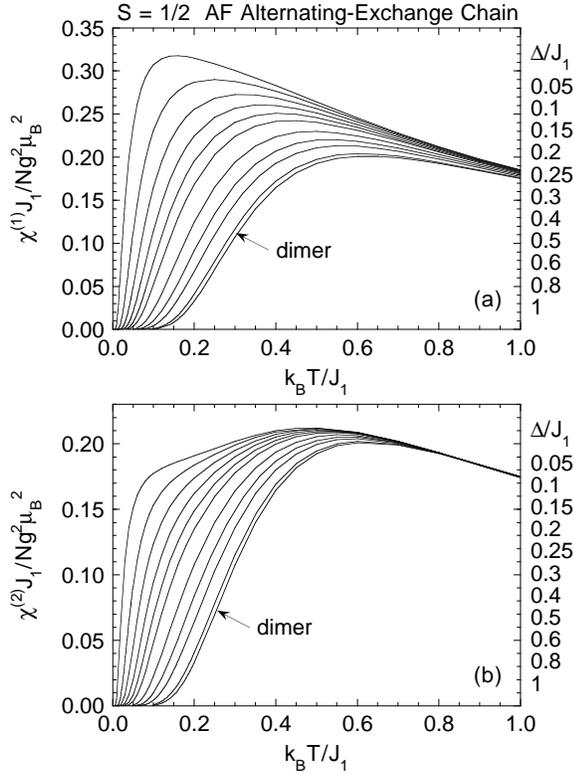}}
\vglue 0.1in
\caption{Magnetic susceptibilities $\chi^{(1)}$ (a) and $\chi^{(2)}$ (b)  vs
temperature $T$ for the spin
$S = 1/2$ antiferromagnetic alternating-exchange Heisenberg chain, calculated using
two different approximations for $\chi^*(t)$, respectively (see text).  Note the
different scales for the ordinates in~(a) and~(b).}
\label{AltChnChi(Gap,T)2}
\end{figure}

\noindent let energy
gaps given by our one-parameter dispersion relation $\varepsilon(\Delta^*,k)$ for
$0\leq\Delta^*\leq 1$ in Eqs.~(\ref{EqMyE(k)}), which takes into account the
interdimer interactions.  Thus from Eq.~(\ref{EqChiDimer:a}) we simply obtain
\begin{equation}
\chi^{*(2)}(\Delta^*,t) = {1\over \pi t}\int_0^\pi {dk\over 3 +
{\rm e}^{\varepsilon(\Delta^*,k)/t}}~.
\label{EqChi(2)}
\end{equation}
Note that we make no assumptions here about the form of the function
$\Delta^*(\alpha)$, since only $\Delta^*$ appears in the expression.  This
$\chi^{*(2)}(\Delta^*,t)$ is exact in both the low- and high-$t$ limits, as is
$\chi^{*(1)}(\Delta^*,t)$, and both reproduce $\chi^*(t)$ for the isolated dimer
($\Delta^* = 1$) exactly, but $\chi^{*(2)}(\Delta^*,t)$ is more accurate at
intermediate temperatures for $\alpha\lesssim 1$ as shown in
Fig.~\ref{AltChnChi(Gap,T)2}(b).  In addition, by comparing
$\chi^{*(1)}(\Delta^*,t)$ and $\chi^{*(2)}(\Delta^*,t)$ with the TMRG
$\chi^*(\alpha,t)$ calculations at low $t$, we found that the low-$t$ corrections to
the low-$t$ limit in Eqs.~(\ref{EqTroyer}) are much more accurately given by
$\chi^{*(2)}(t)$ than by $\chi^{*(1)}(t)$.  We will therefore not discuss
$\chi^{*(1)}(t)$ further here.

At low temperatures, the approximation $\chi^{*(2)}(\Delta^*,t)$ is expected to
accurately describe the leading-order $t$ corrections to the low-$t$ limit only as
long as the average number of magnons $n_{\rm m}$ occupying a state near the minimum
in the one-magnon band is much less than unity.  Using \,the \,expression \,for \,the
\,boson \,occupation \,number 
\begin{figure}
\epsfxsize=3.2in
\centerline{\epsfbox{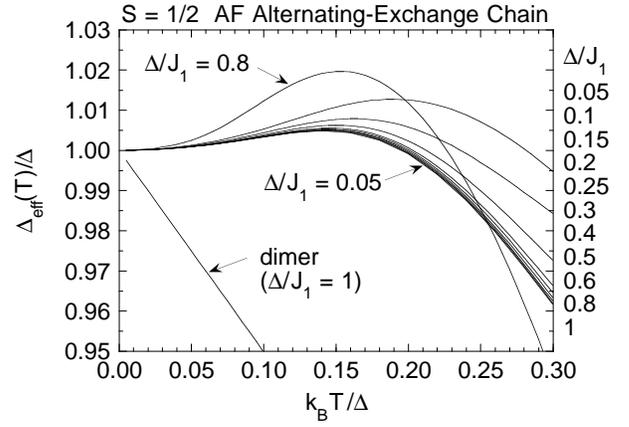}}
\vglue 0.1in
\caption{Effective spin gap $\Delta_{\rm eff}$, defined in
Eq.~(\protect\ref{EqDeff}) and computed using Eq.~(\protect\ref{EqChi(2)}), vs
reduced temperature $k_{\rm B}T/\Delta=t/\Delta^*$ for the $S = 1/2$
antiferromagnetic alternating-exchange Heisenberg chain for various values of
$\Delta^*\equiv \Delta/J_1$, where $t=k_{\rm B}T/J_1$.  A  limiting behavior is seen
for $\Delta^*\to 0$, for which the maximum in $\Delta_{\rm eff}(k_{\rm B}T/\Delta)$
occurs at $k_{\rm B}T/\Delta\approx 0.14$.  The linear in
$T$ behavior of $\Delta_{\rm eff}(k_{\rm B}T/\Delta\to 0)$ for the isolated dimer 
($\Delta^* = 1$) is due to the identically zero width of the one-magnon dispersion
relation for this $\Delta^*$ value.}
\label{AltChnGAPeff}
\end{figure}

\noindent for this case,
\begin{equation}
n_{\rm m} = {1\over {\rm e}^{\Delta^*/t} -1}~,
\end{equation}
yields $t/\Delta^* = 0.22$ and~0.42 for $n_{\rm m} = 0.01$ and~0.1, respectively. 
Thus, when fitting our low-$t$ TMRG $\chi^*(\alpha,t)$ data by the fit function
$\chi^{*(2)}(\Delta^*,t)$ in the following section, we expect 
$\chi^{*(2)}(\Delta^*,t)$ to be sufficiently accurate only for $t/\Delta^*\lesssim
0.4$.  For this reason, our fits will be limited to this maximum scaled temperature. 

 We have computed $\Delta_{\rm eff}(\Delta^*,T)/\Delta$ from
$\chi^{*(2)}(\Delta^*,t)$ in Eq.~(\ref{EqChi(2)}), using the definition in
Eq.~(\ref{EqDeff}), and plot the results vs $k_{\rm B}T/\Delta$ in
Fig.~\ref{AltChnGAPeff}.  For the dimer ($\Delta^*=1$), one finds analytically
that $\Delta_{\rm eff}(T)/\Delta = 1 - 2 k_{\rm B}T/\Delta$ to lowest order in $T$. 
On the other hand, for $0 < \Delta^* < 1$, the initial dependence is positive and
quadratic in $T$, and a maximum is seen in $\Delta_{\rm eff}(T)/\Delta$, which for
$\Delta^*\lesssim 0.4$ occurs at $t/\Delta^* \equiv  k_{\rm B}T/\Delta\approx 0.14$
with a height of $\approx 0.5$\%.  This height is  quantitatively consistent with
the data in Fig.~\ref{TDMRGdln}(b) derived from the TMRG $\chi^*(t)$ for $\alpha =
0.8$.  Thus the weak maximum seen in that figure is not a spurious effect.

\subsection{Fits to the low-$\bbox{t}$ TMRG $\bbox{\chi^*(\alpha,t)}$ data}
\label{SecTMRGLoTFit}

We were tempted to fit $\Delta^*_{\rm eff}(\alpha,t)$ derived from the low-$t$
TMRG $\chi^*(\alpha,t)$ data, as discussed above, to obtain the spin gaps
$\Delta^*(\alpha,t)$.  However, this procedure would have weighted the
$\chi^*(\alpha,t)$ data in an ill advised way.  We therefore decided to do
conventional fits of the low-$t$ $\chi^*(\alpha,t)$ data by the fit function
$\chi^{*(2)}(\Delta^*,t)$ in Eq.~(\ref{EqChi(2)}).  For a given $\alpha$, this is a
one-parameter ($\Delta^*$) fit function and the fits are therefore stringent tests
of both the appropriateness of the fit function and the precision and accuracy of
the data.  Because the temperature dependence of the accuracy of the calculations is
unknown except for the uniform chain data (see Fig.~\ref{Alpha=1TDMRGDev}), we
assumed that all data for a given
$\alpha$ in a given fitted temperature range have the same accuracy.  Thus in the
nonlinear least-squares fits for each $\alpha$ we minimized the square of the
relative rms deviation of the fit from the data
\begin{equation}
\sigma_{\rm rms}^2 = {1\over N_{\rm p}}\sum_{i=1}^{N_{\rm p}} \frac{[\chi^{*(2)}(t_i)
- \chi^*(t_i)]^2}{[\chi^*(t_i)]^2}~,
\label{EqSigma}
\end{equation}
where $N_{\rm p}$ is the number of data points fitted, which was usually 
between 250  and 1500.

Due to the presence of the spin gap $\Delta^*$ in the exponential of the fit
function, $\sigma_{\rm rms}$ is extremely sensitive to the precise value of
$\Delta^*$ when low-$t$ fits are carried out.  For example, close to the optimum
$\Delta^*$ fit value, a change in $\Delta^*$ by only 0.0001 ($\sim 0.1\,\%$) can
change $\sigma_{\rm rms}$ by up to $\sim 300$\,\%.  Thus a few percent accuracy in the
$\chi^*(t)$ data at low~$t$ is sufficient to allow $\Delta^*$ to be determined for
a given fit to a precision better than 0.0001.  For a given $\alpha$, the obtained
$\Delta^*$ was found to be insensitive, typically to within $\approx 0.000$2, to the
$t$ range of the fit, as long as the maximum fitted temperature satisfied
$t/\Delta^*\lesssim 0.4$, consistent with the above discussion of the boson
occupation number.  This lack of sensitivity of the value of the fitted
$\Delta^*$ to the precise fit range demonstrated that the fit function
$\chi^{*(2)}(t)$ is an appropriate one.  Depending on the $\alpha$ value and the $t$
range of the fit, $\sigma_{\rm rms}$ was typically between 0.1\,\% and several
percent.

\begin{figure}
\epsfxsize=3in
\centerline{\epsfbox{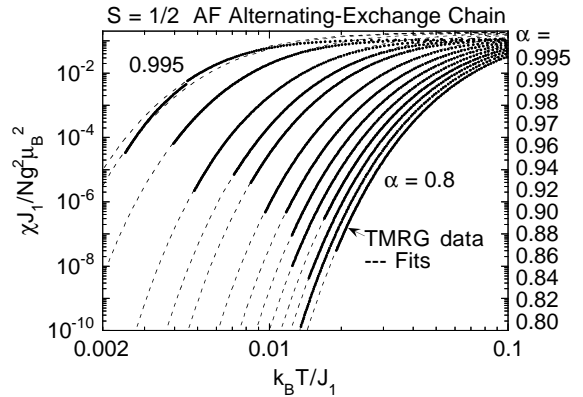}}
\vglue 0.1in
\caption{Log-log plots of reduced magnetic susceptibility $\chi^*\equiv \chi
J_1/Ng^2\mu_{\rm B}^2$ vs reduced temperature $t\equiv k_{\rm B}T/J_1$ (solid
circles) for spin $S = 1/2$ antiferromagnetic alternating-exchange Heisenberg
chains with alternation parameters $\alpha\equiv J_2/J_1$ shown in the figure, 
calculated using TMRG\@.  The corresponding fits to the lower temperature data
[$k_{\rm B}T/\Delta(\alpha)\protect\lesssim 0.4$] by Eq.~(\protect\ref{EqChi(2)}) are
shown as the dashed curves, which are extrapolated to lower and
higher temperatures in the figure.  The discontinuity in the data and fit for $\alpha
= 0.995$ at $t\approx 0.0043$ is due to an increase with increasing $t$ at that $t$
in the number of states kept in the calculations from 150 to 256.  The spin gaps
$\Delta^*(\alpha)\equiv \Delta(\alpha)/J_1$ found from the fits are given in
Table~\protect\ref{TabMoreLit&pw}.}
\label{TDMRGLoTFits}
\end{figure}

The $\Delta^*(\alpha)$ values obtained from the fits are listed in
Table~\ref{TabMoreLit&pw}, together with the estimated accuracies in parentheses. 
Note that a quoted accuracy is associated with variations in $\Delta^*$ in fits to a
specific set of data for a given $\alpha$ over various temperature ranges as
discussed above, and does not include possible systematic errors due to, e.g., the
finite fixed number of states kept in the TMRG calculations.  Also included in
Table~\ref{TabMoreLit&pw} are literature
data\cite{Barnes1998,Ladavac1998,Augier1997,Uhrig1999} which will be compared with
the present results in the next section.  Log-log plots of the low-$t$ data and fits
are shown in Fig.~\ref{TDMRGLoTFits}, where on the scale of this figure, for most
$\alpha$ values the data and fit are identical (cannot be distinguished) within the
fitted temperature range.  Extrapolations of the fits to higher and lower
temperatures are also shown for comparison with the data.

\section{Comparisons of the Calculations with Previous Work}
\label{SecCompPrevWork}

\subsection{Spin gap}
\label{SecCompareSpinGap}
\vglue0.02in
Our $\overline{\Delta^*}(\delta)$ spin gap data determined by fitting our TMRG
$\chi^*(t)$ data by Eq.~(\ref{EqChi(2)}) are plotted in Fig.~\ref{Gap(delta)}(a),

\noindent
along with the results of previous
workers\cite{Barnes1998,Ladavac1998,Augier1997,Uhrig1999} listed in
Table~\ref{TabMoreLit&pw}.  The solid curve is the
function $\overline{\Delta^*}=2\delta^{3/4}$ in Eq.~(\ref{EqDimParams2:b}) proposed
by Barnes, Riera, and Tennant (BRT).\cite{Barnes1998}  The overall behavior of the
data in Fig.~\ref{Gap(delta)}(a) is well described by this function, but significant
deviations of the data from the curve occur as illustrated in the expanded plot for
$\delta\leq 0.1$ in Fig.~\ref{Gap(delta)}(b).  The error bars are included with each
plotted data point in Fig.~\ref{Gap(delta)}(b), except for the data of
Ref.~\onlinecite{Augier1997} which were not available, but they are all
small and not clearly seen.  Our values for $\delta\lesssim 0.1$ are significantly
smaller than those of Uhrig {\it et al.},\cite{Uhrig1999} where the differences are far
outside the combined limits of error, and are larger than those of Augier {\it et
al.}\cite{Augier1997}

As will be seen explicitly in Sec.~\ref{SecNaV2O5Fits} below, our 
$\overline{\chi^*}(\delta,\overline{t})$ fit function allows $\delta(T)$ to be
determined for real materials by using the fit function to model experimental
$\chi(T)$ data.  However, if one would like to determine the spin gap $\Delta(T)$
from the derived $\delta(T)$, an expression is needed for $\Delta(\delta)$ over the
entire range $0\leq\delta\leq 1$ in order to be generally useful and
applicable.  At present, the only extant expression is that of BRT in
Eqs.~(\ref{EqDimParams2:all}).  As seen in Fig.~\ref{Gap(delta)} and in
Table~\ref{TabChiParams} below, this expression is only an approximation that fits
neither BRTs' $\Delta^*(\alpha)$ data for $0.1\leq\alpha\leq0.9$ nor our TMRG
spin gap data for $0.8\leq\alpha\leq0.995$ to within the respective error bars.  To
formulate a more flexible expression, we modify BRTs' formula to read
\begin{mathletters}
\label{EqD(d):all}
\begin{equation}
\overline{\Delta^*}(\delta)\equiv {\Delta(\delta)\over J} = 2\,\delta\,^{y(\delta)}~,
\label{EqD(d):a}
\end{equation}
so the $\delta$-dependent power $y$ is
\newpage
\widetext
\begin{table}
\caption{Spin gaps $\overline{\Delta^*}(\delta)\equiv \Delta(\delta)/J$ and
${\Delta^*}(\alpha)\equiv \Delta(\alpha)/J_1$ for the $S=1/2$ antiferromagnetic
alternating-exchange Heisenberg chain as determined using $T=0$ DMRG calculations by
Uhrig {\it et al.}\ [$\overline{\Delta^*}_{\rm U}(\delta)$ and
${\Delta^*}_{\rm U}(\alpha)$] (Ref.~\protect\onlinecite{Uhrig1999}), by Barnes, Riera,
and Tennant [$\overline{\Delta^*}_{\rm BRT}(\delta)$] using multiprecision
methods (Ref.~\protect\onlinecite{Barnes1998}), by Augier {\it et al.}\  
[$\overline{\Delta^*}_{\rm A}(\delta)$] (Ref.~\protect\onlinecite{Augier1997}), and by
us [$\overline{\Delta^*}_{\rm pw}(\delta)$ and ${\Delta^*}_{\rm pw}(\alpha)$] in the
present work (pw) from our TMRG $\chi^*(\alpha,t)$ data as described in the text.  Two
$\Delta$ values are given for $\alpha=0.995$ in the present work: the first (larger)
value is for the number of states kept in the calculations $m = 150$ at $t <
0.004533$, whereas the second (smaller) value is for $m = 256$ at $t > 0.004533$.  The
${\Delta^*}(\alpha)$ data of Barnes, Riera, and Tennant are given in
Table~\protect\ref{TabChiParams}. Additional literature data include those of
Ladavac {\it et al.}\ obtained using a Green's function Monte Carlo
technique on rings of 6 to 200 spins (Ref.~\protect\onlinecite{Ladavac1998}):
$\overline{\Delta^*}(\delta)$ = 0.1815(5) ($\delta = 0.04$), 0.2156(1) (0.05),
0.301(1) (0.08) and 0.3603(1) (0.10).}
\begin{tabular}{llllllll}
$\delta$ & $\alpha$ & $\overline{\Delta^*}_{\rm U}(\delta)$ &
$\overline{\Delta^*}_{\rm BRT}(\delta)$ & $\overline{\Delta^*}_{\rm A}(\delta)$ &
$\overline{\Delta^*}_{\rm pw}(\delta)$ &
${\Delta^*}_{\rm U}(\alpha)$ & ${\Delta^*}_{\rm pw}(\alpha)$ \\
\hline
0.0025063	&0.995  	&           &          & &0.0268(3)		&           &0.0267(3)  \\
          &        &           &          & &0.0245(1)  &           &0.0244(1)  \\
0.004    	&0.99203	&0.046(1)			&          & &           &0.046(2)	  &           \\
0.0050251	&0.99   	&           &          & &0.0404(2)  &           &0.0402(2)\\
0.006    	&0.98807	&0.058(1)			&          & &           &0.058(2)	  &           \\
0.008    	&0.98413	&0.0685(10)	&          & &           &0.068(2)	  &           \\
0.01     	&0.98020	&0.0785(10)	&          & &           &0.078(2)	  &           \\
0.010101	&0.98   		&           &         &  &0.0667(2)		 &           &0.0660(2)  \\
0.015228	&0.97   		&           &          & &0.0901(2)		 &           &0.0887(2)  \\
0.02    	&0.96078	 &0.1213(1)		&          & &           &0.119(2)	  &           \\
0.020408	&0.96   		&           &          & &0.1116(4)		 &           &0.1094(4)  \\
0.03    	&0.94175	 &0.1559(1)		&       & 0.1269  & &0.151(2)	  &           \\
0.030928	&0.94   		&           &          & &0.1506(3)	&           &0.1461(3)  \\
0.035    &         &           &       &0.1485   &  &          &           \\
0.04    	&0.92308	 &0.1882(1)  &       &0.1686   &  &0.181(2)	  &           \\
0.041667	&0.92   		&           &          & &0.1870(3)		&           &0.1795(3)  \\
0.045    &         &           &       &0.1871   &  &           &           \\
0.05    	&0.90476	 &0.2188(1)		&           &0.2049 &   &0.208(2)	  &           \\
0.052632	&0.9    		&           &0.221(2)	 & &0.2219(3)		 &           &0.2108(3)  \\
0.06    	&0.88679	 &0.2485(1)  &        &0.2383   &           &0.234(2)	  &    \\ 
0.063830	&0.88   		&           &         &  &0.2557(2)		&           &0.2404(2)  \\
0.07    	&0.86916	 &0.2770(1)		&         &  &           &0.259(2)	  &           \\
0.075269	&0.86   		&           &         &  &0.2887(3)		&           &0.2685(3)  \\
0.08    	&0.85185	 &0.3048(1)		&         &  &           &0.282(2)	  &           \\
0.086957	&0.84   		&           &         &  &0.3213(2)		&           &0.2956(2)  \\
0.09    	&0.83486	 &0.3319(1)  &         &  &           &0.305(2)	  &           \\
0.098901	&0.82   		&           &         &  &0.3535(2)		&           &0.3217(2)  \\
0.1    	&0.81818	  &0.3583(1)		&         &  &           &0.326(2)	  &           \\
0.11   	&0.80180	  &0.3842(1)  &         &  &           &0.346(2)	  &           \\
0.11111	&0.8    		 &           &0.3860(3)&	 &0.3852(2)		&           &0.3467(2)  \\
0.12   	&0.78571	  &0.4095(1)		&         &  &           &0.366(2)	  &           \\
0.14   	&0.75439	  &0.4589(1)		&         &  &           &0.403(2)	  &           \\
0.16   	&0.72414	  &0.5066(1)		&         &  &           &0.437(2)	  &           \\
0.17647	&0.7    		 &           &0.54468(6)&	&	          &           &           \\
0.18   	&0.69492	  &0.5530(1)		&         &  &           &0.469(1)	  &           \\
0.2    	&0.66667	  &0.5981(1) &           &           &0.4985(14)	&           \\
0.25   	&0.6    		 &           &0.706620(9)&&		         &           &           \\
0.33333	&0.5    		 &           &0.8766369(7)&&			       &           &           \\
0.4    	&0.42857	  &1.0052(1)		&           & &          &0.718(1)	  &           \\
0.42857	&0.4    		 &           &1.05865915(4)&&			      &           &           \\
0.53846	&0.3    		 &           &1.256683488(2)&&			     &           &           \\
0.6    	&0.25   	  &1.3631(1)		&	          &&           &0.85194(8)	&           \\
0.66667	&0.2    		 &           &1.475349990&&		         &           &           \\
0.7    	&0.17647	  &1.5304(1)		&           &&           &0.90024(7)	&           \\
0.8    	&0.11111	  &1.6917(1)		&           &&           &0.93985(6)	&           \\
0.81818	&0.1    		 &           &1.720507887&&		         &           &           \\
0.85   	&0.081081	 &1.7705(1)		&           &&           &0.95701(6)	&           \\
0.9    	&0.052632	 &1.8480(1)		&           &&           &0.97265(6)	&           \\
\end{tabular}
\label{TabMoreLit&pw}
\end{table}
\newpage
\mbox{ }
\newpage
\narrowtext
\begin{figure}
\epsfxsize=3in
\centerline{\epsfbox{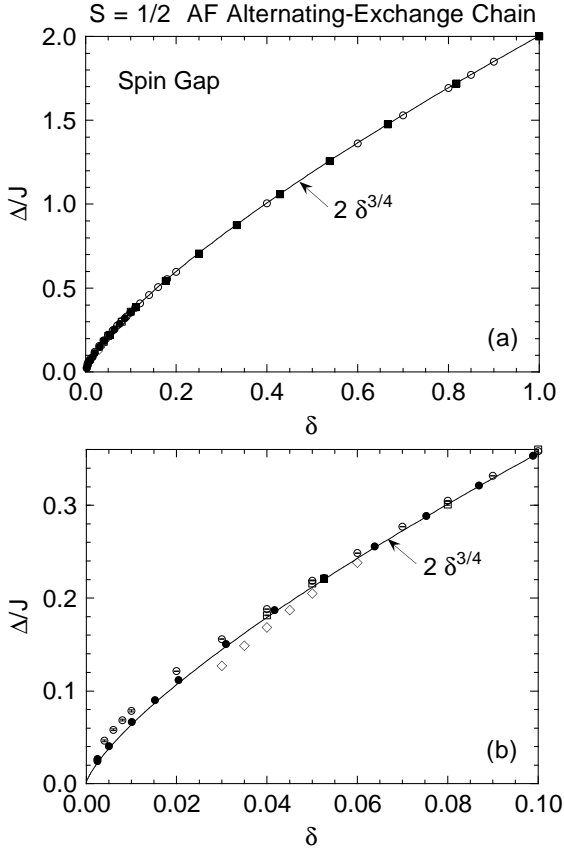}}
\vglue 0.1in
\caption{(a) Spin gap $\Delta/J$ vs alternation parameter $\delta$ for the $S=1/2$
antiferromagnetic alternating-exchange Heisenberg chain.  Our data ($\bullet$)
were determined by fitting our TMRG $\chi^*(t)$ data by
Eq.~(\protect\ref{EqChi(2)}) and are shown along with data of Barnes, Riera, and
Tennant (Ref.~\protect\onlinecite{Barnes1998}) (filled squares), Uhrig {\it et
al.}\ (Ref.~\protect\onlinecite{Uhrig1999}) ($\circ$), Ladavac {\it et
al.}\ (Ref.~\protect\onlinecite{Ladavac1998}) (open squares), and Augier {\it et
al.}\ (Ref.~\protect\onlinecite{Augier1997}) (open diamonds).  The solid curve is the
function (Ref.~\protect\onlinecite{Barnes1998}) $\Delta/J = 2 \delta^{3/4}$. 
(b)~Expanded plot of the data and curve in~(a) for $\delta\leq 0.1$.  Error bars for
the data are not shown in~(a), but are shown in~(b) for all data except for those of
Augier {\it et al.}}
\label{Gap(delta)}
\end{figure}

\begin{equation}
y(\delta) = {\ln[\Delta(\delta)/2J]\over \ln\delta}~.\label{EqD(d):b}
\end{equation}
The numerical prefactor ``2'' in Eq.~(\ref{EqD(d):a}) must be retained in order to
reproduce the exact $\overline{\Delta^*}(\delta=1) = 2$.  Shown in
Fig.~\ref{AltChnLnLnGap}(a) is a semilog plot of
$y$ versus
$\delta$ for the same numerical data as in Fig.~\ref{Gap(delta)}.  This plot [and
Fig.~\ref{AltChnLnLnGap}(b) below] explicitly shows, from BRTs' data, that the
exponent deviates significantly from the value 3/4 even for
$\delta\lesssim 1$.  The plot also clearly differentiates the
various numerical data for small $\delta$ by the different groups, and shows that one
of our two data points from the TMRG for $\delta = 0.0025$ (the one derived from $m =
256$ data at high $t$) is not in agreement with the trend of the remainder of our
data.  This data point will not be included in the plot and fit to be discussed in
the next paragraph.

Our $y(\delta)$ data at small $\delta$ are in agreement with both the magnitude
and trend of BRTs' data at larger $\delta$.  The 
\begin{figure}
\epsfxsize=3.146in
\centerline{\epsfbox{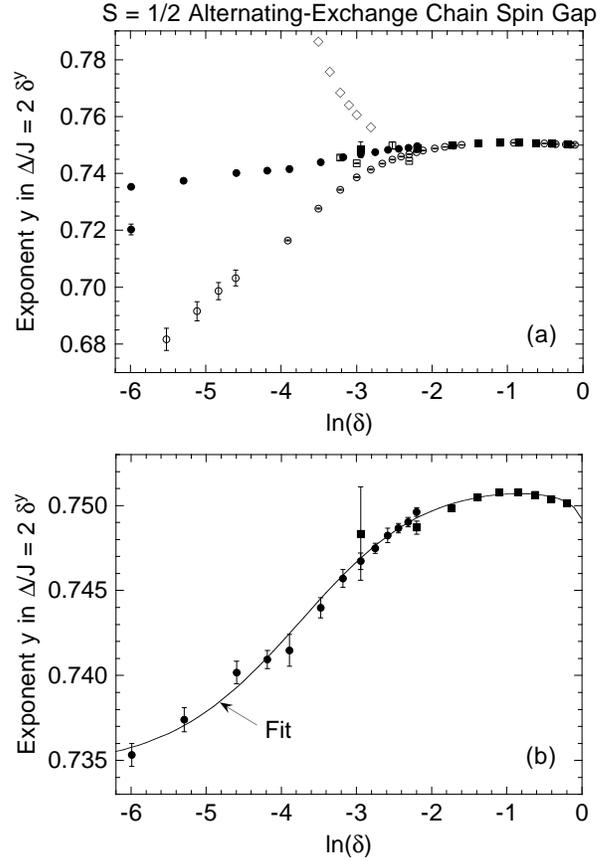}}
\vglue 0.1in
\caption{(a) Semilog plot vs alternation parameter
$\delta$ of the exponent $y = \ln(\Delta/2J)/\ln\delta$ in the
expression $\Delta/J = 2\,\delta^y$ for the spin gap
$\Delta$ of the $S=1/2$ antiferromagnetic alternating-exchange Heisenberg chain.  The
data and symbol references are the same as in Fig.~\protect\ref{Gap(delta)}.  Each
data point has an attached error bar except for those of Augier {\it et
al.}\ (Ref.~\protect\onlinecite{Augier1997}) (open diamonds).  (b)~Expanded view of
our $y(\delta)$ data ($\bullet$) and those derived from the numerical spin gap data of
Barnes, Riera, and Tennant (Ref.~\protect\onlinecite{Barnes1998}) (filled squares),
along with the fit in Eqs.~(\protect\ref{EqD(d):c}) and~(\protect\ref{EqD(d):d}) to
the two combined $y(\delta)$ data sets (solid curve).}
\label{AltChnLnLnGap}
\end{figure}

\noindent $y(\delta)$ for these two data sets
from Fig.~\ref{AltChnLnLnGap}(a), with the exception of one of our two data points
for $\delta = 0.0025$ just noted above, are plotted together on an expanded vertical
scale in Fig.~\ref{AltChnLnLnGap}(b) where a rather smooth behavior of
$y(\delta)$ is seen over the combined range of the two calculations 
$0.0025\leq\delta\lesssim 1$.  With the behavior in Fig.~\ref{AltChnLnLnGap}(b) in
mind, we formulated a five-parameter fit function for these two combined
$y(\delta)$ data sets that yields the correct limits
$\overline{\Delta^*}(\delta\to 0)=0$ and $\overline{\Delta^*}(\delta\to 1)=2$,
with the property $\lim_{\delta\to 0}y(\delta) =$ const, given by
\begin{eqnarray}
y(\delta) = y(1) &+& n_1\tanh\bigg[{\ln\delta\over m_1}\ln\Big({\ln\delta\over
m_2}\Big)\bigg]\nonumber\\
&+& n_2\tanh^2\bigg[{\ln\delta\over m_1}\ln\Big({\ln\delta\over
m_2}\Big)\bigg]~.
\label{EqD(d):c}
\end{eqnarray}
An unweighted fit of this expression to all the data in Fig.~\ref{AltChnLnLnGap}(b)
yielded the parameters
\newpage
\begin{figure}
\epsfxsize=3.1in
\centerline{\epsfbox{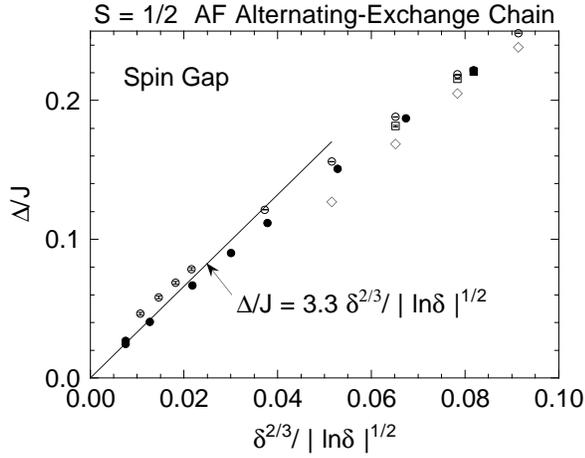}}
\vglue 0.1in
\caption{Spin gap $\Delta/J$ vs $\delta^{2/3}/|\ln\delta|^{1/2}$ obtained from the
data in Fig.~\protect\ref{Gap(delta)}.  The straight line passes through the origin
with slope~3.3.}
\label{Gap(delta^2/3)}
\end{figure}

\[
y(1) = 0.74922~,~~~n_1 = 0.00776~,~~~n_2 = -0.00685~,
\]
\begin{equation}
m_1 = 3.3297~,~~~m_2 = -2.2114~,
\label{EqD(d):d}
\end{equation}
\end{mathletters}
so that
\begin{equation}
\lim_{\delta\to 0}y(\delta) = y(1) - n_1 + n_2 = 0.7346~.
\label{Eqy0}
\end{equation}
The fit is plotted as the solid curve in Fig.~\ref{AltChnLnLnGap}(b).  As can be
seen from the figure, our data are fitted to within our error bars.  In addition,
when the $y(\delta)$ fit function in Eqs.~(\ref{EqD(d):c}) and~(\ref{EqD(d):d}) is
inserted into Eq.~(\ref{EqD(d):a}), the predicted values of
$\overline{\Delta^*}(\delta)$ are in agreement with each of the values of BRT at
larger $\delta$ to within 0.0001, which is sufficient for modeling experimental
data.  The $\delta=0$ limit of $y(\delta)$ in Eq.~(\ref{Eqy0}) is in agreement with
the theoretical {\it effective} value $y(0)=0.72(3)$, which was obtained without the
log correction term by Singh and Weihong\cite{Singh1999} from an
eleventh-order dimer series expansion of the triplet dispersion relation.  We will
use Eqs.~(\ref{EqD(d):all}) to compute $\Delta(T)$ from the experimentally derived
$\delta(T)$ for ${\rm NaV_2O_5}$ in Sec.~\ref{SecNaV2O5Fits} below.  

In order to test the critical behavior prediction
$\overline{\Delta^*}=A\delta^{2/3}/|\ln\delta|^{1/2}$ in Eq.~(\ref{EqCritGap}), which
need only hold in the asymptotic critical regime
$\delta\to 0$ in contrast to the fit function for $0\leq\delta\leq 1$ in
Eqs.~(\ref{EqD(d):all}), in Fig.~\ref{Gap(delta^2/3)} is plotted $\Delta/J$ vs
$\delta^{2/3}/|\ln\delta|^{1/2}$ in the region $\delta\lesssim 0.06$ for the same
data and symbols as in Fig.~\ref{Gap(delta)}.  A proportionality appears to be
developing in our data for $\delta\lesssim 0.005$, as shown by the straight line
with slope $A = 3.3$ passing through the origin of the figure, suggesting that the
asymptotic critical regime begins with decreasing $\delta$ below $\delta\sim 0.005\
(\alpha \gtrsim 0.99)$.  High-accuracy $\overline{\Delta^*}(\delta)$ data for
$\delta\lesssim 0.001$ are needed to test this conjecture.  From
Fig.~\ref{Gap(delta^2/3)}, the slope 3.3 of the line drawn is evidently a lower
limit of the prefactor $A$ within the actual asymptotic critical regime.
\begin{figure}
\epsfxsize=3in
\centerline{\epsfbox{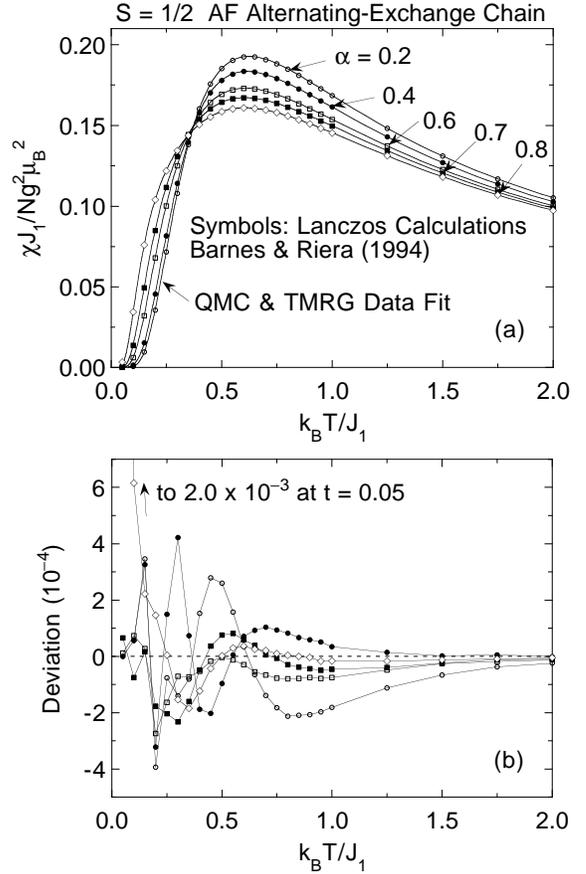}}
\vglue 0.1in
\caption{(a) Magnetic susceptibility $\chi$ versus temperature $T$ for
the spin $S = 1/2$ antiferromagnetic alternating-exchange Heisenberg chain with
alternation parameter $\alpha = 0.2$, 0.4, 0.6, 0.7 and 0.8 (symbols), calculated by
Barnes and Riera (Ref.~\protect\onlinecite{Barnes1994}) using the Lanczos technique. 
Our $\chi^*(t)$ fit function as in Fig.~\protect\ref{QMCDMRGFinalFit} (solid curves)
for the same $\alpha$ values is shown for comparison.  (b)~Deviation of the data of
Barnes and Riera from our fit function vs $T$.}
\label{BarnesQMCCompare}
\end{figure}

\subsection{Numerical $\bbox{\chi^*(\alpha,t)}$ results}
\label{SecBarnesRiera}

Barnes and Riera previously carried out exact diagonalizations of
Hamiltonian~(\ref{EqAltChnHam:all}) for $S = 1/2$ alternating chains of length up
to 16 spins using the Lanczos technique.\cite{Barnes1994}  Their computed
$\chi^*(t)$ values for $\alpha = 0.2,$ 0.4, 0.6, 0.7, and 0.8 were extrapolated to
the bulk limit and the results are shown as the symbols in
Fig.~\ref{BarnesQMCCompare}(a).  Our fit function as in 
Fig.~\ref{QMCDMRGFinalFit} for the same $\alpha$ values is plotted as the solid
curves in Fig.~\ref{BarnesQMCCompare}(a), which are seen to be in good overall
agreement with the calculations of Barnes and Riera.  The deviations of the data of
Barnes and Riera from our fit function are plotted vs temperature in
Fig.~\ref{BarnesQMCCompare}(b).  The average deviation of their data from
our fit function is very small for each data set: $-0.41,\ +0.33,\ -0.40,\ -0.26$, and
$+0.79\times 10^{-4}$ for $\alpha = 0.2$, 0.4, 0.6, 0.7, and 0.8, respectively.  The
absolute rms deviations $\sigma_{\rm rms}$ of their data from our fit function for
$\alpha = 0.2$,  0.4, 0.6, 0.7, and 0.8 are (in units of $10^{-4}$) 1.73, 1.43, 0.73,
0.78, and 3.76, respectively.  We conclude that their data are in good quantitative
agreement with our data and fit, with the exception of their data point for $\alpha
= 0.8$ at their lowest temperature $t = 0.05$.

\subsection{Bulaevskii Theory}
\label{SecBulaevskii}

Bulaevskii\cite{Bulaevskii1969} calculated $\chi^*(t)$ analytically in the
Hartree-Fock approximation.  He first obtained an integral equation for the magnon
spectrum $E(k)$:
\begin{eqnarray}
\varepsilon(k)  \equiv  {E(k)\over J_1} & = & {1\over 2}\bigg[\sqrt{1 +
\alpha^2 - 2\alpha\cos k}\nonumber\\
& + & \frac{C_1 + \alpha C_2 - (\alpha C_1 + C_2)\cos k}{\sqrt{1 + \alpha^2 -
2\alpha\cos k}}\bigg]~,
\label{EqEpsK}
\end{eqnarray}
\begin{eqnarray}
C_1(t) & = & {1\over \pi}\int_0^\pi {\rm d}k\,\frac{1 - \alpha\cos k}{\sqrt{1 +
\alpha^2 - 2\alpha\cos k}}\,\tanh\frac{\varepsilon(k)}{2 t}~,\nonumber\\
\label{EqEpsK2}\\
C_2(t) & = & {1\over \pi}\int_0^\pi {\rm d}k\,\frac{\alpha^2 - \alpha\cos
k}{\sqrt{1 + \alpha^2 - 2\alpha\cos k}}\,\tanh\frac{\varepsilon(k)}{2t}~,\nonumber
\end{eqnarray}
where $k$ is measured in units of $2\pi/d$.  $d\equiv 1$ is the lattice repeat
distance along the chain, which is twice the average distance between spins.  He
then expressed $\chi^*(t)$ in terms of
$\varepsilon(k)$:
\begin{eqnarray}
\chi^*(t) & \equiv & \frac{F(t)}{2 + (1 + \alpha) F(t)}~,\nonumber\\
&&\label{EqChiBul}\\
F(t) & = & {1\over 2\pi t}\int_0^\pi
\frac{{\rm d}k}{\cosh^2[\varepsilon(k)/(2t)]}~.\nonumber
\end{eqnarray}

\begin{figure}
\epsfxsize=2.9in
\centerline{\epsfbox{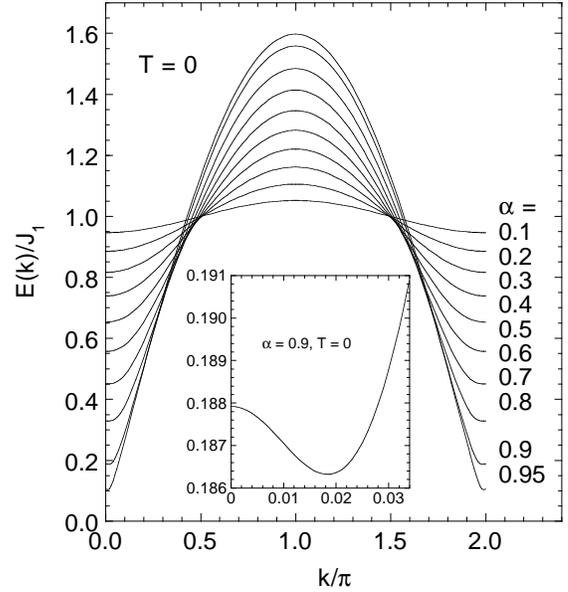}}
\vglue 0.1in
\caption{Dispersion relations $E(k)$ at temperature $T = 0$ from Bulaevskii's
theory (Ref.~\protect\onlinecite{Bulaevskii1969}) in Eqs.~(\protect\ref{EqEpsK})
and~(\protect\ref{EqEpsK3}) for ten values of the alternation parameter $\alpha$. 
The inset shows an expanded plot near $k = 0$ of $E(k)$ for $\alpha = 0.9$.}
\label{BulT=0E(k)}
\end{figure}

\begin{figure}
\epsfxsize=3.1in
\centerline{\epsfbox{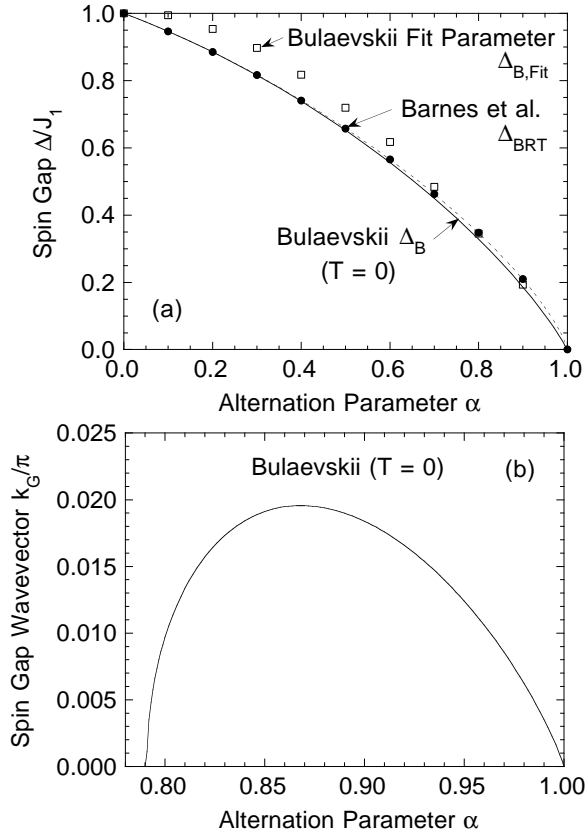}}
\vglue 0.1in
\caption{(a) Energy gap $\Delta$ versus alternation parameter $\alpha$ for the $S
= 1/2$ alternating chain, as calculated by Barnes, Riera, and
Tennant (Ref.~\protect\onlinecite{Barnes1998}) ($\bullet$) and by us using the theory
of  Bulaevskii (Ref.~\protect\onlinecite{Bulaevskii1969}) (solid curve).  The dashed
curve is a plot of $\Delta$ versus $\alpha$ given in
Eq.~(\protect\ref{EqDimParams2:a}).  The values of $\Delta$ obtained by
Bulaevskii (Ref.~\protect\onlinecite{Bulaevskii1969}) by fitting his numerical 
calculations of $\chi(T)$ for $0.033\leq k_{\rm B}T/J_1 \leq 0.25$ according to
Eq.~(\protect\ref{EqBul}) are shown as the open squares.  (b)~Wavevector $k_{\rm
G}$, at which the minimum spin gap occurs in the magnon dispersion relation at $T =
0$, vs alternation parameter $\alpha$.}
\label{BulBarnesT=0Gap}
\end{figure}

\begin{figure}
\epsfxsize=3in
\centerline{\epsfbox{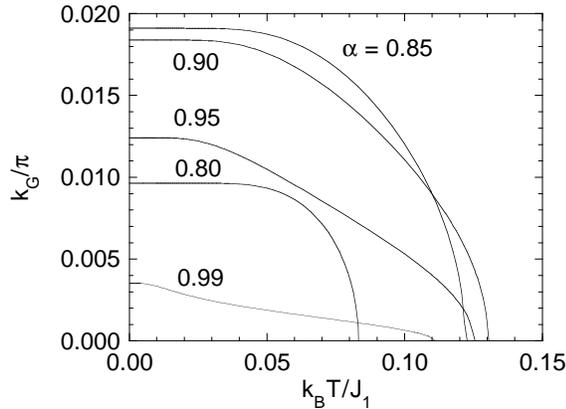}}
\vglue 0.1in
\caption{The temperature $T$ dependence of the wavevector $k_{\rm G}$ at which the
spin-gap occurs in the triplet magnon dispersion relation~(\protect\ref{EqEpsK}),
for five values of the alternation parameter~$\alpha$.}
\label{BulGapK(T)}
\end{figure}

At $t = 0$ and $\alpha\neq 0$, from Eqs.~(\ref{EqEpsK2}) we obtain 
\begin{eqnarray}
C_1(\alpha) &=& {1\over \pi} \bigg\{(1 + \alpha) {\rm E}\bigg[{4\alpha\over (1 +
\alpha)^2}\bigg]\nonumber\\
 &&+\ (1 - \alpha) {\rm K}\bigg[{4\alpha\over (1 +
\alpha)^2}\bigg]\bigg\}~,\nonumber\\
\label{EqEpsK3}\\
C_2(\alpha)&=& {1\over \pi} \bigg\{(1 + \alpha) {\rm E}\bigg[{4\alpha\over (1 +
\alpha)^2}\bigg]\nonumber\\
 &&-\ (1 - \alpha) {\rm K}\bigg[{4\alpha\over (1 +
\alpha)^2}\bigg]\bigg\}~,\nonumber
\end{eqnarray}
where K$(y)$ and E$(y)$ are, respectively, the complete elliptic integrals of the
first and second kinds.  The dispersion relations versus $\alpha$ at $t = 0$ are
obtained by inserting Eqs.~(\ref{EqEpsK3}) into~(\ref{EqEpsK}) and a selection of
results is shown in Fig.~\ref{BulT=0E(k)}.  From Eqs.~(\ref{EqEpsK})
and~(\ref{EqEpsK3}), at $t = 0$ the spin-gap $\Delta_{k=0}(\alpha)$ at $k = 0$ is
given by

\begin{equation}
\Delta_{k=0}(\alpha) = {1 - \alpha\over 2}\Bigg\{1 + \frac{2(1 - \alpha)}{\pi}\,{\rm
K}\bigg[\frac{4 \alpha}{(1 + \alpha)^2}\bigg]\Bigg\}~.
\label{EqBulGap}
\end{equation}
This expression gives the actual spin-gap for $0 < \alpha \leq 0.79$.  However,
for $0.79 \lesssim \alpha < 1$, the minimum in the dispersion relation
does not occur at $k = 0$, as illustrated in an expanded plot of $E(k)$ for
$\alpha = 0.9$ in the inset to Fig.~\ref{BulT=0E(k)}.  The wave vector $k_{\rm G}$
at which the minimum spin gap $\Delta_{\rm B}$ occurs is plotted versus
$\alpha$ in Fig.~\ref{BulBarnesT=0Gap}(b).  The $\Delta_{\rm B}$ from Bulaevskii's
theory at $t = 0$ is plotted versus $\alpha$ as the solid curve
in Fig.~\ref{BulBarnesT=0Gap}(a), and a few representative values are given in
Table~\ref{TabChiParams}.  The predictions of Bulaevskii's theory are in very good
agreement with those of Barnes, Riera, and Tennant\cite{Barnes1998} for
$\alpha\lesssim 0.3$, but the agreement becomes progressively worse as $\alpha$
increases further.  From Eqs.~(\ref{EqEpsK}) and~(\ref{EqEpsK2}), $E(k)$ is
temperature dependent.  In addition, in the range $0.79 \lesssim \alpha < 1$ for
which $k_{\rm G}\neq 0$ at $t = 0$, we find that $k_{\rm G}$ depends on $t$, as
shown in Fig.~\ref{BulGapK(T)}.  From Fig.~\ref{BulGapK(T)}, $k_{\rm G} \to 0$ at
$t \approx 0.083$, 0.122, 0.131, 0.125, and 0.111 for $\alpha = 0.8$, 0.85, 0.9,
0.95, and 0.99, respectively.

\begin{figure}
\epsfxsize=3in
\centerline{\epsfbox{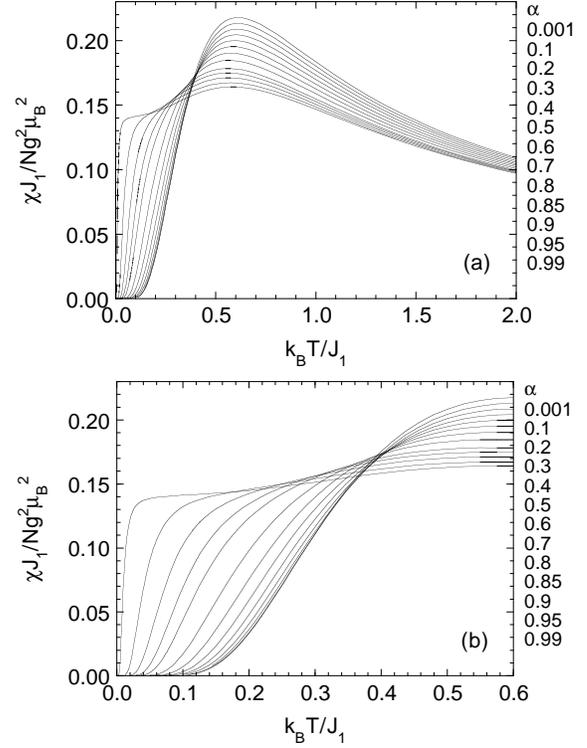}}
\vglue 0.1in
\caption{Magnetic susceptibility $\chi$ versus temperature $T$ for
$0.001\leq\alpha\leq 0.99$ as predicted by the theory of
Bulaevskii (Ref.~\protect\onlinecite{Bulaevskii1969}) in
Eqs.~(\protect\ref{EqEpsK})--(\protect\ref{EqChiBul}).}
\label{BulChi(t)2}
\end{figure}

We computed $\chi^*(t)$ by inserting $\varepsilon(k)$ in Eq.~(\ref{EqEpsK}) into
Eqs.~(\ref{EqEpsK2}), numerically solving the latter two 
simultaneous equations for $C_1$ and $C_2$ at each $t$, and then inserting the
resulting $\varepsilon(k)$ into Eqs.~(\ref{EqChiBul}).  The progression of
$\chi^*(t)$ with increasing $\alpha$ from 0.001 to 0.99 is shown in
Fig.~\ref{BulChi(t)2}.  As noted by Bulaevskii,\cite{Bulaevskii1969} the values of
$\chi^*$ at the maxima are too large and the temperatures at which these occur are
too small by $\sim 5$--10\% (compare Fig.~\ref{BulChi(t)2} with
Fig.~\ref{QMCDMRGFinalFit}).

At low temperatures $0.033 \leq t \leq 1/4$, Bulaevskii fitted
$\chi^*(\alpha,t)$, calculated from Eqs.~(\ref{EqChiBul}), by the two-parameter form

\begin{equation}
\chi^*(\alpha,t) = {a(\alpha)\over t}\,{\rm e}^{-\Delta(\alpha)/(J_1 t)}~,
\label{EqBul}
\end{equation}
and obtained values of $a_{\rm B}$ and $\Delta_{\rm B,Fit}/J_1$ for $0 \leq \alpha
\leq 0.9$ which are reproduced in Table~\ref{TabChiParams}; $\Delta_{\rm
B,Fit}(\alpha)/J_1$ is plotted as the open squares
in Fig.~\ref{BulBarnesT=0Gap}(a).  Note that the temperature exponent in the
prefactor to the exponential is $\gamma = 1$, contrary to the $\gamma = 1/2$ in 
Eq.~(\ref{EqTroyer:b}) which is expected in the low-$t$ limit for any 1D $S = 1/2$
Heisenberg spin system with a spin gap (and with a nondegenerate one-magnon band
with a parabolic minimum).  We have confirmed that over the temperature range fitted
by Bulaevskii, one indeed obtains
$\gamma\sim 1$ for the best fit of Eq.~(\ref{EqBul}) to numerical
calculations of $\chi^*(\alpha,t)$.  We infer that the discrepancy between
\mbox{Bulaevskii's} $\gamma = 1$ and the expected $\gamma = 1/2$ arises because the
fits were not carried out completely within the low-$t$ limit.  This issue is
discussed in more detail below.

Equation~(\ref{EqBul}) together with Bulaevskii's table of
\{$a_{\rm B}(\alpha),\Delta_{\rm B,Fit}(\alpha)/J_1$\} values were subsequently used
extensively in the analysis of experimental $\chi(T)$ data for compounds exhibiting
spin-Peierls transitions to determine the alternation parameter $\alpha$ at low
temperatures $T \ll T_{\rm c}$ where the experimental spin-gap is nearly independent
of $T$.  However, from Table~\ref{TabChiParams} and Fig.~\ref{BulBarnesT=0Gap}(a),
the $\Delta_{\rm B,Fit}(\alpha)$ values of Bulaevskii\cite{Bulaevskii1969} are in
generally poor agreement with the actual spin gaps $\Delta_{\rm B}(\alpha)$ of his
theory and with those [$\Delta_{\rm BRT}(\alpha)$] calculated for the same $\alpha$
values by Barnes, Riera, and Tennant.\cite{Barnes1998}  Therefore, one should consider
the $\Delta_{\rm B,Fit}$ parameters as fitting parameters only, with no direct
relation to the actual spin gap.

\begin{figure}
\epsfxsize=3.5in
\centerline{\epsfbox{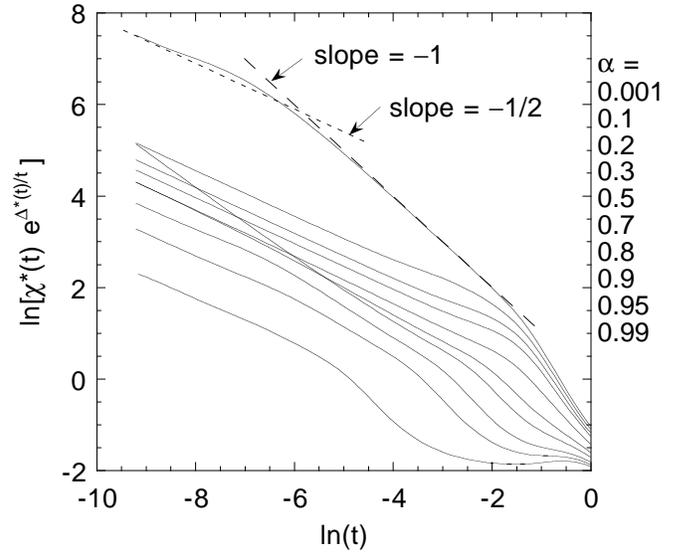}}
\vglue 0.1in
\caption{Log-log plot of $\chi^*(t)\,{\rm e}^{\Delta^*(t)/t}$ at low $t$
versus reduced temperature $t$ for $0.001\leq\alpha\leq 0.99$ [see
Eq.~(\protect\ref{EqLogChi})] as predicted by the theory of
Bulaevskii (Ref.~\protect\onlinecite{Bulaevskii1969}) in
Eqs.~(\protect\ref{EqEpsK})--(\protect\ref{EqChiBul}).}
\label{BulLog[ChiExp]2}
\end{figure}

According to Eq.~(\ref{EqChiDimer:a}) for $\chi^*(t)$ of the isolated dimer which
is a zero-dimensional spin system, the form of $\chi^*(t)$ in Eq.~(\ref{EqBul})
with $\gamma = 1$ is correct for $\alpha = 0$ and $t \to 0$.  On the other hand, for
one-dimensional spin systems such as the two-leg spin ladder (and the
alternating-exchange chain) at temperatures $k_{\rm B}T \ll \Delta$ and $k_{\rm B}T
\ll$ one-magnon bandwidth, Eqs.~(\ref{EqTroyer}) apply, with $\gamma = 1/2$,
assuming that the triplet one-magnon dispersion relation $E(k)$ is parabolic at the
minimum.  In this case one expects  $\gamma = 1/2$ at sufficiently low $t$ for any
finite $\alpha$.   Thus, in the temperature region of validity of
Eq.~(\ref{EqTroyer:a}), a plot of the left-hand-side of Eq.~(\ref{EqLogChi}) vs
ln$t$ should give a straight line with slope $-\gamma$.  Shown in
Fig.~\ref{BulLog[ChiExp]2} are such plots, obtained using our $\chi^*(\alpha,t)$
calculated from Bulaevskii's theory as described above, for $\alpha = 0.001$
to~0.99.  For $\alpha = 0.001$, a crossover is clearly evident from $\gamma = 1$ to
$\gamma = 1/2$ with decreasing $t$.  The other curves also exhibit signs of a
crossover, with $\gamma \approx 1/2$ at the lowest temperatures, with the exception
of the curve for $\alpha = 0.8$.  For this $\alpha$ value, which is just above the
value $\alpha\approx 0.79$ at which $k_{\rm G}$ becomes nonzero at
$t = 0$ [see Fig.~\ref{BulBarnesT=0Gap}(b)], the $\gamma$ at the lowest $t$ is
intermediate between the values of 1/2 and 1, and the assumption of a parabolic form
for $E(k)$ at the band minimum is evidently not satisfied (see
Fig.~\ref{BulT=0E(k)}). In fact, Troyer, Tsunetsugu and W\"urtz\cite{Troyer1994}
calculated the low-$t$ limit of $\chi^*(t)$ for 1D systems with general dispersion
relation $\varepsilon(k) = \Delta^* + c^*|ka|^n$, where $k$ is the deviation of the
wave vector from that at the band minimum.  They found the same form $\chi^*(t) =
(A_n/t^\gamma)\exp(-\Delta^*/t)$ as for the parabolic case $n = 2$, but where
$\gamma = 1 - (1/n)$.  Thus, e.g., $\gamma = 2/3$ and 3/4 for $n = 3$ and~4,
respectively.  This range of $\gamma$ is consistent with the slope of the data at
the lowest temperatures for $\alpha = 0.8$ in Fig.~\ref{BulLog[ChiExp]2}.

The predictions of Bulaevskii's theory for $\chi^*(t)$ from Fig.~\ref{BulChi(t)2}
are compared with our $\chi^*(\alpha,t)$ fit function (solid
\widetext
\begin{table}
\caption{Prefactor $a$ and spin-gap $\Delta$ describing the low-temperature
spin susceptibility in Eq.~(\protect\ref{EqBul}) of the $S = 1/2$ alternating
chain with alternation parameter $\alpha$ [and $\delta = (1-\alpha)/(1+\alpha)$, 
see Eqs.~(\protect\ref{EqAltChnHam:all})].  Fit parameters given by
Bulaevskii (Ref.~\protect\onlinecite{Bulaevskii1969}) ($a_{\rm B},\ \Delta_{\rm
B,Fit}$) that he obtained by fitting to his low-$t$ $\chi^*(t)$ calculations using
Eq.~(\protect\ref{EqBul}) are shown.  We obtained the actual spin gap values
$\Delta_{\rm B}/J_1$ in Bulaevskii's theory by numerically solving
Eqs.~(\protect\ref{EqEpsK}) and~(\protect\ref{EqEpsK2}).  Also included are the
accurate calculations of the spin gap $\Delta_{\rm BRT}/J_1$ by Barnes, Riera, and
Tennant (Ref.~\protect\onlinecite{Barnes1998}), which are compared with numerical
values of their approximate form $\Delta(\alpha)/J_1\approx (1 - \alpha)^{3/4}(1 +
\alpha)^{1/4}$  [Eq.~(\protect\ref{EqDimParams2:a})].}
\begin{tabular}{dlcccdc}
$\alpha$ & $\delta$& $a_{\rm B}$ & $\Delta_{\rm B,Fit}/J_1$ & $\Delta_{\rm B}/J_1$ &
$\Delta_{\rm BRT}/J_1$ & $(1 - \alpha)^{3/4}(1 + \alpha)^{1/4}$\\
\hline
0.0 & 1 & 1 & 1 & 1 & 1 & 1 \\
0.1 & 0.81818 & 0.980 & 0.995 & 0.946245 & 0.946279339 & 0.94630 \\
0.2 & 0.66667 & 0.873 & 0.954 & 0.884911 & 0.885209996 & 0.88535 \\
0.3 & 0.53846 & 0.733 & 0.897 & 0.815791 & 0.816844275(1)  & 0.81716 \\
0.4 & 0.42857 & 0.582 & 0.818 & 0.738504 & 0.74106141(3) & 0.74156 \\ 
0.5 & 0.33333 & 0.427 & 0.720 & 0.652443 & 0.6574777(5)  & 0.65804 \\
0.6 & 0.25 & 0.346 & 0.617 & 0.556661 & 0.565296(7) & 0.56569 \\
0.7 & 0.17647 & 0.224 & 0.484 & 0.449626 & 0.46298(5) & 0.46286 \\
0.8 & 0.11111 & 0.138 & 0.345 & 0.328631 & 0.3474(3) & 0.34641 \\
0.9 & 0.05263 & 0.076 & 0.193 & 0.186319 & 0.2098(17)  & 0.20878 \\
1.0 & 0 &     &       & 0 & 0 & 0 \\
\end{tabular}
\label{TabChiParams}
\end{table}

\begin{figure}
\epsfxsize=3.04in
\centerline{\epsfbox{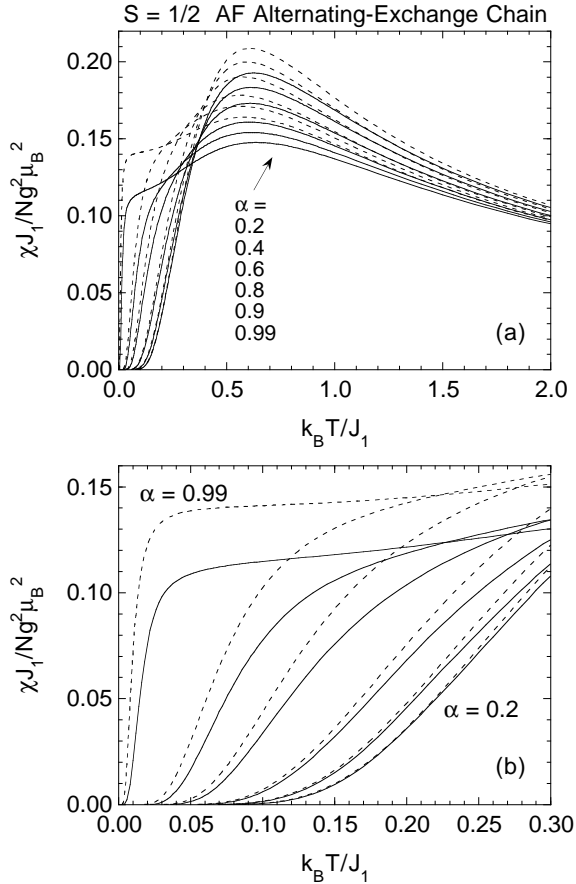}}
\vglue 0.1in
\caption{(a) Magnetic susceptibility $\chi$ versus temperature $T$ for
alternating chains with $\alpha = 0.2$, 0.4, 0.6, 0.8, 0.9, and 0.99 (solid curves)
generated using our $\chi^*(\alpha,t)$ fit function as in
Fig.~\protect\ref{QMCDMRGFinalFit}.  These are compared with the predictions of the
theory of Bulaevskii (Ref.~\protect\onlinecite{Bulaevskii1969}) (corresponding dashed
curves).  (b)~Expanded plots at low $T$ from~(a).}
\label{BulQMCComp}
\end{figure}

\narrowtext
\vglue0.2in
\noindent curves) for
$\alpha = 0.2$, 0.4, 0.6, 0.8 and 0.99 (as in Fig.~\ref{QMCDMRGFinalFit}) in
Fig.~\ref{BulQMCComp}, where the Bulaevskii prediction for each of these
$\alpha$ values is shown as the corresponding dashed curve.  The disagreement
between the two calculations becomes progressively more severe as temperature
decreases and as the uniform chain limit is approached with increasing
$\alpha$.  Therefore, the accuracies of the $\alpha$ and
$J_1$ values previously extracted from experimental data at low $T$ for compounds
with $\alpha \lesssim 1$ using Bulaevskii's theory are unclear.  Our
$\chi^*(\alpha,t)$ fit function now provides a much more accurate and reliable means
of extracting exchange constants and spin gaps from experimental
$\chi(T)$ data.
\newpage

\section{Magnetic Susceptibility of N\lowercase{a}V$_2$O$_5$}
\label{SecNaV2O5}

Crystals of $\rm Na_{0.996(3)}V_2O_{5.00(6)}$ were grown at the
Max-Planck-Institut f\"ur Festk\"orperforschung, Stuttgart, in a Pt crucible in
flowing Ar atmosphere by a self-flux method from a 5:1:1 mixture of $\rm NaVO_3$,
$\rm V_2O_3$ and $\rm V_2O_5$.\cite{Isobe1997}  The flux was dissolved by boiling the
solidified melt in distilled water. X-ray powder diffraction patterns collected with
a STOE diffractometer yielded the lattice  parameters $a = 11.3187(8)$\,\AA, $b =
3.6111(3)$\,\AA, and $c = 4.8007(5)$\,\AA.  Chemical analyses on two independent
representative samples of the batch were performed with a standard AAS analysis
technique for V and Pt and ICP emission spectroscopy for the Na content. The oxygen
content was determined by measuring with IR spectroscopy the amount of CO generated
when the sample is fused in a graphite crucible at 2700\,$\rm ^oC$ {\it in vacuo}. 
Platinum impurities above the level of sensitivity of the analysis (500\,ppm with
respect to V) could not be detected.  
\begin{figure}
\epsfxsize=3in
\centerline{\epsfbox{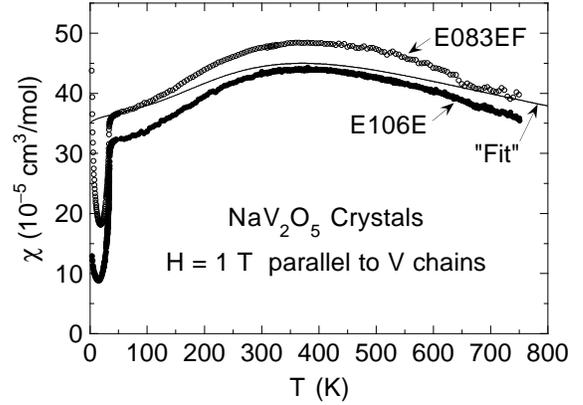}}
\vglue 0.1in
\caption{Magnetic susceptibility $\chi$ in a field $H = 1$\,T parallel to the V
chains ($b$ axis) versus temperature $T$ for two crystals of NaV$_2$O$_5$ as
indicated.  The solid curve is a ``Fit'' of the data by the theoretical prediction
for the $S=1/2$ uniform Heisenberg chain with parameters in
Eq.~(\protect\ref{EqFit1Pars}).}
\label{NaV2O5XtalsChi2}
\end{figure}
\vglue0.11in

At Ames Laboratory, single crystals of ${\rm NaV_2O_5}$ were grown out of the ternary
melt.\cite{Isobe1997} Powders of ${\rm V_2O_5}$ and V$_2$O$_3$ were prepared by
oxidizing and reducing NH$_4$VO$_3$ at 600$\,^\circ$C and 900\,$^\circ$C,
respectively.  The resulting ${\rm V_2O_5}$ is reacted with Na$_2$CO$_3$ at
550\,$^\circ$C yeilding NaVO$_3$.  About 10 grams of NaVO$_3$, ${\rm V_2O_5}$, and
V$_2$O$_3$ in the molar ratio 32:1:1 were placed in a Pt crucible and sealed in an
evacuated quartz tube.  The melt was then slowly cooled from 800 to
660\,$^\circ$C over 50\,h and the remaining liquid was decanted.  Small amounts of
solidified melt remaining on the crystals were dissolved with hot water.  Typical
dimensions of the ribbon-shaped crystals grown in this manner are $0.5 \times 1.5
\times 11$\,mm$^3$ with the $c$ axis perpendicular to the plane of the ribbon,
the $b$ axis along the length of the ribbon and the $a$ axis along the
width of the ribbon, with lattice parameters $a \approx 11.303$\,\AA, $b
\approx 3.611$\,\AA, and $c \approx 4.752$\,\AA.  The crystal denoted as AL1 has a
mass of 8.2\,mg and approximate planar dimensions $1.5 \times 2.5$\,mm$^2$.    

The magnetic susceptibility $\chi(T) \equiv M(T)/H$ of the crystals was measured
using Quantum Design SQUID magnetometers at Stuttgart and Ames.  The measurements on
eight crystals of NaV$_2$O$_5$ in Stuttgart were carried out in a field $H = 1$\,T
along the V ladder ($b$) axis direction in various temperature ranges between 2 and
750\,K\@.  Measurements of the anisotropy of $\chi(T)$ along the $a$, $b$, and
$c$ axis directions were carried out from 2 to 300\,K in Ames on crystal AL1 in $H =
2$\,T\@.

The results for two of the crystals up to 750\,K are shown in
Fig.~\ref{NaV2O5XtalsChi2}.  The data illustrate the variabilities we have observed
between measurements along the same axis on different crystals.  Above $\sim 50$\,K,
the two data sets are nearly parallel, with the difference between them being
$\approx 3-4\times 10^{-5}$\,cm$^3$/mol; we have no explanation for this difference,
and no comments have been made in the literature about such variabilities and/or
their origins in $\chi(T)$ along the same axis in different crystals that we
are aware of.  The data from $T_{\rm c} \approx 33$--34\,K up to 300\,K are in
approximate agreement with the single crystal data of Isobe, Kagumi and Ueda taken
in this $T$ range along the same axis in $H = 5$\,T\@.\cite{Isobe1997}  A variable
Curie-Weiss-like contribution $\chi^{\rm CW}(T)$ to $\chi(T)$ occurs below $\sim
20$\,K which is attributed to paramagnetic defects, impurities, inclusions and/or
intergrowths in the crystals.  The ``Fit'' shown in the figure will be discussed
later in Sec.~\ref{SecNaV2O5Modeling}.

The experimental data are analyzed with the general expression
\begin{mathletters}
\label{EqExpChi:all}
\begin{equation}
\chi(T) = \chi_0 + \chi^{\rm CW}(T) + \chi^{\rm spin}(T)~,
\label{EqExpChi:a}
\end{equation}
\begin{equation}
\chi_0 = \chi^{\rm core} + \chi^{\rm VV}~,
\label{EqExpChi:b}
\end{equation}
\begin{equation}
\chi^{\rm CW}(T) = \frac{C_{\rm imp}}{T - \theta}~,\label{EqExpChi:c}
\end{equation}
\begin{equation}
\chi^{\rm spin}(T) = {Ng^2\mu_{\rm B}^2\over J}\,\overline{\chi^*}\Big({k_{\rm
B}T\over J}\Big)~,
\label{EqExpChi:d}
\end{equation}
where $\chi_0$ is the sum of a temperature independent and (nearly) isotropic
orbital diamagnetic core contribution and a usually anisotropic and temperature
independent orbital paramagnetic Van Vleck contribution.  We
estimate $\chi^{\rm core}$ using the values $-$5, $-$7, $-$4, and $-12 \times
10^{-6}$\,cm$^3$/mol for Na$^{+1}$, V$^{+4}$, V$^{+5}$, and O$^{-2}$,
respectively,\cite{Selwood1956} yielding the isotropic value
\begin{equation}
\chi^{\rm core} = -7.8 \times 10^{-5}\,{\rm\frac{cm^3}{mol\,NaV_2O_5}}~.
\label{EqExpChi:e}
\end{equation}
\end{mathletters}
The second term in Eq.~(\ref{EqExpChi:a}) is the above-noted Curie-Weiss impurity
and/or defect contribution and the last term is the intrinsic spin susceptibility,
each of which may or may not be anisotropic.  For a Heisenberg spin system,
$\overline{\chi^*}$ is isotropic, and therefore so is $\chi^{\rm spin}$ apart from
anisotropy in the $g$ factor.  The impurity Curie-Weiss term $\chi^{\rm CW}(T)$ can
be anisotropic if the impurities are defects or intergrowths in the crystals which
have atomic coordination principal axes which are fixed with respect to the
crystal axes rather than being randomly oriented.   We model our
$\chi(T)$ data according to Eq.~(\ref{EqExpChi:a}) in terms of the
$\overline{\chi^*}(t)$ in Eq.~(\ref{EqExpChi:d}), which are (fit functions to)
theoretical susceptibility calculations presented in previous sections.  Before
moving on to do that, we first experimentally examine the anisotropy in $\chi(T)$ of
${\rm NaV_2O_5}$ and its implications in the next section.

\subsection{Anisotropy of the magnetic susceptibility}
\label{SecMagAnis}

\begin{figure}
\epsfxsize=3.46in
\centerline{\epsfbox{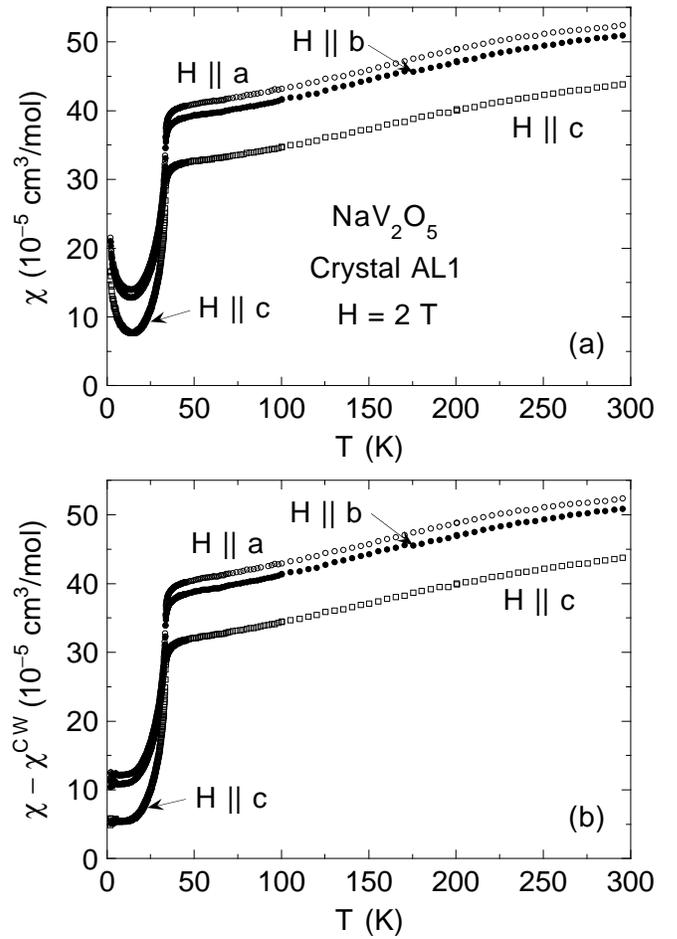}}
\vglue 0.1in
\caption{(a) Magnetic susceptibility $\chi$ versus temperature $T$ in a field $H =
2$\,T parallel ($\bbox{H}||b$) and perpendicular ($\bbox{H}||a$, $\bbox{H}||c$) to
the V chains in NaV$_2$O$_5$ crystal AL1.  (b) The data in (a) corrected for the
respective Curie-Weiss contributions $\chi^{\rm CW} = C_{\rm imp}/(T-\theta)$
attributed to paramagnetic defects or impurities.}
\label{NaV2O5Chi_abc}
\end{figure}

The magnetic susceptibilities of ${\rm NaV_2O_5}$ crystal AL1 along the $a$, $b$,
and $c$ axes are plotted vs temperature in Fig.~\ref{NaV2O5Chi_abc}(a), where the
$a$ and $c$ axes are perpendicular to the V chains which run along the $b$ axis,
and the $c$ axis is perpendicular to the trellis layers that the V chain/ladders
reside in.  The data are similar to the aniso-

\noindent tropic $\chi(T)$ data reported by Isobe,
Kagami and Ueda,\cite{Isobe1997} although the anisotropies we measure at both room
temperature and at low temperatures are somewhat larger than they reported.  The
anisotropies at low temperatures are seen more clearly if the respective impurity
term $\chi^{\rm CW}(T)$ in Eq.~(\ref{EqExpChi:c}) is subtracted from each data set,
as shown in Fig.~\ref{NaV2O5Chi_abc}(b).  The impurity Curie constant
$C_{\rm imp}$ and Weiss temperature $\theta$ for each direction of the applied field
were determined by the requirement that $\chi(T)$ become independent of $T$ for
$T\to 0$.  The fitted  values of $C_{\rm imp}$ were found to be slightly anisotropic
and are given in Table~\ref{TableLowTFitParams} below.  The values of $C_{\rm imp}$
are equivalent to the contribution of only 0.07\,mol\% of $S = 1/2$ impurities with
$g = 2$; if the impurity spin is actually greater than 1/2, the concentration of
paramagnetic impurities could be much less than this estimate.  From a comparison of
Figs.~\ref{NaV2O5Chi_abc}(a) and~\ref{NaV2O5Chi_abc}(b), $\chi^{\rm CW}(T)$ is seen
to make a negligible contribution to the measured $\chi(T)$ above $\sim 100$\,K\@. 
Since in the presence of a spin gap $\chi^{\rm spin} = 0$ at the lowest
temperatures, from Eqs.~(\ref{EqExpChi:all}) \ and \ Fig.~\ref{NaV2O5Chi_abc}(b) \
we obtain
\newpage
\begin{eqnarray}
\chi_b^{\rm VV} &=& 18.7\times 10^{-5}\ {\rm cm^3\over mol}~,~~~
\chi_c^{\rm VV} = 13.3\times 10^{-5}\ {\rm cm^3\over mol}~,\nonumber\\
\nonumber\\
\chi_a^{\rm VV} &=& 20.0\times 10^{-5}\ {\rm cm^3\over mol}~~~~(T\ll T_{\rm c})~.
\label{EqOurChiVV}
\end{eqnarray}

From Fig.~\ref{NaV2O5Chi_abc}(b), the anisotropies of $\chi(T)$ are seen to be quite
temperature-dependent upon heating through $T_{\rm c}=33.4$\,K\@.  These
results are  surprising, because $\chi^{\rm spin}$ is expected to be isotropic
(apart from the small anisotropy due to the anisotropic $g$ factor), with $\chi^{\rm
spin}(T\to 0) = 0$ because of the spin gap, and the anisotropic $\chi^{\rm VV}$
values are expected to be temperature independent for our $S = 1/2$ system over the
temperature range of our measurements.  Thus one expects the difference
$\chi_\alpha(T) - \chi_\beta(T)\ (\alpha,\ \beta = a,\ b,\ c)$ to be nearly
independent of temperature compared with the magnitude of either, where a subscript
refers to the crystallographic axis along which the magnetic field is applied.

\begin{figure}
\epsfxsize=3.09in
\centerline{\epsfbox{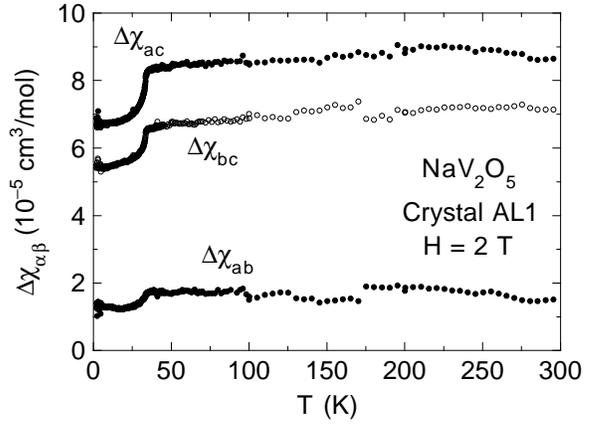}}
\vglue 0.1in
\caption{Temperature $T$ dependences of the intrinsic anisotropy differences 
$\Delta\chi_{\alpha\beta}\ (\alpha\beta = ac,\ bc,\ ab)$ in the magnetic
susceptibilities along the $a$, $b$, and $c$ axes in ${\rm NaV_2O_5}$, as defined
in Eq.~(\protect\ref{EqChiAnis}).  These data were obtained from the respective
differences between the three pairs of data sets in
Fig.~\protect\ref{NaV2O5Chi_abc}(b).}
\label{NaV2O5DeltaChi_abc}
\end{figure}

To be more quantitative, we define the anisotropy in the intrinsic susceptibility as
\begin{equation}
\Delta\chi_{\alpha\beta}(T) \equiv [\chi_\alpha - \chi_\alpha^{\rm CW}](T) -
[\chi_\beta - \chi_\beta^{\rm CW}](T)~,
\label{EqChiAnis}
\end{equation}
which eliminates extrinsic anisotropy in the Curie-Weiss impurity contribution
from the values calculated from the experimental data.   For example, according to
this definition, $\Delta\chi_{ac}(T)$ is the difference between the uppermost and
lowermost data sets in Fig.~\protect\ref{NaV2O5Chi_abc}(b).  The three
$\Delta\chi_{\alpha\beta}(T)$ anisotropies are plotted
in Fig.~\ref{NaV2O5DeltaChi_abc}.  It seems to
us that the only reasonable explanation for the strong temperature-dependent
anisotropies in Fig.~\ref{NaV2O5DeltaChi_abc} for two of the three data sets is that
one or more of the $\chi_\alpha^{\rm VV}$ susceptibilities is strongly temperature
dependent near $T_{\rm c}$, contrary to our initial expectations. 
Such a temperature dependence may be associated with the crystallographic and
charge-ordering transitions which occur at or near the same temperature as the spin
dimerization transition, as discussed in the Introduction.

One can make rather strong general statements about the magnetic
susceptibility anisotropies and their temperature dependences as follows.  Defining
the Van Vleck susceptibility anisotropy $\Delta\chi_{\alpha\beta}^{\rm VV} =
\chi_\alpha^{\rm VV} - \chi_\beta^{\rm VV}$ and similarly the spin
susceptibility anisotropy $\Delta\chi_{\alpha\beta}^{\rm spin} =
\chi_\alpha^{\rm spin} - \chi_\beta^{\rm spin}$, from Eqs.~(\ref{EqExpChi:all}) one
obtains an expression for the anisotropy $\Delta\chi_{\alpha\beta}(T)$ for a spin
system in which the only anisotropy in $\chi^{\rm spin}$ arises from anisotropy in
the $g$ factor, given by
\begin{equation}
\Delta\chi_{\alpha\beta}(T) = \Delta\chi_{\alpha\beta}^{\rm VV} + {N\mu_{\rm
B}^2(g_\alpha^2 - g_\beta^2)\over J}\,\overline{\chi^*}\Big({k_{\rm B}T\over
J}\Big)~.
\label{EqDeltaChiCalc}
\end{equation}
The reduced spin susceptibility $\overline{\chi^*}(\overline{t})$ is necessarily
positive, and it is isotropic for a Heisenberg spin system as noted above.  Thus, if
$\chi_\alpha^{\rm VV}$ and $\chi_\beta^{\rm VV}$ and therefore
$\Delta\chi_{\alpha\beta}^{\rm VV}$ are independent of temperature, the slope
$\partial\Delta\chi_{\alpha\beta}(T)/\partial T$ must \  have \ the \ same \ sign \ as
\ the difference $g_\alpha^2 - g_\beta^2$.  As \ discussed \ in \ the \ next
\ subsection, \ for
\ ${\rm NaV_2O_5}$,  this \ difference \ has \ been \ reported \ to \ be \ positive
\ for 

\noindent
$\alpha\beta = ac$
and $bc$ and near zero for $\alpha\beta = ab$, 
consistent with the slopes in Fig.~\ref{NaV2O5DeltaChi_abc}.  However, in a simple
ionic crystalline electric field model and with a positive spin-orbit coupling
parameter for V one would predict that a $\chi_\alpha^{\rm VV}$ should increase with
the negative deviation of $g_\alpha$ from the free electron value $g = 2$.  Thus, a
particularly visible and puzzling discrepancy is that since $(2-g_a) \approx (2-g_b)
< (2-g_c)$ according to the reported $g_\alpha$ values below, on this basis one
strongly expects $\chi_a^{\rm VV} \approx \chi_b^{\rm VV} < \chi_c^{\rm VV}$; thus
two of the three $\chi_\alpha^{\rm VV}$ values should be about the same and {\it
smaller} than the third one.  Qualitatively contrary to this expectation, for $T
\ll T_{\rm c}$ we observe in Eq.~(\ref{EqOurChiVV}) that $\chi_a^{\rm VV} \approx
\chi_b^{\rm VV} > \chi_c^{\rm VV}$.  

We will not emphasize or further discuss these puzzling discrepancies with
expectation with respect to their possible influence on our theoretical modeling of
our $\chi(T)$ data in Secs.~\ref{SecNaV2O5Modeling} and~\ref{SecNaV2O5Fits}, since
at present there is no way to model, e.g., a temperature dependent Van Vleck
susceptibility which changes rapidly near $T_{\rm c}$, but the anisotropic
susceptibility results and the above discussion should be kept in mind.  In the
following two subsections the reported anisotropies in the
$g$ factor as measured using electron spin resonance (ESR) and in the Van Vleck
susceptibility as deduced from nuclear magnetic resonance (NMR) measurements will be
discussed, respectively, in light of our anisotropic $\chi(T)$ data.

\paragraph*{Anisotropy in the $g$ factor from ESR\@.}

Many ESR measurements have
recently been reported for 
NaV$_2$O$_5$.\cite{Onoda1999,Vasilev1997,Lohmann1997,Schmidt1998,%
Hemberger1998,Palme1998}  Each study found a signal with $g \approx 2$ which was
attributed to bulk V species, and the $g$ values found in the various studies were
the same within the errors, e.g.,\cite{Vasilev1997}
\begin{equation}
g_a = g_b = 1.972(2),~~~g_c = 1.938(2)~.
\label{EqGs}
\end{equation}
The powder-average value is $g = \sqrt{(g_a^2+g_b^2+g_c^2)/3} = 1.961(2)$.  The
$g$ values were found to be independent of $T$ down to 20\,K, which is below
$T_{\rm c}$.  From all these measurements, there is no indication that the $S = 1/2$
Heisenberg Hamiltonian is not appropriate to the spin system in NaV$_2$O$_5$. 
Unfortunately, given the sensitivity of the ESR technique, we cannot be certain that
these ESR results are representative of the bulk spin species in NaV$_2$O$_5$,
because no quantitative measurements of the concentration of spin species observed
in these measurements were reported.  Although the (uncalibrated) ESR intensity
versus temperature measurements approximately mirror the bulk susceptibility
behavior in most (but not all) of these studies, it is still possible that the
signal arises from a minority spin species that is coupled to the bulk spin system. 
An interesting related issue which has not been discussed in the literature is why
the presumed bulk $S = 1/2$ species in NaV$_2$O$_5$ are observable to low
temperatures $T \lesssim 0.03 J/k_{\rm B}$ by ESR, where the AF exchange constant is
$J/k_{\rm B} \sim 580$\,K (see below), whereas the bulk Cu$^{+2}$ spins 1/2 in the
high-$T_{\rm c}$ cuprates are not observable by ESR up to 1100\,K, which is $\approx
0.7 J/k_{\rm B}$ where $J/k_{\rm B}\approx 1600$\,K is only a factor of 2.8
larger.\cite{Johnston1997}

In Ref.~\onlinecite{Ohama1997b} the authors estimated the $\chi^{\rm VV}$ values
using the reported anisotropic $g$ values obtained from ESR measurements, obtaining 
$\chi^{\rm VV}_a = \chi^{\rm VV}_b = 2.4 \times 10^{-5}$\,cm$^3$/mol and  
$\chi^{\rm VV}_c = 6.6 \times 10^{-5}$\,cm$^3$/mol, which were stated to be in
agreement with the values from their $K$-$\chi$ analysis discussed in the
following subsection.  These values do not agree with our
$T=0$ values in Eq.~(\ref{EqOurChiVV}). In addition, from the $\chi^{\rm VV}$
values of Ohama {\it et al.},\cite{Ohama1997b} one obtains $\Delta\chi_{ca}^{\rm VV} =
\Delta\chi_{cb}^{\rm VV} = 4.2 \times 10^{-5}$\,cm$^3$/mol, which are similar in
magnitude but opposite in sign to our data in Eq.~(\ref{EqOurChiVV}).  If
the strong change in each of $\Delta\chi_{ac}$ and $\Delta\chi_{bc}$ below $T_{\rm
c}$ in Fig.~\ref{NaV2O5DeltaChi_abc} is due to a respective
$\Delta\chi_{\alpha\beta}^{\rm VV}$ which is strongly temperature dependent in this
temperature range, an effect similar to that reported to occur from NMR measurements
discussed in the next subsection, it is hard to understand why this change is not
reflected in a distinct change in the reported temperature dependent anisotropy of
the $g$ values at $T_{\rm c}$ if these $g$-value measurements are recording the
characteristics of the bulk phase.

\paragraph*{Anisotropy in the Van Vleck susceptibility from NMR.}

From a so-called $K$-$\chi$ analysis using NMR paramagnetic nuclear
resonance shift $K(T)$ data, combined with $\chi(T)$ measurements, under certain
assumptions $\chi^{\rm VV}$ can be obtained if $K$ is proportional to $\chi$, with
$T$ as an implicit parameter.  In this way, $\chi^{\rm VV}$ values have been
obtained by Ohama and coworkers for NaV$_2$O$_5$ using $^{23}$Na
(Ref.~\onlinecite{Ohama1997a}) and $^{51}$V (Ref.~\onlinecite{Ohama1997b}) NMR
measurements on the same aligned powder sample.  The former $^{23}$Na study yielded
$\chi^{\rm VV}_b = 23 \times 10^{-5}$\,cm$^3$/mol below $T_{\rm c}$ and $16 \times
10^{-5}$\,cm$^3$/mol above $T_{\rm c}$, corresponding to a decrease of $7\times
10^{-5}$\,cm$^3$/mol at $T_{\rm c}$.  Their low temperature value is quite
similar to our value in Eq.~(\ref{EqOurChiVV}).

The $^{51}$V NMR study,\cite{Ohama1997b} carried out above $T_{\rm c}$, yielded
$\chi^{\rm VV}_b = 2(1) \times 10^{-5}$\,cm$^3$/mol, roughly an order of magnitude
smaller than obtained in the authors' first study (no comment was made about this
discrepancy), and in addition gave $\chi^{\rm VV}_a = 1(1) \times
10^{-5}$\,cm$^3$/mol and $\chi^{\rm VV}_c = 4(1) \times 10^{-5}$\,cm$^3$/mol. 
These values are significantly smaller than our values.  We note that a $K$-$\chi$
analysis on the $d^1\ {\rm V}^{+4}$ compound VO$_2$ yielded  $\chi^{\rm VV} = 6.5
\times 10^{-5}$\,cm$^3$/mol.\cite{Pouget1972} 

\subsection{Modeling the susceptibility of $\bbox{\rm NaV_2O_5}$ above $\bbox{T_{\rm
c}}$}
\label{SecNaV2O5Modeling}
\vglue0.13in
Turning now to the experimental $\chi(T)$ data in Fig.~\ref{NaV2O5XtalsChi2},
we have $T^{\rm max} \approx 370$\,K.  Assuming the validity of the
Hamiltonian~(\ref{EqHeisChn}), Eq.~(\ref{EqChainPars:a}) for the uniform
Heisenberg chain yields the exchange constant $J/k_{\rm B} \approx 580$\,K\@.  Then
the $g_b$ value in Eq.~(\ref{EqGs}) and our $\chi_0$ values at $T = 0$ in
Table~\ref{TableLowTFitParams} below, together with Eqs.~(\ref{EqChimaxTmax}) and
(\ref{EqExpChi:all}), predict that the measured
$\chi^{\rm max} \sim 40 \times 10^{-5}$\,cm$^3$/mol, which is
similar to the measured values of $\approx 44$ and $48\times 10^{-5}$\,cm$^3$/mol for
the two crystals in Fig.~\ref{NaV2O5XtalsChi2}, respectively.  We therefore
proceeded to try to fit the data by the uniform chain model.  The ``Fit'', shown
as the solid curve in Fig.~\ref{NaV2O5XtalsChi2}, is a plot of
Eqs.~(\ref{EqExpChi:all}), with $\chi^*(t)$ being the susceptibility of the uniform
chain (Fit~2 above) and with the parameters
\begin{eqnarray}
\chi_0 = 8\times 10^{-5}\,{\rm\frac{cm^3}{mol}},
~~~C_{\rm imp} = 0,\nonumber\\
\label{EqFit1Pars}\\
g = 1.972~,~~~{J\over k_{\rm B}} = 580\,{\rm K}~.~~~{\rm (``Fit")}\nonumber
\end{eqnarray}
This ``Fit'' is not really a fit, since we just set the $g$ and $J$
values to those estimated above and then set $\chi_0$ so that the calculated curve
is in the vicinity of the data, because no small change in the
parameters can bring the theory in agreement with the data.  It is clear that
adjusting $\chi_0$ further will not improve the agreement, nor will including a
nonzero impurity Curie constant $C_{\rm imp}$.  However, the shapes of the curve and
the data are similar, so the agreement can be improved considerably (not shown) by
simultaneously decreasing $\chi_0$ to $\approx -10 \times 10^{-5}\,{\rm cm^3/mol}$,
which is not possible physically according to Eqs.~(\ref{EqExpChi:all}) because it
would require the Van Vleck susceptibility to be negative, and increasing $g$ to the
unphysically large value of $\approx 2.4$, while keeping $J$ constant.  These results
are in disagreement with the conclusion of Isobe and Ueda who found that the
Bonner-Fisher prediction\cite{Bonner1964} fitted their powder susceptibility data
from 50 to 700\,K very well assuming $g = 2$.\cite{Isobe1996}  We can only note that
their $\chi(T)$ data have not been quantitatively reproduced in either
their\cite{Isobe1997a,Isobe1997} or others' subsequent measurements on NaV$_2$O$_5$,
including ours, and that the Bonner-Fisher prediction is not accurate at
temperatures below $\sim J/(4k_{\rm B})\approx 145$\,K as discussed in the
Introduction.

Lohmann {\it et al.}\cite{Lohmann1997} and Hemberger {\it et al.}\cite{Hemberger1998}  also
previously concluded that the $\chi(T)$ of NaV$_2$O$_5$ is not described (below
250\,K) by the prediction for the $S = 1/2$ Heisenberg chain, based on their fits
by the Bonner-Fisher prediction\cite{Bonner1964} to their $\chi(T)$ deduced from
ESR measurements up to 650\,K\@.  They suggested that additional exchange couplings
may be required to explain the observed $\chi(T)$.  We consider this possibility
here by modeling the influence of possible interchain spin coupling.  Because
there are no accurate and generally applicable numerical calculations for this
case that we are aware of, we utilize the following simple  molecular field theory
(MFT)  prediction for the spin susceptibility\cite{Johnston1997,Miyahara1998}
\begin{equation}
\frac{1}{\chi^*(t)} = \frac{1}{\chi^*_{\rm chain}(t)} + {z^\prime J^\prime\over
J}~,
\label{EqMFT}
\end{equation}
where $\chi^*_{\rm chain}(t)$ is the reduced spin susceptibility of the isolated
quantum $S = 1/2$ uniform Heisenberg chain (our Fit~2 above).  The parameter
$z^\prime$ is the effective number of spins on other chains to which a spin in a
given chain is coupled with effective (or average) exchange constant
$J^\prime$.  To be consistent with our sign convention for the intrachain
exchange constant $J$, $J^\prime$ is positive for AF interactions and negative for
ferromagnetic (FM) interactions.  Equation~(\ref{EqMFT}) is very accurate when
$|z^\prime J^\prime/J| \ll 1$.\cite{Johnston1997,Miyahara1998}

\begin{figure}
\epsfxsize=3.17in
\centerline{\epsfbox{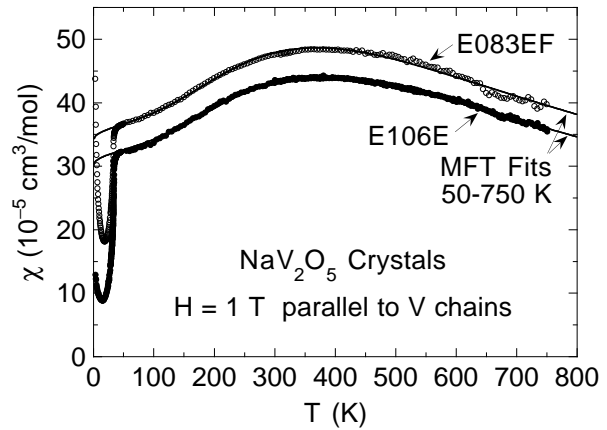}}
\vglue 0.1in
\caption{Fits of the magnetic susceptibility $\chi$ vs temperature $T$ from 50\,K 
to 750\,K for the two crystals in Fig.~\protect\ref{NaV2O5XtalsChi2} by
Eqs.~(\protect\ref{EqExpChi:all}) with
$C_{\rm imp} = 0$,  $g = 1.972$ and the spin susceptibility $\chi^{\rm spin}$ given
by the molecular field theory (MFT) prediction~(\protect\ref{EqMFT}) for coupled
quantum $S = 1/2$ uniform Heisenber chains.  The fits are shown as the solid curves
and the fit parameters are given in Table~\protect\ref{TabMFTPars}.  The fits
overlap the data so they are difficult to see; consequently they have been
extrapolated to higher and lower temperatures to show where they are.}
\label{NaV2O5HiTMFTFits}
\end{figure}

We fitted the $\chi(T)$ data above 50\,K for the two crystals in
Fig.~\ref{NaV2O5XtalsChi2} by Eqs.~(\ref{EqExpChi:all}) and~(\ref{EqMFT}),
where we fixed $g_b = 1.972$ and $C_{\rm imp} = 0$ and allowed $\chi_0$, $J$ and
the product $z^\prime J^\prime$ to vary.  Very good fits were obtained, for
which the fitting parameters are given in Table~\ref{TabMFTPars}.  The fits are
plotted as the solid curves in Fig.~\ref{NaV2O5HiTMFTFits}.  For the parameters of
the two crystals taken together, the fitted $J/k_{\rm B}$ = 584(9)\,K is
the same as deduced above (580\,K) from the temperature of the maximum
in $\chi(T)$, and the fitted $\chi_0 = 1.4(16) \times 10^{-5}$\,cm$^3$/mol is similar 
to our results at low temperatures in Fig.~\ref{NaV2O5Chi_abc}(b).  A moderately
large and negative (FM) interchain coupling $z^\prime J^\prime/J = -1.26(5)$ was
obtained.  This coupling is sufficiently strong that long-range magnetic ordering
might be expected, but which is not observed, possibly due to magnetic frustration
effects.  If the present mean-field interchain coupling analysis is correct, this
interchain coupling should be evident in the magnon dispersion relations observable
by inelastic magnetic neutron scattering measurements.  Indeed, moderately strong
dispersions of 1.4\,meV in each of two bands perpendicular to the chains have in
fact been observed by Yosihama {\it et al.}\cite{Yosihama1998} in such measurements on
single crystals.  It remains to be seen whether the magnitude and sign of the
interchain exchange coupling that we infer in the mean-field analysis are consistent
with the dispersion relations deduced from the neutron scattering data.

An alternative and/or additional mechanism which can produce a strong deviation of
the measured $\chi(T)$ of a uniform chain compound from that predicted for
Heisenberg uniform and alternating chains is the spin-phonon
interaction.\cite{Sandvik1997,Wellein1998,Kuhne1999,Augier1998,Klumper1998b,%
Bursill1998} 
At low $T$ this interaction can lead to a  spin-Peierls transition and can strongly
modify $\chi(T)$ above $T_{\rm c}$ from that expected for the Heisenberg
\,chain.\cite{Sandvik1997,Kuhne1999}  \ Sandvik, \,Singh, \,and Campbell carried 

\noindent out a
detailed QMC investigation of a spin-Peierls model in which the spin 1/2
interactions were modified by the presence of dynamical (quantum mechanical)
dispersionless Einstein phonons.\cite{Sandvik1997}  For  particular values of the
spin-phonon coupling constant and phonon frequency, they found that the effective
exchange constant $J_{\rm eff}$ decreases strongly with increasing $T$, and at $T=0$
is about 27.3\% larger than the bare $J$.  Perhaps surprisingly, they found however
that if the bare $g$ factor is reduced by $\approx 7$\% and the bare $J$ by
$\approx 18$\% in the $\chi^*(t)$ predicted for the Heisenberg model, this model
was then in good agreement with their QMC simulations for temperatures above
$T_{\rm c}$.  A recent important extensive study of many finite-temperature
properties of the same model using QMC simulations was carried out by K\"uhne and
L\"ow.\cite{Kuhne1999}  They found that for not too low temperatures, the
susceptibilities for various Einstein phonon frequencies and spin-phonon coupling
constants can all be scaled onto a universal curve, given by that for the uniform
Heisenberg chain, using only a suitably defined effective exchange constant $J_{\rm
eff} > J$.  Contrary to the result of Ref.~\onlinecite{Sandvik1997}, they found that
a rescaling of the $g$ factor was not necessary.  
\begin{table}
\caption{Fit parameters for the magnetic susceptibility of two
NaV$_2$O$_5$ crystals according to Eq.~(\protect\ref{EqExpChi:all}) with
$C_{\rm imp} = 0$,  $g = 1.972$ and the spin susceptibility $\chi^{\rm spin}$ given
by the molecular field theory expression~(\protect\ref{EqMFT}) for coupled quantum 
$S = 1/2$ uniform Heisenberg chains.}
\begin{tabular}{lccc}
      & $\chi_0$ & $J/k_{\rm B}$ & $z^\prime J^\prime/J$ \\
crystal & $\big(10^{-5}{\rm {cm^3\over mol}}\big)$ & (K) \\
\hline
E083EF & $+$2.8(2) & 577(2) & $-$1.28(3) \\
E106E  & $-$0.1(1) & 592(1) & $-$1.23(2) \\
\end{tabular}
\label{TabMFTPars}
\end{table}
\noindent Our experimental results for
${\rm NaV_2O_5}$ are not consistent with either of these theoretical studies,
because as discussed below Eq.~(\ref{EqFit1Pars}) above, to force-fit the Heisenberg
chain $\chi(T)$ prediction onto the data requires an unphysically large negative
value of $\chi_0$, as well as an unphysically large increase in $g$.  

On the other hand, our observed
$\chi(T)$ does not agree with the Heisenberg chain
model (with a temperature-independent $J$), and in the next section we simultaneously
model the data both above and below $T_{\rm c}$ within the context of the Heisenberg
chain model with a temperature-dependent $J$, where we find that $J(T)$ above
$T_{\rm c}$ is  very similar in form to that deduced in the calculations of
Refs.~\onlinecite{Sandvik1997} and~\onlinecite{Kuhne1999}.  Thus it may be the case
that the spin-phonon interaction is indeed important to determining
$\chi(T)$ in NaV$_2$O$_5$, but where the effects on $\chi(T)$ are somewhat different
than calculated in the models.  In particular, the theoretical predictions may be
substantially modified if phonon spectra appropriate to real materials
were to be used in the calculations instead of dispersionless Einstein phonons.

\subsection{Simultaneous modeling of the susceptibility of NaV$_2$O$_5$ below and
above $\bbox{T_{\rm c}}$}
\label{SecNaV2O5Fits}

Previous modeling of $\chi(T)$ of NaV$_2$O$_5$ to extract the spin gap has usually
been done at the lowest temperatures without reference to the magnitude of $\chi$
above $T_{\rm c}$.  Here we utilize our fit to the $\chi^*(t)$ for the
Heisenberg chain to extract $J$ above $T_{\rm c}$ from the experimental data. 
Clearly, since the measured $\chi(T)$ above $T_{\rm c}$ cannot be modeled within
this framework using a temperature-independent $J$ as shown in the previous
section, it follows that if we are to remain within this framework, $J$, which is
then evidently an effective exchange constant incorporating additional physics of
the material, must be temperature dependent.  Then with $J(T)$ fixed, we derive
the $T$-dependent spin gap $\Delta(T)$ and exchange alternation parameter
$\delta(T)$ near and below $T_{\rm c}$ directly from the measured $\chi(T)$ data,
which has not, to our knowledge, been carried out before for any system showing a
spin-dimerization transition, using our $\chi^*(\alpha,t)$ fit function for
the alternating chain.

The specific procedure we adopted for modeling the $\chi_b(T)$ measurement on
each crystal consists of the following six steps, where we fixed 
$g_b =1.972$ in steps~3--5.

(1) The $\chi(T)$ from 2 to 10\,K is fitted by
Eqs.~(\ref{EqExpChi:all}), setting $\chi^{\rm spin}=0$ because of the presence of the
spin gap, thereby obtaining the parameters $\chi_0,\ C_{\rm imp}$, and $\theta$.

(2) Using these $\chi_0,\ C_{\rm imp}$ and $\theta$ parameters, we solve for $J(T)$
for $T\geq 60$\,K, or for $T = 50$\,K only, using our ``Fit~1'' function for
$\overline{\chi^*}(\overline{t})$ of the Heisenberg chain, which is one end-point
function of our $\overline{\chi^*}(\delta,\overline{t})$ fit function, and fit the
resulting $J(T)$ by a polynomial in $T$ for extrapolation below $T_{\rm c}$; we used
the extrapolation function $J(T) = J(0) + a T^2 + b T^3$.

\begin{figure}
\epsfxsize=3.15in
\centerline{\epsfbox{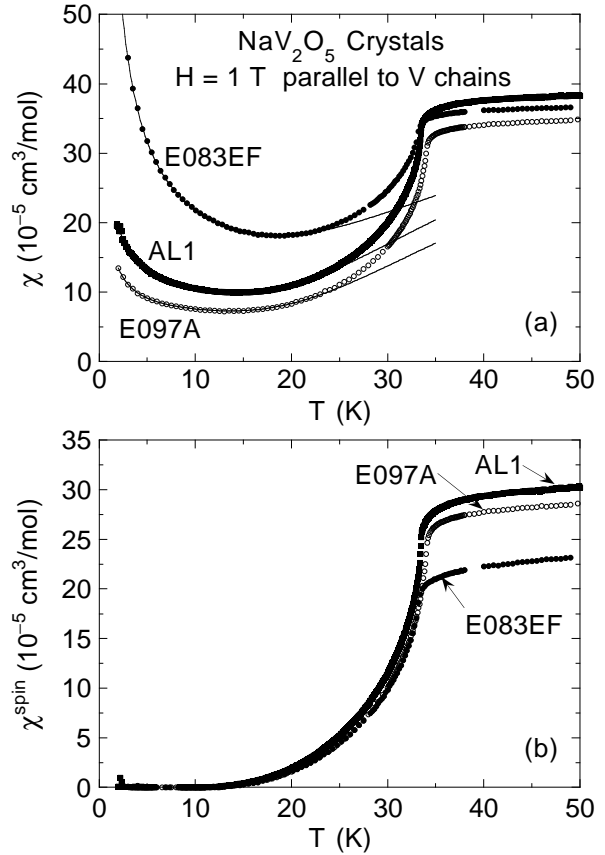}}
\vglue 0.1in
\caption{(a) Magnetic susceptibility $\chi$ versus temperature
$T$ for three crystals of NaV$_2$O$_5$ in the low-$T$ regime near the
dimerization transition temperature $T_{\rm c} \approx 33$--34\,K\@.  The
crystal symbol designations are E083EF ($\bullet$), E097A ($\circ$), AL1 (filled
squares).  The solid curves are fits to the data below 20\,K by
Eq.~(\protect\ref{EqExpChi:a}), where the spin gap is assumed independent of $T$,
and have been extrapolated to higher temperatures.  (b) Magnetic spin susceptibility
$\chi^{\rm spin}(T)$, obtained from the data in (a) by subtracting [$\chi_0 +
C_{\rm imp}/(T-\theta)$] appropriate to each crystal according to
Eq.~(\protect\ref{EqExpChi:a}).}
\label{NaV2O5ChiLoT}
\end{figure}

(3) With this $J(T)$, or using $J(50$\,K) only, we fitted $\chi(T)$ from 2
to 20\,K, now including $\chi^{\rm spin}(T)$ for the alternating-exchange chain
[i.e., using our alternating chain $\overline{\chi^*}(\delta,\overline{t})$ fit
function]  assuming a $T$-independent $\delta$ (and $\Delta$), and obtain a new set
of $\chi_0,\ C_{\rm imp}$ and $\theta$ parameters [in addition to $\delta(0)$].

(4) Steps 2 and~3 are repeated until convergence is achieved, which takes in practice
only one additional iteration.  Note that we implicitly assume that $\chi_0,\ C_{\rm
imp}$ and $\theta$ are independent of $T$, i.e., that the transition(s) at $T_{\rm
c}$ do not affect them.

(5) The experimentally determined molar spin susceptibility $\chi^{\rm spin}(T)$ is
now computed by inserting the final $\chi_0,\ C_{\rm imp}$ and $\theta$ fit
parameters into Eq.~(\ref{EqExpChi:a}).  Then using the fitted $J(T)$ or $J(50$\,K),
the $\delta(T)$ is computed using our fit function
$\overline{\chi^*}(\delta,\overline{t})$ for the alternating-exchange chain by
finding the root for $\delta$, at each data point temperature $T$, of
\begin{figure}
\epsfxsize=3in
\centerline{\epsfbox{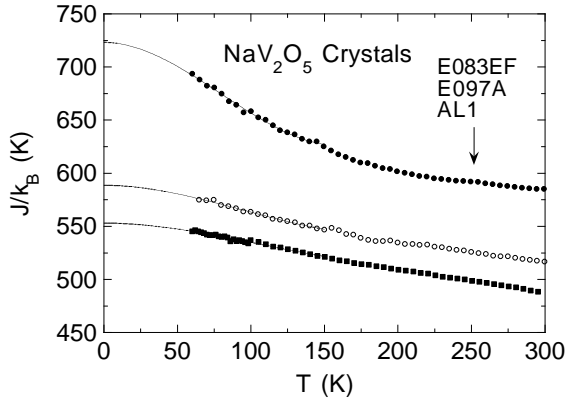}}
\vglue 0.1in
\caption{Exchange constant $J$ versus temperature $T$ for the three crystals of
NaV$_2$O$_5$ in Fig.~\protect\ref{NaV2O5ChiLoT} in the temperature regime $T \geq
60$\,K which is above the dimerization temperature $T_{\rm c} \approx 34$\,K\@. 
The solid curves are polynomial fits to the data between 60 and 150\,K for the
respective samples, which are extrapolated to $T = 0$ as shown.}
\label{NaV2O5J(T)}
\end{figure}
\vglue0.05in
\begin{equation}
\chi_b^{\rm spin}(T) = \frac{N_{\rm A}g_b^2\mu_{\rm
B}^2}{J_b(T)}\ \overline{\chi^*}\Big[\delta,\frac{k_{\rm B}T}{J_b(T)}\Big]~.
\end{equation}
(6) In a separate step not associated with the fitting procedure in
steps~1--5, the spin gap $\Delta(T)$ is computed from $\delta(T)$ determined in
step~5 using an independently known function $\overline{\Delta^*}(\delta)\equiv
\Delta(\delta)/J$ and our $J(T)$ or $J(50$\,K).  We used our
$\overline{\Delta^*}(\delta)$ fit function in Eqs.~(\ref{EqD(d):all}) for this
purpose.

We measured $\chi_b(T)$ for nine different crystals from four different batches of
NaV$_2$O$_5$ and now present illustrative results \,obtained \,in \,each \,of \,the
\,above \,modeling steps 2 
to~4 (final iteration), 5 and 6 for three representative crystals. 
We will follow in graphical form the data modeling through successive steps for these
three crystals to show how differences in one property between the crystals may or
may not propagate through the next step(s) 

\widetext
\begin{table}
\caption{Fitted parameters in Eqs.~(\protect\ref{EqExpChi:all}), using the
$\chi^*(\alpha,t)$ fit function~(\protect\ref{EqChi*Gen:all}) for the
alternating-exchange chain, obtained by fitting the $\chi(T)$ data in the range
2--20\,K for nine crystals of NaV$_2$O$_5$, for an assumed $g$ factor of 1.972.  If
an arror bar is not given for $J(0)$, this value is $J(50$\,K) which was determined
from a single data point near 50\,K\@.  The spin gap $\Delta(0)$ is not a fitted
parameter, but is rather computed from the fitted alternation parameter
$\delta(0)$ using Eqs.~(\protect\ref{EqD(d):all}).  Similarly, the alternation
parameter $\alpha(0)$ is computed from $\delta(0)$ using
Eq.~(\protect\ref{EqDimParams:b}).  Note that all three measurements for crystal AL1
were carried out in a field of 2\,T, whereas all the other crystals were measured in
a field of 1\,T\@.}
\begin{tabular}{ldddcccd}
Crystal & $\chi_0$ & $C_{\rm imp}$ & $-\theta$ & $J(0)/k_{\rm B}$ &
$\delta(0)$ & $\alpha(0)$ & $\Delta(0)/k_{\rm B}$ \\
 &  $\big(10^{-5}\,{{\rm cm}^3\over {\rm mol}}\big)$ & 
$\big(10^{-3}\,{{\rm cm}^3\,{\rm K}\over {\rm mol}}\big)$ & (K) & (K) & & &(K) \\
\hline
E082E  & 6.82(4) & 1.123(5) & 0.46(1) & 710(4) & 0.0287(2) & 0.9442(4) & 101.1(10) \\
E083B  & 4.56(9) & 0.81(1) & 0.30(5) & 654(3) & 0.0327(3) & 0.9366(6) & 102.6(13) \\
E083EF & 11.2(1) & 1.11(1) & 0.43(4) & 723(2) & 0.0279(4) & 0.9458(8) & 100.8(14) \\
E083G  & 6.55(3) & 1.112(3) & 0.45(1) & 688 & 0.0298(1) & 0.9421(2) & 100.7(2)\\
E083H  & 4.24(3) & 0.780(3) & 0.33(1) & 650 & 0.0332(2) & 0.9357(4) & 102.9(5) \\
E083I  & 4.18(6) & 0.946(7) & 0.32(2) & 657 & 0.0329(4) & 0.9363(8) & 103.3(9) \\
E097A  & 5.92(3) & 0.170(3) & 0.25(4) & 589(2) & 0.0389(2) & 0.9251(4) & 104.6(6) \\
E106E  & 5.67(8) & 0.134(8)  & 0.31(1)  & 662(2) & 0.0332(5) & 0.9358(9)  &104.8(13)
\\ 
AL1 $(\bbox{H}||a)$& 12.22(6) & 0.221(7) & 0.46(6) & 598(1) & 0.0366(3) & 0.9294(6) &
101.5(8)\\
AL1 $(\bbox{H}||b)$& 10.94(6) & 0.240(7) & 0.46(6) & 607(1) & 0.0352(3) & 0.9320(6) &
100.4(8)\\
AL1 $(\bbox{H}||c)$& 5.49(7) & 0.298(8) & 0.75(7) & 635(1) & 0.0337(4) & 0.9348(7) &
101.7(9)\\
\end{tabular}
\label{TableLowTFitParams}
\end{table}
\narrowtext
\noindent of the analysis, but we present the
fitting parameters for all of the crystals in Table~\ref{TableLowTFitParams}.

The measured $\chi(T)$ data below 50\,K for the three crystals are shown in
Fig.~\ref{NaV2O5ChiLoT}(a), where the fits below 20\,K in step~4 are shown as the
solid curves with parameters in Table~\ref{TableLowTFitParams}.  Crystals E097A and
AL1 are seen to have much lower levels of paramagnetic impurities than E083EF, as
reflected in the impurity Curie constant, i.e., the magnitude of the impurity
Curie-Weiss upturn at low $T$.  By subtracting the $\chi_0$ and impurity Curie-Weiss
terms from the data, the spin susceptibility $\chi^{\rm spin}(T)$ is obtained for
each crystal as shown in Fig.~\ref{NaV2O5ChiLoT}(b).  These data show good
consistency below $T_{\rm c}$ for the three crystals, despite the differences in the
$\chi_0$ values, the magnitudes of the Curie-Weiss impurity term and in the
$\chi(T)$ above $T_{\rm c}$.  The $J(T)$ determined for the three crystals in step~2
are shown up to 300\,K in Fig.~\ref{NaV2O5J(T)}.  $J$ is found to decrease by $\sim
10$--20\,\% upon increasing $T$ from 60 to 300\,K, which when $T$ is scaled by $J$ is
similar to the fractional decrease predicted by Sandvik {\it et al.}\cite{Sandvik1997} due
to the spin-phonon interaction.  It is noteworthy that crystal E083EF, with by far
the highest level of paramagnetic defects and/or impurities, also has the largest
$J(T)$ and the largest change in $J$ with $T$.

Figures~\ref{NaV2O5Delta,Gap(T)}(a) and~\ref{NaV2O5Delta,Gap(T)}(b) show the spin
dimerization parameter $\delta(T)$ and spin gap $\Delta(T)$ determined for each of the
three crystals in the final modeling steps~5 and~6, respectively.  Several features of
these data are of note.  First, there is a rather large variation in the
dimerization parameter,  $\delta(0) = 0.028$--0.040, between the three crystals,
despite the fact that $T_{\rm c} = 33$--34\,K is nearly the same for the different
crystals; the most impure crystal E083EF has the smallest $\delta(0)$, as might have
been expected.  Despite this variability, these $\delta(0)$ values are
all significantly smaller than the three values reported for various samples by
different groups as determined using different techniques, which \,are \,listed \,in
Table~\ref{TabLitDat} along 
\newpage
\begin{figure}
\epsfxsize=3in
\centerline{\epsfbox{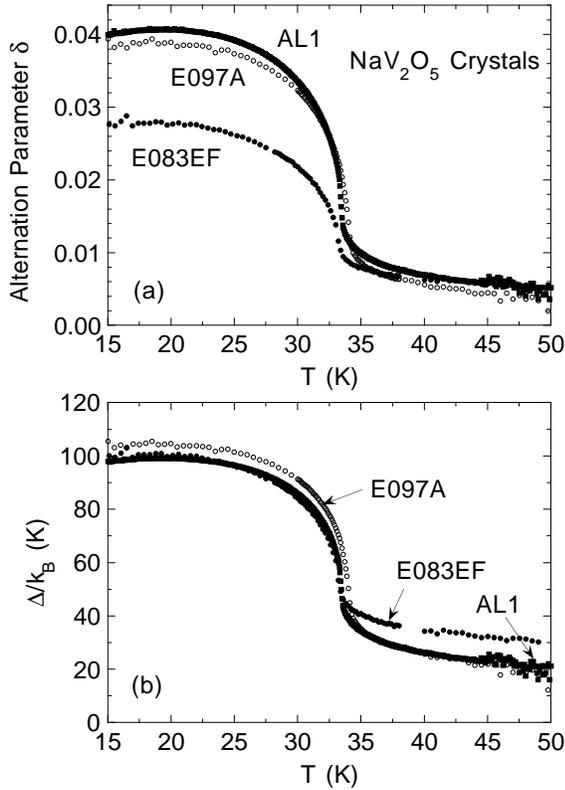}}
\vglue 0.1in
\caption{Alternation parameter $\delta$ (a) and spin gap $\Delta$ (b) versus
temperature $T$ below 50\,K for the three crystals of NaV$_2$O$_5$ in
Fig.~\protect\ref{NaV2O5ChiLoT}.  The nonzero
$\delta$ and $\Delta$ above the transition temperature
$\sim 33$--34\,K are presumed to arise from spin dimerization fluctuations and
concurrent spin gap fluctuations, respectively.}
\label{NaV2O5Delta,Gap(T)}
\end{figure}
\vglue-0.01in
\noindent 
\mbox{with~~~~~~~~~~~ other~~~~~~~~~~~ related~~~~~~~~~~~ information.}  \mbox{}
\cite{Ohama1999,Lemmens1998,Isobe1996,Fujii1997,Weiden1997,Koppen1998,Ladavac1998,%
Vasilev1997,%
Lohmann1997,Schmidt1998,Hemberger1998,Ohama1997a,Mila1996,%
Luther1998,Schnelle1998}
\ \ On \,the \,other \,hand, \,the 
\widetext
\begin{table}
\caption{Exchange constant $J$, spin gap $\Delta$, and alternation parameter
$\delta$ for NaV$_2$O$_5$ at the temperature(s) $T$ as determined by the listed
method for the sample with transition temperature $T_{\rm c}$.  The
literature reference is given in the last column.  Method abbreviations: $\chi$,
magnetic susceptibility; Neutrons, neutron scattering; NMR, nuclear magnetic
resonance; ESR, electron spin resonance; $C_{\rm p}$, specific heat; Raman, Raman
light scattering.}
\begin{tabular}{lccccccr}
sample & $T_{\rm c}$\,(K) &  $J/k_{\rm B}$\,(K) & $T$(K) & $\Delta/k_{\rm
B}$\,(K) & $\delta$ & Method & Ref. \\
\hline
Powder & none  & 529  &  350   &       &      & $\chi$  &\onlinecite{Mila1996}\\
Powder & 33.9   &  560 & 35--700&       &      & $\chi$  &\onlinecite{Isobe1996}\\
Powder & 35.3 &       &   7   & 114   &      & Neutrons & \onlinecite{Fujii1997}\\
Aligned powder &      &       &10--20 & 98    &   &$^{23}$Na
NMR&\onlinecite{Ohama1997a}\\ Crystals& 33  & 441   & 2--30 & 85(15)&   &$\chi$  
&\onlinecite{Weiden1997}\\ Crystal&35  &  560 &2--34& 92(20) &0.10(2) &ESR,
$\chi$&\onlinecite{Vasilev1997}\\ Crystal & 33.5 & 560  & 15--30  & 100(2) & 0.107
& ESR &\onlinecite{Lohmann1997}\\ Crystal & 34  &  578  &250--650&100   &  &ESR,
$\chi$&\onlinecite{Hemberger1998}\\ Crystal &32.7,33.0&  & 1.8--12 & 84(10)&
&$C_{\rm p}$&\onlinecite{Koppen1998}\\  Crystal &     &       &   5   & 88(2) &  
& Raman     &\onlinecite{Lemmens1998}\\ Crystal & 35  &       & 10--35& 85(20)&  
&  ESR     &\onlinecite{Schmidt1998}\\ Crystal & 34  &  455  &  15   &      
&0.047 & Raman   &\onlinecite{Ladavac1998}\\ Crystal &     &       &  4.2  &  94  
&     &ESR&\onlinecite{Luther1998}\\ Crystals& 33  &       &  7--15&  67(5)&  
&$C_{\rm p}$&\onlinecite{Schnelle1998}\\ Aligned powder& 34.0 & & 11--20& 108   &  
&$^{51}$V NMR&\onlinecite{Ohama1999}\\
Powder & 34.0 & 491 & 4--30 & 77 &  & $\chi$ & \onlinecite{Onoda1999}\\
\end{tabular}
\label{TabLitDat}
\end{table}

\narrowtext
\newpage
\noindent corresponding range of $\Delta(0)/k_{\rm B} =103(2)$\,K
for the three crystals is fractionally much smaller than that of $\delta(0)$.  We
infer that some of the discrepancies between the $\Delta(0)$
values in Table~\ref{TabLitDat} reported for NaV$_2$O$_5$ by different groups may
arise from differences in, e.g., the types of measurements which are used to
determine $\Delta(0)$ and in the different analyses of those data, rather than from
different $\Delta(0)$ values in the samples.  The variability in
$\delta(0)$ between the crystals in Fig.~\ref{NaV2O5Delta,Gap(T)}(a), compared with
the lack of much variability in $\Delta(0)$ in Fig.~\ref{NaV2O5Delta,Gap(T)}(b),
evidently arises because  $\delta$ must be combined with $J$ to obtain $\Delta$, and
the variations in the first two parameters must largely cancel.  Thus, not
surprisingly, the low-$T$ $\chi^{\rm spin}(T)$ is governed by the spin
gap $\Delta$ and not by $\delta$ or $J$ separately.

The $\delta(T)$ data for our best crystals show very sharp, nearly vertical increases
with decreasing $T$ at $T_{\rm c}$.  We cannot extract a precise critical exponent
$\beta$ from our $\delta(T)$ data due to the large temperature-dependent 
background above $T_{\rm c}$, to be discussed shortly.  However, rough fits below
$T_{\rm c}$ by the expression $\delta(T) \sim (1 - T/T_{\rm c})^\beta$ gave $\beta$
values consistent with  the values $\beta = 0.25(10)$
(Ref.~\onlinecite{Smirnov1998}) from infrared reflectivity measurements, 0.34(8)
(Ref.~\onlinecite{Fertey1998}) from sound velocity measurements along the chain
axis and 0.35(8) (Ref.~\onlinecite{Koppen1998}) from thermal expansion
measurements along that axis.  We note that these values are a factor of two
larger than the value of
$\sim 0.15$ (Ref.~\onlinecite{Ravy1998}) inferred from x-ray diffuse scattering
measurements.

The data in Figs.~\ref{NaV2O5Delta,Gap(T)}(a) and~\ref{NaV2O5Delta,Gap(T)}(b) clearly
show the existence of spin dimerization fluctuations and a spin pseudogap above
$T_{\rm c}$ for each crystal, respectively, irrespective of the crystal quality as
judging from the Curie-Weiss impurity term in the low-$T$ $\chi(T)$, with magnitudes
just above $T_{\rm c}$ \ of \ about \ 20\,\% \ and \ 40\,\% of $\delta(0)$ and
$\Delta(0)$, respectively.  This is a robust result, which was obtained
\newpage
\noindent for each of the nine
crystals we measured, which does not depend on the precise value of $J$ [and
resultant $\chi^{\rm spin}(T,\delta = 0)$] or the details of how $J$ is determined
above $T_{\rm c}$, or even on the detailed formulation of the $\chi^*(\alpha,t)$ fit
function for the alternating-exchange chain.  For example, setting $J$ to be a
constant, equal to the value at 50\,K, yields nearly the same $\Delta(T)$ near
$T_{\rm c}$ as determined using a $T$-dependent $J$.  Similarly, deleting the
impurity Curie-Weiss term in the fit to the data above $T_{\rm c}$ changes the
derived $\chi_0$ and $J(T)$ or
$J$(50\,K) somewhat as well as the detailed temperature dependence of the pseudogap
$\Delta(T)$ above $T_{\rm c}$ but has little influence on the magnitude of $\Delta$
near $T_{\rm c}$. Further, in a previous version of the QMC and TMRG
$\chi^*(\alpha,t)$ fit function (not otherwise discussed in this paper), we did not
enforce the requirement~(iii) in Sec.~\ref{SecAltChnFitFcn} that the transformed
$\overline{\chi^*}(\delta,\overline{t})$ satisfy
$\partial\overline{\chi^*}(\delta,\overline{t})/\partial\delta|_{\delta=0}=0$,
and the same fluctuation effects above $T_{\rm c}$ were found using that fit
function as using the present one, although these fluctuations were somewhat reduced
in magnitude compared to the present results.  Finally, these fluctuations are
observable directly in the measured $\chi(T)$ data in Fig.~\ref{NaV2O5ChiLoT}(a) as a
rounding of the susceptibility curves above $T_{\rm c}$.

From Fig.~\ref{NaV2O5Delta,Gap(T)}, the fluctuation effects persist up to high
temperatures $T> 50$\,K, although the fluctuation amplitudes decrease with
increasing $T$.  Precursor effects above $T_{\rm c}$ have been reported in x-ray
diffuse scattering measurements\cite{Ravy1998} up to $\sim 90$\,K, in ultrasonic
sound velocity\cite{Fertey1998} and optical
absorption\cite{Damascelli1998,Damascelli1999,Smirnov1999} measurements up to $\sim
70$\,K, and in specific heat measurements\cite{Hemberger1998,Powell1998} up to
$\sim 40$--50\,K, so it is not surprising that spin dimerization parameter
fluctuations in Fig.~\ref{NaV2O5Delta,Gap(T)}(a), and a spin pseudogap in
Fig.~\ref{NaV2O5Delta,Gap(T)}(b) reflecting fluctuations in the spin gap, are 
found above $T_{\rm c}$.

\subsection{Specific heat of NaV$_2$O$_5$}
\label{SecCpDatModeling}

\begin{figure}
\epsfxsize=3in
\centerline{\epsfbox{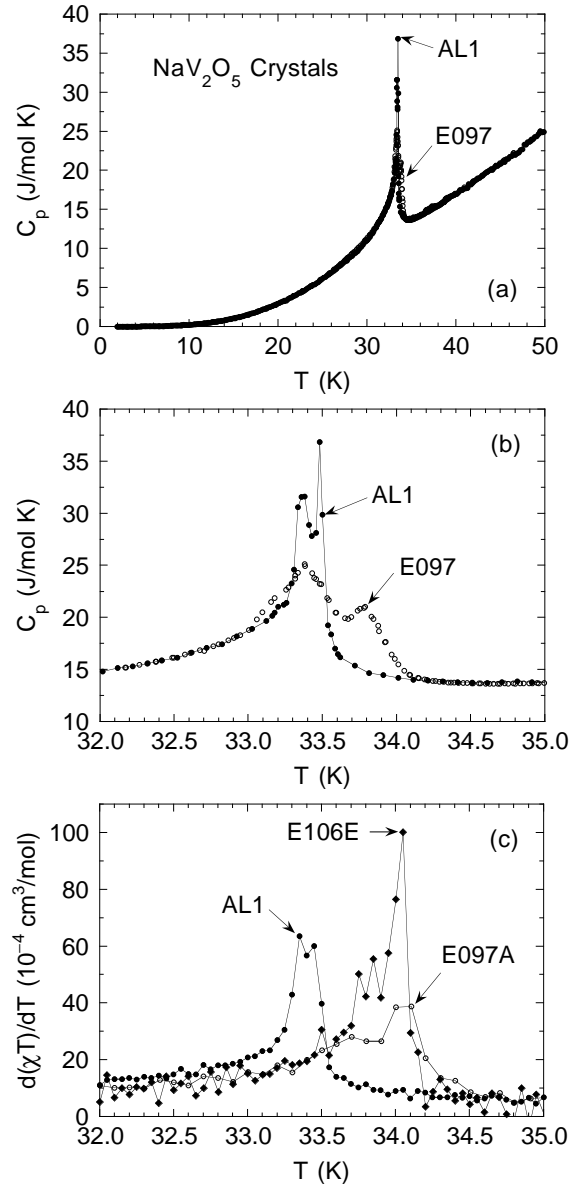}}
\vglue 0.1in
\caption{(a) Specific heat $C_{\rm p}$ vs temperature $T$ for NaV$_2$O$_5$
crystals E097 and AL1.  (b) Expanded plots of the data in the vicinity of the
transition temperatures of the two crystals.  (c) Temperature derivative of $\chi T$
vs $T$ for the same crystal AL1 as in (a) and (b) plus data for crystals E097A (from
the same batch as E097) and E106E\@.  The lines connecting the data points are
guides to the eye.}
\label{NaV2O5XtalsCp}
\end{figure}

\begin{figure}
\epsfxsize=3.06in
\centerline{\epsfbox{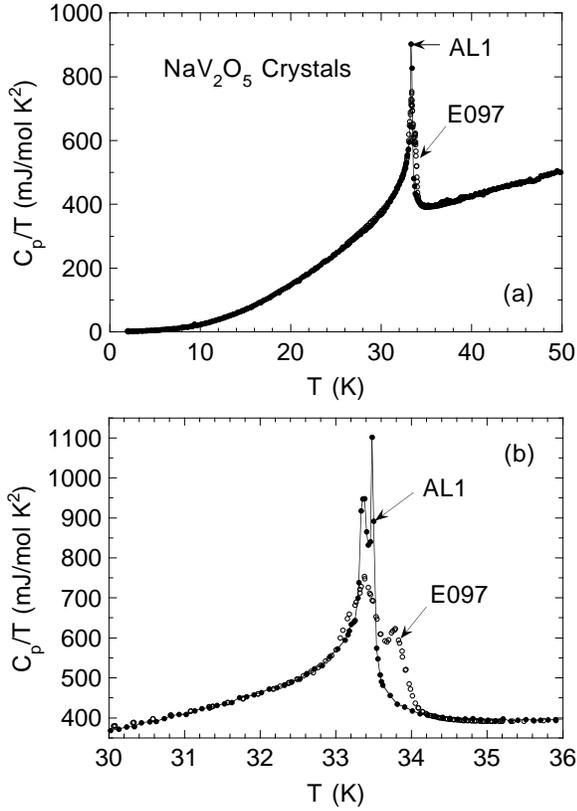}}
\vglue 0.1in
\caption{(a) Specific heat $C_{\rm p}$ divided by temperature $T$ versus $T$ for
NaV$_2$O$_5$ crystals E097 and AL1.  (b) Expanded plots of the data in the vicinity
of the transition temperatures of the two crystals.  The lines connecting the data
points for crystal AL1 are guides to the eye.}
\label{NaV2O5XtalsCp/T}
\end{figure}

In order to correlate the magnetic effects discussed above in
NaV$_2$O$_5$ with thermal effects, we have carried out specific heat vs temperature
$C_{\rm p}(T)$ measurements on the same crystal AL1, and a crystal E097 from the
same batch as E097A, for which $\chi(T)$ data were presented and modeled
above.  The results from 2\,K to 50\,K for crystals E097 and AL1 are shown in
Fig.~\ref{NaV2O5XtalsCp}(a).  Over this temperature range, the $C_{\rm p}(T)$ data
for the two crystals agree extremely well, except in the range 33.0--34.2\,K, i.e.,
in the vicinity of the transitions as will be discussed shortly. The shapes of the
specific heat anomalies at $T_{\rm c}$ are not mean-field-like specific heat jumps as
observed in, e.g., conventional superconductors, but instead are $\lambda$-shaped
anomalies.  Thus, any attempt to define a (mean-field) ``specific heat jump at
$T_{\rm c}$'' is fraught with ambiguity.  These shapes are retained in plots of
$C_{\rm p}(T)/T$ vs $T$ as shown in Fig.~\ref{NaV2O5XtalsCp/T}(a).  This
$\lambda$ shape has been observed previously, and variously attributed to
fluctuation effects or a possible smeared-out first order transition.  \ In view 
\vglue-0.01in
\noindent
of the coupled structural, charge-ordering and spin dimerization transitions at
$T_{\rm c}$ in NaV$_2$O$_5$ as discussed in the Introduction, their relative
contributions to the specific heat anomalies are not clear, if indeed their
contributions can be uniquely distinguished.

Expanded plots of $C_{\rm p}(T)$ and $C_{\rm p}(T)/T$ versus $T$, shown in
Figs.~\ref{NaV2O5XtalsCp}(b) and~\ref{NaV2O5XtalsCp/T}(b), respectively, reveal a
sharp high peak at 33.4\,K for crystal AL1, which is slightly split by $\approx
0.1$\,K in spite of the fact that \ the \ overall \,height \,of \,the \,anomaly \,is
\,much \,larger \,than \,pre-

\noindent viously reported for any crystal of NaV$_2$O$_5$.  Two peaks are also
observed for crystal E097, at 33.4\,K and 33.8\,K, which are more widely separated
than for crystal AL1.  From Fig.~\ref{NaV2O5XtalsCp/T}(b), the entropy under the
anomaly(ies) for each crystal is about the same (see below).  Comparing these
results with the $\delta(T)$ and $\Delta(T)$ data in Fig.~\ref{NaV2O5Delta,Gap(T)},
the larger splitting of the $C_{\rm p}(T)$ peak for crystal E097 does not result in
any major difference in the magnetic order parameter properties between the two
crystals, although the transition onset is slightly rounded for crystal E097A
compared to AL1.  By using the Fisher relation,\cite{Fisher1962}
$\partial[\chi(T)T]/\partial T\sim C(T)$ where $C(T)$ is the magnetic contribution
to the specific heat, one obtains results which show the same features near $T_{\rm
c}$ as does the specific heat, as shown in Fig.~\ref{NaV2O5XtalsCp}(c).  Thus,
careful scrutiny of the magnetic properties can reveal the fine detail observed in
the specific heat near $T_{\rm c}$.  In particular, this comparison suggests that
both anomalies in the specific heat near $T_{\rm c}$ for each crystal are
associated with and/or reflected by magnetic effects.

The splitting of the transition into two apparent transitions that we report here was
previously observed in thermal expansion, but not seen in their specific heat,
measurements of a crystal by K\"oppen {\it et al.}\cite{Koppen1998}  The detailed origin
of the transition splitting, and more fundamentally whether the splitting is
instrinsic to ideal crystallographically ordered NaV$_2$O$_5$, remain to be
clarified.  An essential feature that any explanation must account for is that the
temperature splitting between the two transitions in a crystal varies from crystal
to crystal.

\paragraph*{Modeling.}

In this section we will only consider the model utilized above for
analyzing our $\chi(T)$ data, in which ${\rm NaV_2O_5}$ consists, effectively, of
isolated $S = 1/2$ uniform or (below $T_{\rm c}$) alternating-exchange
Heisenberg chains, where the (average) exchange constant $J$ shows,
at most, only a smooth and relatively small change below $T_{\rm c}$.  For reasons
which will become clear below, unfortunately we cannot use our specific heat data
to extract detailed information about the magnetic subsystem in ${\rm NaV_2O_5}$. 
However, other types of important information about the thermodynamics will be
derived using various of the theoretical results presented and discussed previously
in this paper.

There have been two reports\cite{Koppen1998,Schnelle1998} deriving the spin gap from
$C_{\rm p}(T)$ data at $T\lesssim 15$\,K\@.  We first discuss the limits of this type
of analysis.  Using $J(0)/k_{\rm B} = 600$\,K, $\delta(0) = 0.040$, and
$\Delta/k_{\rm B} = 100$\,K (see Table~\ref{TableLowTFitParams}),
Eqs.~(\ref{EqsChiCfbar:b}) and~(\ref{EqsChiCfbar:c}) predict that the magnetic
specific heat $C(T)$ in the dimerized phase at low temperatures $T \ll (\Delta/k_{\rm
B},\,T_{\rm c})$ is
\begin{eqnarray}
C(T) = &&1.0{\rm J\over mol\,K}\Big({100\over T}\Big)^{3/2}\nonumber\\
&&\times\Big[1 + {T\over 100} + {3\over 4}\Big({T\over100}\Big)^2\Big]\,{\rm
e}^{-100/T}~,
\label{EqCPred}
\end{eqnarray}
\vglue0.03in
\noindent with $T$ in units of K\@.  Equation~(\ref{EqCPred}) predicts that
$C(15\,{\rm K}) = 0.026$\,J/mol\,K, which is about 40 times smaller than the
observed $C_{\rm p}(15\,{\rm K}) \approx 1$\,J/mol\,K (which must therefore be due to
the lattice contribution) and hence is unresolvable at such low temperatures.  Within
this model, we must therefore conclude that the previous estimates of the spin gap
based upon modeling the low temperature specific heat were most likely artifacts of
modeling the lattice specific heat.  This can happen if one does not utilize the fact
that the prefactor to the activated exponential term of the magnetic contribution
$C(T)$ is not an independently adjustable parameter, but is rather determined by the
spin gap itself as we have previously demonstrated and emphasized in
Sec.~\ref{SecAltChnDispRlns}.

A related question is whether the entropy associated with the transition(s) at
$T_{\rm c}$ can be associated solely with the magnetic subsystem.  The minimum
possible estimate of the entropy of the transition is obtained from the $C_{\rm
p}(T)/T$ vs $T$ data in Fig.~\ref{NaV2O5XtalsCp/T}(b) by drawing a horizontal line
from the data at the $C_{\rm p}(T)/T$ minimum at $\approx 35.0$\,K, just above
$T_{\rm c}$, to the data that the line intersects with below $T_{\rm c}$ at $\approx
30.6$\,K, and then computing the area between the line and the peak(s) above the
line.  In this way we obtain a value of 0.397\,J/mol\,K for crystal E097 and
0.375\,J/mol\,K for crystal AL1.  On the other hand, the maximum magnetic entropy of
the  $S = 1/2$ uniform chain subsystem at $T_{\rm c}$, using rough values $J/k_{\rm B}
= 600$\,K 
\begin{figure}
\epsfxsize=3.1in
\centerline{\epsfbox{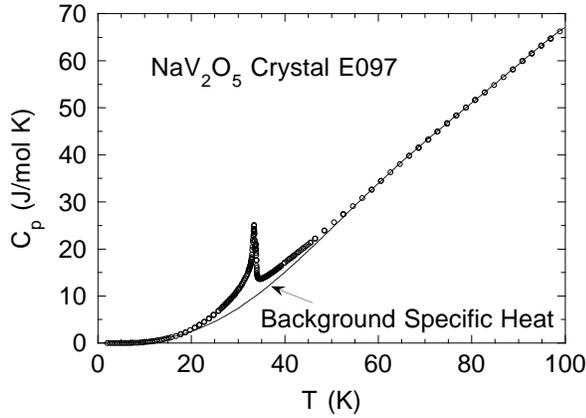}}
\vglue 0.1in
\caption{Measured specific heat $C_{\rm p}$ vs temperature $T$ up to 100\,K for
NaV$_2$O$_5$ crystal E097 ($\circ$).  The solid curve is the background
specific heat, which is the specific heat that would have been observed had no
transitions or order parameter fluctuations occurred, determined as described in the
text.}
\label{E097Cp}
\end{figure}

\noindent  and $T_{\rm c} = 34$\,K, is roughly $S(T_{\rm c}) \approx (2R/3)(k_{\rm B}T_{\rm c}/J) =
0.31$\,J/mol\,K\@.  Thus, the specific heat $\lambda$ anomaly at $T_{\rm c}$ cannot
arise solely from the magnetic subsystem, since the minimum possible entropy of the
transition is significantly greater than the maximum possible magnetic entropy at
$T_{\rm c}$.  At the least, the remaining entropy must therefore be due to the
crystallographic and/or charge-ordering transitions which occur at or close to the
spin dimerization transition temperature as discussed in the Introduction.

A potentially definitive and effective way to proceed from this point would be to
{\it quantitatively} determine the magnetic contribution $C(T)$ to the measured
specific heat $C_{\rm p}(T)$ from $\chi(T)$ at and near $T_{\rm c}$, using a
relationship between $\chi(T)$ and $C(T)$ such as the Fisher relation cited above,
and then compare this result with $C_{\rm p}(T)$.  From a comparison of
Figs.~\ref{NaV2O5XtalsCp}(b) and~\ref{NaV2O5XtalsCp}(c), it seems clear that such a
relation must exist, at least for temperatures near $T_{\rm c}$, but the
relationship between $\chi(T)$ and $C(T)$ near spin dimerization transitions has not
yet been worked out theoretically.

In the absence of such a formulation, we proceed to estimate the {\it change} in
the specific heat associated with the transition(s).  In order to do this
modeling, we must fit $C_{\rm p}(T)$ to higher temperatures than we have
been discussing so far.  The $C_{\rm p}(T)$ data from 2 to~100\,K for ${\rm
NaV_2O_5}$ crystal E097 are shown as the open circles in Fig.~\ref{E097Cp}.  As
noted above, except in the immediate vicinity of $T_{\rm c}$ the $C_{\rm p}(T)$ data
for crystal AL1 are nearly identical with those for crystal E097 up to at least
50\,K, so it will suffice to model the data for crystal E097.  The four modeling
steps and the assumptions we employed are as follows.

(1) We assume that critical and other order parameter fluctuations
associated with the transition(s) at $T_{\rm c}$ make a negligible contribution to
$C_{\rm p}(T)$ over some specified high temperature ($T\gg
T_{\rm c}$) range.  By subtracting the known magnetic contribution $C(T)$ due to
isolated Heisenberg chains [obtained using our fit function for
$C(k_{\rm B}T/J)$] in this temperature range, we obtain the background lattice
contribution $C^{\rm lat}(T)$ in the high temperature region.  Also, since we have
shown that $C(T)$ is negligible for $T \lesssim 15$\,K, the measured
$C_{\rm p}(T)$ in this $T$ range is assumed to be identical to $C^{\rm lat}(T)$ at
these temperatures (we again neglect the possible but unknown specific heats
associated with possible order parameter fluctuations in this range).  Thus we
obtain background lattice specific heats $C^{\rm lat}(T)$ in high and low temperature
ranges which are assumed unaffected by the transition(s) and associated order
parameter fluctuations.

(2) We interpolate between the $C^{\rm lat}(T)$ determined in step~1 in the low-
and high-temperature ranges to obtain, in the intermediate temperature range, what
$C^{\rm lat}(T)$ would have been in the absence of the transition(s) and associated
order parameter fluctuations.

(3) We add the $C(T)$ for isolated chains, used in step~1, back to the $C^{\rm
lat}(T)$ derived in step 2 over the full temperature range of the measurements. 
This is the total background specific heat that would have occurred in the absence
of the transition(s) and associated order parameter fluctuations.  Then we subtract
the total calculated background specific heat from the measured $C_{\rm p}(T)$ data. 
This difference $\Delta C(T)$ should hopefully be a reasonable estimate of the
change in the specific heat associated with the transition and order parameter
fluctuations, including all lattice, charge and spin contributions.  $\Delta C(T)$
must go to zero, by construction, at temperatures above the lower end of the high
temperature region fitted in step~2.

(4) Finally we integrate $\Delta C/T$ with $T$ up to and beyond $T_{\rm c}$ to
obtain the change in entropy $\Delta S(T)$ associated with the transition and order
parameter fluctuations.  $\Delta S(T)$ must become constant, by construction, at
temperatures above the lower end of the high temperature region fitted in step~2.

In the following we will present and discuss the results in each of the four steps of
our modeling program described above.

\begin{figure}
\epsfxsize=3in
\centerline{\epsfbox{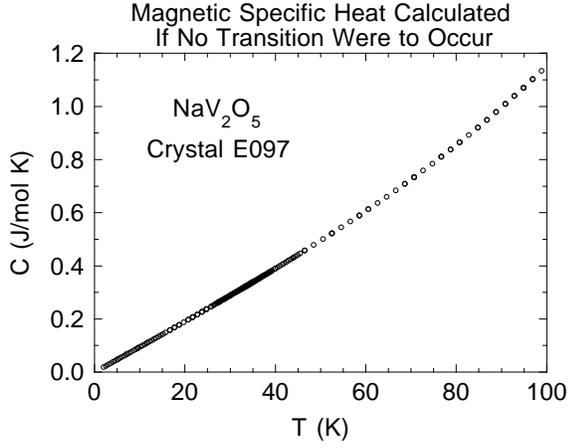}}
\vglue 0.1in
\caption{Magnetic specific heat $C$ vs temperature $T$ for
NaV$_2$O$_5$ crystal E097, calculated for uniform chains with exchange constant
$J(T)$ determined from the analysis of the susceptibility data for crystal E097A. 
The values are those which are assumed to have been observed had no transitions or
associated order parameter fluctuations occurred.}
\label{E097CmagNoXtion}
\end{figure}

\begin{figure}
\epsfxsize=3.11in
\centerline{\epsfbox{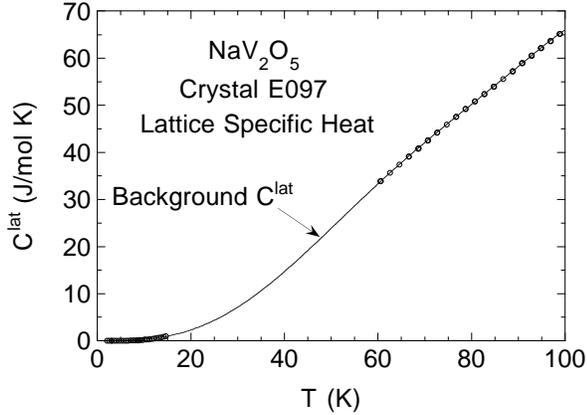}}
\vglue 0.1in
\caption{Background lattice specific heat $C^{\rm lat}$ vs temperature $T$ for
NaV$_2$O$_5$ crystal E097 ($\circ$).  The data shown, in the temperature ranges
2--15\,K and 60--101\,K, were fitted by a polynomial; this interpolation fit is
shown as the solid curve.  This background, including the curve in the interpolated
temperature region, is the lattice specific heat assumed to have been observed had
no transitions or associated order parameter fluctuations occurred.}
\label{E097Clat}
\end{figure}

\paragraph*{Step 1.}Here we first use our $C(T)$ fit function for the numerical
$C(T)$ data,\cite{Klumper1998} which was given in Eqs.~(\ref{EqCFit:all}), to
extract $C^{\rm lat}(T)$ in the high-temperature region above $T_{\rm c}$.  For
consistency with our analysis of the susceptibility in Sec.~\ref{SecNaV2O5Fits}, we
use the temperature-dependent $J(T)$ derived in that section for crystal E097A (see
Fig.~\ref{NaV2O5J(T)}) when computing $C(T)$. The background $C(T)$ thus
estimated for crystal E097, i.e., the values which would have been observed if no
transition(s) at $T_{\rm c}$ or associated order parameter fluctuations had occurred,
is shown in Fig.~\ref{E097CmagNoXtion}.  Comparison of these data with the measured
$C_{\rm p}(T)$ data in Figs.~\ref{NaV2O5XtalsCp} and~\ref{E097Cp} shows that this
$C(T)$ is a small, but non-negligible ($\gtrsim 1$\%), fraction of $C_{\rm p}(T)$
above $T_{\rm c}$.  On the other hand, $C(T)$ is much larger than the observed
$C_{\rm p}(T)$ at low temperatures, because in this temperature range $C\propto T$
whereas $C_{\rm p}(T) \equiv C^{\rm lat}(T)\propto T^3$.  The $C^{\rm lat}(T) =
C_{\rm p}(T) - C(T)$ \ in \ the \ high \ temperature (60--100\,K) \ region \ is \
shown \ in \ Fig.~\ref{E097Clat}, \ together \ with

\noindent $C^{\rm lat}(T)\equiv C_{\rm p}(T)$ in the low temperature (2--15\,K)
region.
\paragraph*{Step 2.}In this step we must interpolate $C^{\rm lat}(T)$ between
the low- and high-temperature regions, i.e., in a broad temperature range spanning the
transition region.  The best way to do this would be to determine $C^{\rm lat}(T)$
directly from $C_{\rm p}(T)$ measurements on a suitably chosen reference compound,
but such measurements have not yet been done.  At first sight, a physically realistic
possibility might be to interpolate the low and high temperature  $C^{\rm lat}(T)$
data using the Debye specific heat function; however, this method is questionable
because the Debye temperature $\Theta_{\rm D}$ in real materials can be rather
strongly temperature dependent within the temperature range of interest here.  The
Debye function for the molar lattice specific heat at constant volume
$C^{\rm Debye}(T)$ is given by\cite{Kittel1971}
\begin{equation}
C^{\rm Debye}(T) = 9 r R \Big({T\over \Theta_{\rm D}}\Big)^3 \int_0^{\Theta_{\rm
D}/T} {x^4\,{\rm e}^x\over ({\rm e}^x - 1)^2}\,dx~,
\label{EqDebyeCv}
\end{equation}
where $r$ is the number of atoms per formula unit ($r = 8$ here) and $R$ is the molar
gas constant.  We attempted to fit our $C^{\rm lat}(T)$ data for the temperature
ranges 2--15\,K and 40--100\,K to 80--100\,K by Eq.~(\ref{EqDebyeCv}).  The fits
parametrized the data very poorly.  We obtained a more reasonable fit by allowing
$r$ to be a fitting parameter, yielding a fitted value $r\approx 4$, but the data
were still poorly fitted, due to too much curvature in the Debye function in the high
temperature region.  Therefore, we were led  to interpolating between the low- and
high-temperature regions using a simple polynomial interpolation function.

To obtain the background lattice specific heat interpolation function, we fitted the
combined $C^{\rm lat}(T)$ data (a total of 141 data points) in the low and high
temperature ranges 2--15\,K and 60--101\,K, respectively, by polynomials of the form
\begin{equation}
C^{\rm lat}(T) = \sum_{n=3}^{n^{\rm max}} c_n T^n~.
\end{equation}
\vglue-0.05in
\noindent The minimum summation index $n = 3$ is set by the expected Debye
low-temperature $T^3$ behavior of the lattice specific heat.  The maximum value
$n^{\rm max}$ was varied to see how the fit parameters and variance changed.  In
addition, for checking the final fits we fitted the $C^{\rm lat}(T)$ data in the
2--15\,K low-$T$ range together with $C^{\rm lat}(T)$ data in a high-temperature
range varying from 40--101\,K to 90--101\,K\@.  We found that the most stable fits
were for $n^{\rm max} = 7$ and~8.  For both values, the fit did not visibly change
when the lower limit of the upper temperature range of the fitted data was varied
from 60 to 70\,K\@.  We chose to use the fit for $n^{\rm max} = 7$ because in this
case the fit was also stable for lower limits of 50 and 80\,K\@.  This stability
allows one to be confident that the interpolation of the fit between the fitted low-
and high-temperature ranges is an accurate representation of the background lattice
specific heat in the interpolated intermediate temperature range.  The fit for the
temperature ranges 2--15\,K and 60--101\,K is shown as the solid curve in
Fig.~\ref{E097Clat}.  The absolute rms deviation of this fit from the fitted data is
quite small, $\sigma_{\rm rms} = 0.046$\,J/mol\,K\@.  The curve over the full
temperature range 2--101\,K represents the background lattice specific heat $C^{\rm
lat}(T)$ expected in the absence of any transitions or order parameter fluctuations.

\paragraph*{Step 3.}Adding the magnetic background specific heat contribution
$C(T)$ obtained in step~1 to the lattice background specific heat contribution
$C^{\rm lat}(T)$ obtained in step~2 gives the total background specific heat,
which is plotted as the solid curve in Fig.~\ref{E097Cp}.  We reiterate that this
background is interpreted as the specific heat that would \ have \ been \ observed
\ had the transition(s) at  $T_{\rm c}$ 
\begin{figure}
\epsfxsize=3.05in
\centerline{\epsfbox{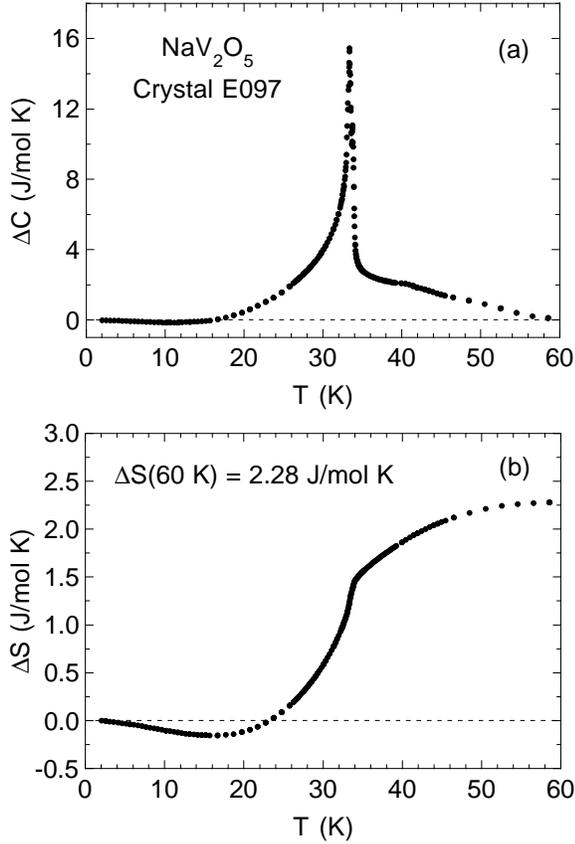}}
\vglue 0.1in
\caption{Temperature $T$ dependence of the change in the specific heat $\Delta C$ (a)
and in the entropy $\Delta S$ (b) in NaV$_2$O$_5$ crystal E097 due to the
transition(s) at $T_{\rm c}\approx 34$\,K as well as to crystallographic, magnetic
and charge order parameter fluctuations associated with this (these) transition(s). 
The occurrances of negative $\Delta C$ and $\Delta S$ values at low temperatures are
real effects due to loss of magnetic specific heat and magnetic entropy,
respectively, at these temperatures due to the opening of the spin gap at $T_{\rm
c}$.  By construction, \bbox{$\Delta C(T>60$\,K) = 0} and $\Delta S(T>60$\,K) =
constant.  The actual order parameter fluctuation effects likely extend to
temperatures higher than 60\,K\@.}
\label{NaV2O5E097DeltaC_S}
\end{figure}

\noindent and associated order parameter fluctuations not
occurred.  The difference $\Delta C$ between the measured $C_{\rm p}(T)$ and the
total background specific heat is plotted versus temperature in
Fig.~\ref{NaV2O5E097DeltaC_S}(a).  As would have been qualitatively anticipated,
$\Delta C$ is negative below about 16\,K due to the loss of magnetic
specific heat at low temperatures arising from the opening of the spin gap at $T_{\rm
c}$.  This negative $\Delta C$ does not arise from a problem in our polynomial
interpolation $C^{\rm lat}(T)$ fit function or from our $C(T)$ function; these
functions are both positive for all $T > 0$.  Since the magnetic background
contribution is proportional to $T$ and the lattice background contribution [which
is assumed not to change below 15\,K due to the occurrence of the transition(s)] is
proportional to $T^3$ at low $T$, opening a spin gap at $T_{\rm c}$ must necessarily
lead to a negative $\Delta C$ at sufficiently low temperatures since the magnetic
contribution then becomes exponentially small there.

\paragraph*{Step 4.}Finally, we can compute the change $\Delta S$ in the total
entropy of the system versus temperature due to the transition(s) and associated
order parameter fluctuations by integrating $\Delta C(T)$ from step~3 according to
$\Delta S(T) = \int_0^T [\Delta C(T)/T]\,dT$.  The result is shown in
Fig.~\ref{NaV2O5E097DeltaC_S}(b).  The entropy change is negative below about 22\,K,
due to the loss of magnetic entropy at low temperatures associated with the loss of
magnetic specific heat as just discussed.  From conservation of magnetic entropy,
this lost entropy must reappear at higher temperatures.

By construction, step~2 requires that $\Delta C(T > 60$\,K) =~0 and consequently
$\Delta S(T > 60$\,K) = const.  This requirement is not desirable, but we had to
enforce it to ensure that the $\Delta C(T)$ and $\Delta S(T)$ derived at lower
temperatures were accurate.  Since the effects of the order parameter fluctuations
are likely to continue to be present at temperatures higher than 60\,K, the $\Delta
C(T)$ and $\Delta S(T)$ at temperatures at and near 60\,K in
Fig.~\ref{NaV2O5E097DeltaC_S} are lower limits.

The net change in the entropy at 60\,K in Fig.~\ref{NaV2O5E097DeltaC_S}(b) due to the
occurrence of the transition(s) at $T_{\rm c}\approx 34$\,K and associated order
parameter fluctuations above and below $T_{\rm c}$ is $\Delta S(60$\,K) =
2.28\,J/mol\,K\@.  This is far larger than the maximum possible change $\Delta S_{\rm
mag}^{\rm max} = 0.556$\,J/mol\,K in the magnetic entropy at this temperature
obtained from Fig.~\ref{E097CmagNoXtion}, where this value is just the maximum
possible entropy of the magnetic subsystem at this temperature, confirming our
qualitative conclusion above based on very rough arguments.  In particular, our
quantitative analysis indicates that at least 76\% of the entropy change at 60\,K 
must arise from the lattice and charge degrees of freedom, and only a minor fraction
($< 24$\,\%) from the magnetic degrees of freedom.  Similarly, at $T_{\rm c} =
33.7$\,K, we obtain $\Delta S = 1.38$\,J/mol\,K and
$\Delta S_{\rm mag}^{\rm max} = 0.311$\,J/mol\,K, yielding $\Delta S_{\rm mag}^{\rm
max}/\Delta S \leq 23$\,\% at $T_{\rm c}$.

As a closing remark for this section, it is clear from Fig.~\ref{NaV2O5E097DeltaC_S}
and the discussion in the above two paragraphs that $\Delta C$ and $\Delta S$ do not
saturate to their respective high temperature limiting values until a temperature of
at least 60\,K is reached, which   is almost twice $T_{\rm c}$.  The present
analysis of the thermal behavior of ${\rm NaV_2O_5}$ thus lends strong support to
our independent  analysis and interpretation of our magnetic susceptibility data for
this compound in Sec.~\ref{SecNaV2O5Fits}.

\section{Summary and Concluding Discussion}
\label{SecSummary}

We have shown that the high-accuracy numerical Bethe ansatz calculations of the
magnetic susceptibility $\chi^*(t)$ for the $S=1/2$ uniform Heisenberg
chain by Kl\"umper and Johnston\cite{Klumper1998} are in excellent agreement with the
theory of Lukyanov\cite{Lukyanov1997} over 18 decades of temperature at low
temperatures. An independent  high precision empirical fit to these data was obtained
over 25 decades of temperature which we found useful to determine the accuracy of our
TMRG $\chi^*(t)$ calculations.  The magnetic specific heat data\cite{Klumper1998} for
the uniform chain at very low temperatures was also compared with the theoretical
predictions of Lukyanov, and extremely good agreement was found over many decades in
temperature.  We formulated an empirical fit function for these data which is highly
accurate over a temperature range spanning 25 orders of magnitude; the infinite
temperature entropy calculated using this fit function is within 8 parts in $10^8$
of the exact value.  We used both of the above fit functions to model our respective
experimental data for ${\rm NaV_2O_5}$ in later sections of the paper.  We expect
that they will be useful to other theorists and experimentalists as well.

We have carried out extensive QMC simulations and TMRG calculations of 
$\chi^*(\alpha,t)$ for the spin $S = 1/2$ antiferromagnetic alternating-exchange
Heisenberg chain for reduced temperatures $t
\equiv k_{\rm B}T/J_1$ from 0.002 to~10 and alternation parameters $\alpha \equiv
J_2/J_1$ from 0.05 to~1, where
$J_1~ (J_2)$ is the larger (smaller) of the two alternating exchange constants.  An
accurate global two-dimensional ($\alpha,t$) fit to these combined data was obtained,
constrained by the fitting parameters for the accurately known
$\chi^*(t)$ for the $\alpha$ parameter end points, the dimer ($\alpha = 0$) and the
uniform chain ($\alpha = 1$), resulting in an accurate fit function over the
entire range $0 \leq \alpha \leq 1$ of the alternation  parameter.  Our fit
function incorporates the first four terms of the exact high-temperature series
expansion in powers of $1/t$, which allows accurate extrapolation to arbitrarily
high temperatures.  This function should prove useful for many applications
including the modeling of experimental $\chi(T)$  data as we have shown.

Our $\chi^*(\alpha,t)$ fit function for the alternating chain can be easily
transformed (as we have done) into an equivalent fit function
$\overline{\chi^*}(\delta,\overline{t})$ in the two variables
$\delta \equiv (J_1 - J_2)/(2J)$ and $\overline{t}\equiv k_{\rm B}T/J$, where the
average exchange constant is $J = (J_1 + J_2)/2$.  This is a more appropriate
function for analyzing experimental $\chi(T)$ data for $S=1/2$ Heisenberg chain
compounds showing dimerization transitions (such as a spin-Peierls transition) which
result in an alternating-exchange chain with a small value of
$\delta$ at low temperatures.  Once $J$ has been determined by fitting our function for $\delta = 0$ to
the experimentally determined spin susceptibility $\chi^{\rm spin}(T)$ data above the
transition temperature, the alternation parameter $\delta$ is uniquely determined by
our fit function at each temperature below the transition temperature from the value
of $\chi^{\rm spin}$ at that temperature.  One can then find the spin gap
$\Delta(T)$ using an independently known $\overline{\Delta^*}(\delta)$.

Our QMC and TMRG data and fit for $\chi^*(\alpha,t)$ are in good agreement
with previous calculations based on exact diagonalization of the nearest neighbor
Heisenberg Hamiltonian for short chains with $\alpha = 0.2$, 0.4, 0.6, 0.7, and 0.8,
extrapolated to the thermodynamic limit, by Barnes and Riera.\cite{Barnes1994} 
However, the numerical and analytical theoretical predictions of
Bulaevskii,\cite{Bulaevskii1969} which have been used extensively in the past by
experimentalists to model their
$\chi(T)$ data for weakly-dimerized chain compounds, are found to be in poor
agreement with our results and should be abandoned for such use in favor of our fit
function.  Similarly, the previously used fit function\cite{Hatfield1981} for the
Bonner-Fisher calculation of $\chi^*(t)$ for the uniform chain ($\alpha = 1$)
should be replaced by one of our two fit functions for the most accurate calculation
to date\cite{Klumper1998} of $\chi^*(t)$ for the uniform chain.

An important theoretical issue in the study of the alternating exchange chain is how
the spin gap $\overline{\Delta^*}(\delta)$ evolves as the uniform chain
limit is approached ($\delta\to 0,\ \alpha\to 1$).  We formulated a fit function for
the temperature dependence of our TMRG susceptibility $\chi^*(\alpha,t)$ calculations
at low temperatures, which was used to extract the dependence
$\overline{\Delta^*}(\delta)$ in this regime.  We find that the asymptotic critical
regime is not entered until, at least, $\delta \lesssim 0.005$ ($\alpha\gtrsim
0.99$).  We compared our spin gap data with many literature data.  We
formulated a fit function for our spin gap data together with those of Barnes, Riera,
and Tennant\cite{Barnes1998} which quite accurately covers the entire range $0\leq
\delta\leq 1$.

In the remainder of this paper, we showed how the above theoretical results could be
used to obtain detailed information about real systems.  As a specific illustration,
we carried out a detailed analysis of our experimental $\chi(T)$ and specific
heat $C_{\rm p}(T)$ data for ${\rm NaV_2O_5}$ crystals. This compound shows a
transition to a spin dimerized state below the transition temperature $T_{\rm
c}\approx 34$\,K\@.   We used one of our two $\chi^*(t)$ fit functions for the
uniform Heisenberg chain to model the $\chi(T)$ above $T_{\rm c}$, where we found
that the experimental $\chi(T)$ is not in quantitative agreement with the prediction
for the uniform Heisenberg chain.  A model incorporating a mean-field ferromagnetic
interchain coupling between quantum $S = 1/2$ Heisenberg chains fits the experimental
data very well with reasonable parameters.  It remains to be seen whether the
inelastic neutron scattering measurements of the magnon dispersion
relations\cite{Yosihama1998} are consistent with our derived intrachain and
interchain exchange constants.

In an alternate description, we modeled the deviation in the measured $\chi(T)$ of
NaV$_2$O$_5$ above 60\,K $> T_{\rm c}$ from the Heisenberg chain model (with fixed
exchange constant $J$) as due to a temperature-dependent $J$.  We found that this 
$J$ decreases with increasing $T$ up to 300\,K in a manner very similar to $J_{\rm
eff}(T)$  predicted by Sandvik, Singh and Campbell\cite{Sandvik1997} and K\"uhne and
L\"ow\cite{Kuhne1999} for the spin-Peierls chain.  Our $J(T)$ cannot however be
compared directly with their $J_{\rm eff}(T)$ because the two quantities are defined
differently.  They found that by defining an appropriate effective exchange constant
$J_{\rm eff}$, their resulting susceptibility $\chi(k_{\rm B}T/J_{\rm eff})$
is universal at the higher temperatures for various Einstein phonon
frequencies and spin-phonon coupling constants.  This function agrees well with the
$\chi(k_{\rm B}T/J)$ for the $S = 1/2$ AF uniform Heisenberg chain at these
temperatures.  As we discussed, these $\chi(T)$ calculations are not applicable to
NaV$_2$O$_5$, possibly because the calculations do not incorporate realistic phonon
spectra.

Below $T_{\rm c}$, we used the $J(T)$ extrapolated from above
60\,K and our global $\chi^*(\alpha,t)$ fit function for the alternating
Heisenberg chain to determine the temperature-dependent alternation parameter
$\delta(T)$, and then the spin gap $\Delta(T)$ from $\delta(T)$, directly from the
$\chi(T)$ data.  We find that the $\Delta(0)/k_{\rm B}$ values for nine single
crystals of NaV$_2$O$_5$ are in the range 103(2)\,K\@.  This result is in
agreement, within the errors, with many previous analyses of data from various types
of measurements for this compound by other groups.  However, our values of
$\delta(0) = 0.034(6)$ for various crystals are significantly smaller than previous
estimates.  We note that the two estimates with $\delta(0) \approx 0.1$ in
Table~\ref{TabLitDat} were obtained using Bulaevskii's theory\cite{Bulaevskii1969}
for the alternating-exchange chain, which we have shown is not accurate at low
temperatures in the relevant alternation parameter range.

The dispersion of two one-magnon branches perpendicular to the chains
observed in the neutron scattering measurements has been recently explained
quantitatively by Gros and Valenti assuming that a zig-zag charge ordering
transition occurs at $T_{\rm c}$.\cite{Gros1999}  They also predict that
$\delta(0)\sim 0.034$.  This is within our range of $\delta(0)$ values in spite of the
fact that we assumed that $J(T)$ is either constant or increases slightly with
decreasing $T$ below $T_{\rm c}$, contrary to their prediction that $J$ decreases
below $T_{\rm c}$.  Gros and Valenti made no predictions for $\chi(T)$, $\delta(T)$,
$\Delta(T)$ or $C(T)$, so comparisons with our results for these quantities are not
possible.  We note that Kl\"umper, Raupach, and Sch\"onfeld\cite{Klumper1998b}
obtained a good fit to the $\chi(T)$ data below $T_{\rm c}$ for the spin-Peierls
compound CuGeO$_3$ within the context of a spin-Peierls model containing frustrating
second-neighbor interactions and static spin-phonon coupling.

We discovered that $\Delta(T)$ [and $\delta(T)$] of NaV$_2$O$_5$ does not go to zero
at $T_{\rm c}$, indicating the existence of a spin pseudogap above $T_{\rm c}$ with
a large magnitude just above $T_{\rm c}$ of $\approx 40$\% of $\Delta(0)$; the
pseudogap is present up to at least 50\,K with a magnitude decreasing with
increasing $T$ above $T_{\rm c}$.  To our knowledge, this pseudogap has not been
reported  previously, and there are as yet no theoretical predictions for the
magnitude or temperature dependence of this pseudogap.  The pseudogap is strongly
reminiscent of the spin pseudogap derived by one of us using $\chi(T)$ measurements
above the transition temperature of inorganic quasi-one-dimensional charge density
wave compounds,\cite{Johnston1984} as predicted theoretically by Lee, Rice, and
Anderson\cite{Lee1973} long before those observations were made.  Similar to that
case, in the present system one may think of the pseudogap as the rms fluctuation in
the spin gap above $T_{\rm c}$, with an associated reduction in the magnon density of
states at low energy.  In this interpretation, the pseudogap in NaV$_2$O$_5$ should be
observable in high resolution quasielastic neutron scattering and other
spectroscopic measurements probing the low energy magnetic excitations.

Finally, we carried out an extensive modeling study of our specific heat data for 
NaV$_2$O$_5$ crystals, using the same model that we used to analyze our
susceptibility data.  The most important part of this study is that we have been
able to determine a limit on the relative contributions of the magnetic and
lattice/charge degrees of freedom to the entropy associated with
the transition(s) at $T_{\rm c}$.  We find that at least 77\,\% of the change in the
entropy at $T_{\rm c}$ must arise from the lattice and/or charge degrees of freedom,
to which the spin degrees of freedom must of course be coupled, and that the spin
degrees of freedom themselves contribute less than 23\,\% of this entropy change. 
Our results also indicate that order parameter fluctuation effects are important in
the specific heat up to at least 60\,K, strongly confirming the above similar and
independent conclusion based on our modeling of our magnetic susceptibility data for
the same crystals.

\section*{Acknowledgments}

We thank E. Br\"ucher and C. Lin for help with the sample preparation, and C. Song
for assistance with Laue x-ray diffraction measurements.  We are grateful to S.
Eggert and T. Barnes for providing the numerical
$\chi^*(t)$ calculation results for the uniform chain in
Ref.~\onlinecite{Eggert1994} and the alternating chain in
Ref.~\onlinecite{Barnes1994}, respectively, and to D.~Poilblanc and G.~S.~Uhrig for
sending us their $\overline{\Delta^*}(\delta)$ data in Refs.~\onlinecite{Augier1997}
and~\onlinecite{Uhrig1999}, respectively.  We are grateful to M.~Greven and X.~Zotos
for helpful discussions, and to A.~A.~Zvyagin for helpful correspondence.  One of us
(D.C.J.) thanks the Max-Planck-Institut f\"ur Festk\"orperforschung, Stuttgart, where
this work was started, for kind hospitality.  Ames Laboratory is operated for the U.S.
Department of Energy by Iowa State University under Contract No.\ W-7405-Eng-82.  The
work at Ames was supported by the Director for Energy Research, Office of Basic Energy
Sciences.  The QMC program was written in C++ using a parallelizing Monte Carlo
library developed by one of the authors.\cite{Troyer1998}  The QMC simulations by
M.T. were performed on the Hitachi SR2201 massively parallel computer of the
University of Tokyo and on the IBM SP-2 of the Competence Center for Computational
Chemistry of ETH Z\"urich.  X.W. acknowledges Swiss National Funding Grant No.\
20-49486.96.   A.K. acknowledges financial support by the {\it Deutsche
Forschungsgemeinschaft} under Grant No.~Kl~645/3 and support by the research program
of the Sonderforschungsbereich 341, K\"oln-Aachen-J\"ulich.
\vglue-0.1in

\end{document}